\documentclass[fleqn,usenatbib]{mnras}

\usepackage{newtxtext,newtxmath}

\usepackage[T1]{fontenc}

\DeclareRobustCommand{\VAN}[3]{#2}
\let\VANthebibliography\thebibliography
\def\thebibliography{\DeclareRobustCommand{\VAN}[3]{##3}\VANthebibliography}

\usepackage{graphicx}	
\usepackage{deluxetable}
\usepackage{tabularx}
\usepackage{lscape}
\usepackage{rotating}
\usepackage{multirow}
\usepackage{subcaption}
\usepackage{longtable}

\graphicspath{ {./figures/} }

\title[QPOs from a ULX in M\,101]{Quasi periodic whispers from a transient ULX in M\,101: signatures of a fast-spinning neutron star?}

\author[R.~T.~Urquhart et al.]{
Ryan T.~Urquhart$^{1}$\thanks{E-mail: 
urquha20@msu.edu},
Roberto Soria$^{2,3}$,
Rosanne Di Stefano$^4$,
Kaiming Cui$^5$,
Paolo Esposito$^{6,7}$,\newauthor
Gian Luca Israel$^8$,
Sammarth Kumar$^9$,
Sara Motta$^{10}$,
Fabio Pintore$^{11,7}$,
Giacomo Riva$^{12}$
\\
$^{1}$Center for Data Intensive and Time Domain Astronomy, Department of Physics and Astronomy, Michigan State University, East Lansing, MI 48824, USA\\
$^{2}$College of Astronomy and Space Sciences, University of the Chinese Academy of Sciences, Beijing 100049, China\\ 
$^{3}$Sydney Institute for Astronomy, School of Physics A28, The University of Sydney, Sydney, NSW 2006, Australia\\
$^{4}$Institute for Theory and Computation, Harvard--Smithsonian Center for Astrophysics, 60 Garden Street, Cambridge, MA 02138, USA\\
$^{5}$Tsung-Dao Lee Institute, Shanghai Jiao Tong University, 800 Dongchuan Road, Shanghai 200240, People’s Republic of China\\
$^6$Scuola Universitaria Superiore IUSS Pavia, Palazzo del Broletto, piazza della Vittoria 15, I-27100 Pavia, Italy\\ 
$^{7}$INAF--Istituto di Astrofisica Spaziale e Fisica Cosmica di Milano, via A. Corti 12, I-20133 Milano, Italy\\
$^8$INAF--Osservatorio Astronomico di Roma, via Frascati 33, I-00078 Monteporzio Catone, Italy\\
$^{9}$ Step By Step School, 201303, Noida, India \\
$^{10}$INAF--Osservatorio Astronomico di Brera
via E.~Bianchi 46, I-23807 Merate, Italy \\
$^{11}$INAF--Istituto di Astrofisica Spaziale e Fisica Cosmica di Palermo, via U. La Malfa 153, I-90146 Palermo, Italy\\
$^{12}$Dipartimento di Fisica, Università degli Studi di Milano, via G. Celoria 16, I-20133 Milano, Italy
}

\date{Accepted 2022 January 17. Received 2022 January 17; in original form 2021 October 29}

\pubyear{2021}

\begin{document}
\label{firstpage}
\pagerange{\pageref{firstpage}--\pageref{lastpage}}
\maketitle

\begin{abstract}
We have studied the unusual time variability of an ultraluminous X-ray source in M\,101, 4XMM\,J140314.2$+$541806 (henceforth, J1403), using {\it Chandra} and {\it XMM-Newton} data. Over the last two decades, J1403 has shown short-duration outbursts with an X-ray luminosity $\sim$1--3 $\times 10^{39}$ erg s$^{-1}$, and longer intervals at luminosities $\sim$0.5--1 $\times 10^{38}$ erg s$^{-1}$. The bimodal behaviour and fast outburst evolution (sometimes only a few days) are more consistent with an accretor/propeller scenario for a neutron star than with the canonical outburst cycles of stellar-mass black holes. If this scenario is correct, the luminosities in the accretor and propeller states suggest a fast spin ($P \approx$ 5 ms) and a low surface magnetic field ($B \sim 10^{10}$ G), despite our identification of J1403 as a high-mass X-ray binary. The most striking property of J1403 is the presence of strong $\sim$600-s quasi periodic oscillations (QPOs), mostly around  frequencies of $\approx$1.3--1.8 mHz, found at several epochs during the ultraluminous regime. We illustrate the properties of such QPOs, in particular their frequency and amplitude changes between and within observations, with a variety of techniques (Fast Fourier Transforms, Lomb-Scargle periodograms, weighted wavelet Z-transform analysis). The QPO frequency range $<$10 mHz is an almost unexplored regime in X-ray binaries and ultraluminous X-ray sources. We compare our findings with the (few) examples of very low frequency variability found in other accreting sources, and discuss possible explanations (Lense-Thirring precession of the inner flow or outflow; radiation pressure limit-cycle instability; marginally stable He burning on the neutron star surface).
\end{abstract}




\section{Introduction}


Short-term time variability in luminous X-ray binaries (in particular those that at some point reach or exceed their Eddington luminosity) is still largely unexplained or unexplored. We cannot assume that we have already discovered all possible types of time variability from the well-studied but limited sample of X-ray binaries in the Milky Way and Local Group. X-ray population studies in nearby galaxies have revealed the existence of entire classes of (rare) sources not present or not yet observed in the Milky Way: for example super-Eddington neutron stars (NSs), usually referred to as pulsar ultraluminous X-ray sources (PULXs), that reach luminosities $\sim$10$^{40}$--$10^{41}$ erg s$^{-1}$ \citep{bachetti14,israel17a,israel17b,sathyaprakash19}, or supersoft ULXs with characteristic temperatures $\sim$0.1 keV and luminosities of a few $10^{39}$ erg s$^{-1}$ \citep{urquhart16}. Thus, it is not surprising to discover also new variability behaviours. 

Even the nearest galaxies outside the Local Group have been observed only a handful of times by {\it Chandra} and {\it XMM-Newton}: this hinders the study of state transitions and duty cycles and the search for X-ray eclipses; in addition, the low count rate usually limits the study of short-term variability and quasi periodic oscillations (QPOs). Among the nearest large galaxies with an abundant population of X-ray binaries, M\,101 (median Cepheid distance of 6.9 Mpc, from the NASA Extragalactic Database) is one of the targets with the largest number of archival observations (about 30 between {\it{Chandra}} and {\it XMM-Newton}) and is therefore one of the best places to look for such investigations.

\begin{figure}
    \centering
    \includegraphics[width=0.48\textwidth]{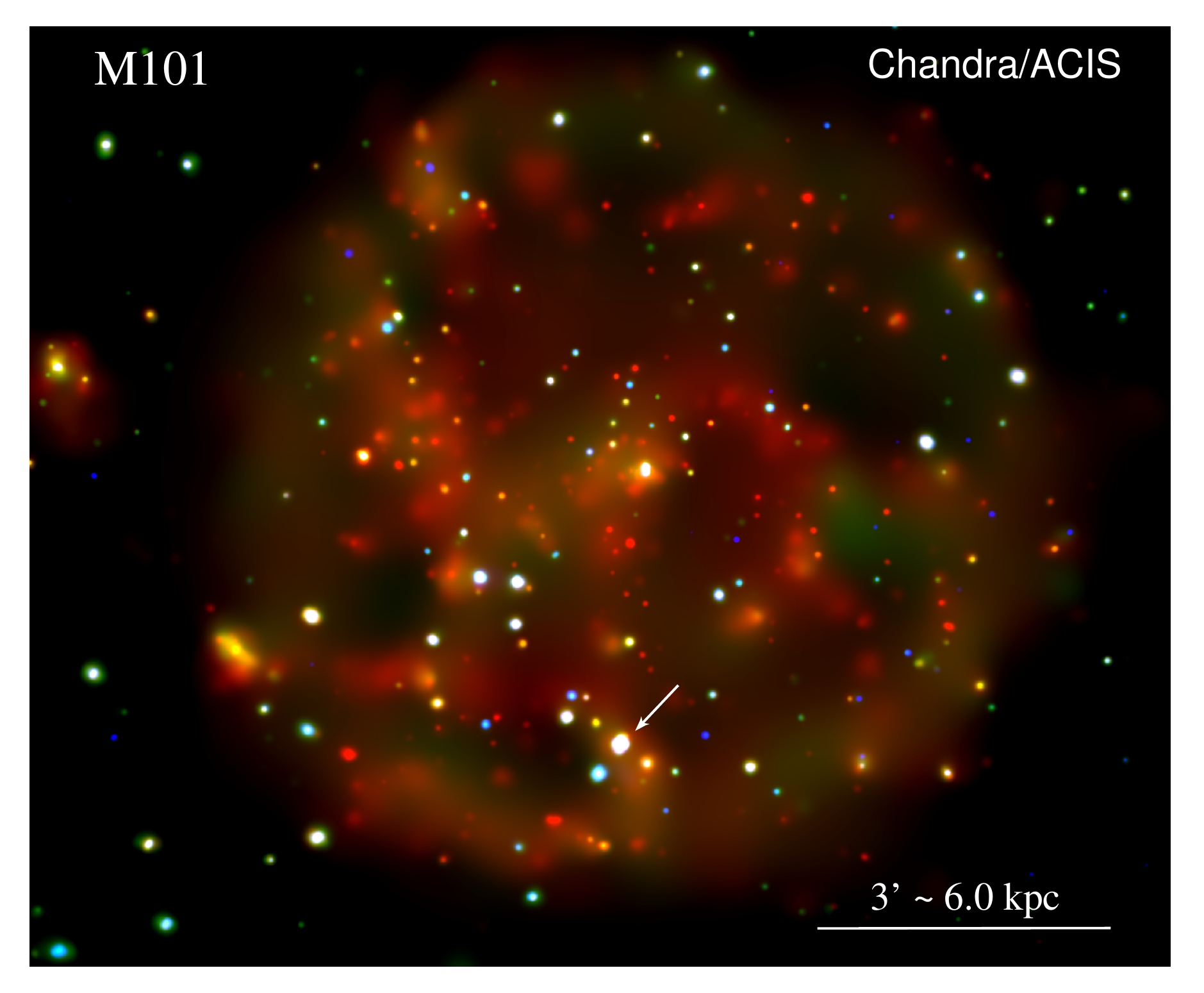}
    
    \caption{True-colour X-ray image of M\,101, obtained by aligning and stacking all the available {\it Chandra}/ACIS observations of this galaxy. Colours are: red for 0.3--1 keV; green for 1--2 keV; blue for 2--7 keV. The individual colour images were adaptively smoothed with the {\sc ciao} task {\it csmooth}. North is up, east is to the left. The target of our study, J1403, is marked with a small white arrow.} 
    \label{fig:location}
\end{figure}

One of the X-ray sources with the highest peak luminous in M\,101, 2CXO\,J140314.3+541806 \citep{vizier:IX/57,evans10} = 4XMM\,J140314.2$+$541806 \citep{webb20,traulsen20}, 
henceforth, J1403 (Figure \ref{fig:location}), is an obvious candidate for such studies. Although previously recognized as a transient ULX \citep{2014MNRAS.442.1054H,wang16,2020MNRAS.497..917L}, J1403 has received little attention so far in terms of individual X-ray studies.
Simple flux estimates from a long sequence of {\it Chandra} and {\it XMM-Newton} observations between 2000 and 2017 (Section 2.1 for details) show that J1403 was in the ULX regime at $L_{\rm X} \sim$ a few $10^{39}$ erg s$^{-1}$ some of the times, and was barely detected at $L_{\rm X} \sim$ a few $10^{37}$ erg s$^{-1}$ on other occasions. 
Furthermore, we noticed that its short-term variability properties are even more remarkable than its state transitions. In this study, we will illustrate the presence of strong quasi-periodic oscillations (QPOs) on a characteristic timescale of $\sim$600 s ($\sim$1.7 mHz) in several of the epochs in which J1403 was in the ultraluminous regime. 

The origin of so-called very-low-frequency QPOs in X-ray binaries and in particular in ULXs is still unexplained, with several alternative possibilities (see \citealt{ingram19} for a review). Milli-Hz frequencies are too low to be consistent with Keplerian rotation from the region responsible for the X-ray emission (inner part of the accretion disk and/or hot spots on the surface of the NS). Instead, they might be associated with the Lense--Thirring precession of the inner disk \citep{Mottalt2018MNRAS.473..431M}, or of the outflow \citep{middleton18}. 
An alternative explanation for milli-Hz QPOs is repeated episodes of marginally-steady He burning on the surface of the NS in low-mass X-ray binaries \citep{heger07,tse21}; however, such oscillations are expected to have a characteristic frequency $\sim$10 milli-Hz, and indeed the frequencies observed in the most promising candidates are $\sim$3--15 mHz \citep{tse21}.
Some NS high mass X-ray binaries exhibit milli-Hz QPOs, too, but typically at frequencies $\gtrsim$10 mHz \citep{james10}.
So far, only two NS high mass X-ray binaries have shown QPOs with similar frequency to J1403. In LMC X-4, a QPO with a frequency varying between $\approx$0.65--1.35 mHz was seen \citep{moon01} in correspondence of high-luminosity (likely super-Eddington) flares; this was interpreted as beating between the spin frequency of the NS and the Keplerian frequency of accreting clumps near the corotation radius. In IGR J19140$+$0951, a 1.46-mHz QPO was detected by \citet{sidoli16}; they suggested that it could be produced by large-scale convective motion of a hot shell of gas that accumulates just outside the magnetospheric radius (``settling accretion model'', \citealt{shakura12}). However, IGR J19140$+$0951 was in a low-luminosity state ($L_{\rm X} \sim 10^{34}$ erg s$^{-1}$), much different from the ULX regime of J1403.


In this paper, we present the observational results of our study of J1403, based on the {\it Chandra} and {\it XMM-Newton} data analysis. We show how the source switches between high and low states, and discuss whether its luminosity evolution is more consistent with canonical outbursts of stellar-mass black holes (BHs) \citep{fender04,remillard06,fender16} or with accretor/propeller transitions in NSs and PULXs \citep{corbet96,campana02,tsygankov16,campana18}. 
We use a variety of techniques to show that the QPOs are significant and to measure their frequencies. In particular, we use wavelet decomposition to build dynamical power spectra and show how the oscillating components change in amplitude and frequency during individual observations. 
We also derive the spin period and magnetic field that the NS in J1403 would have if the accretion/propeller scenario is applicable. We place those estimates in the context of younger and older NS populations. Finally, we refine the astrometric position and identify the optical counterpart in {\it Hubble Space Telescope} images. 




\section{Data Analysis}

\begin{table}
    \caption{Log of the {\it Chandra} observations}
    \centering
    \begin{tabular}{rccc}
        \hline                                      \\[-5pt]
        ObsID & Observation Epoch$^a$   & Exp.~Time & Count Rate$^b$\\
              &                     & (ks)     & $\left(10^{-3}\right.$ ct s$\left.^{-1}\right)$     \\[5pt]
        \hline                                      \\[-5pt]
        934   & 2000-03-26 00:20:55 & 98.38 & $1.6^{+0.3}_{-0.2}$ \\[3pt]
        2065  & 2000-10-29 08:05:35 & 9.63   & $1.5^{+0.8}_{-0.6}$  \\[3pt]
        4731  & 2004-01-19 05:57:05 & 56.24  &  $1.6^{+0.4}_{-0.4}$ \\[3pt]
        5296  & 2004-01-21 10:57:22 & 3.09   &  $5.1^{+2.8}_{-2.0}$\\[3pt]
        5297  & 2004-01-24 01:34:02 & 21.69  &  $18.1^{+1.6}_{-1.6}$\\[3pt]
        5300  & 2004-03-07 09:29:06 & 52.09  & $<0.2$\\[3pt]
        5309  & 2004-03-14 01:41:21 & 70.77   & $54.3^{+1.5}_{-1.5}$\\[3pt]
        4732  & 2004-03-19 08:15:25 & 69.79  & $54.9^{+1.5}_{-1.5}$\\[3pt]
        5322  & 2004-05-03 07:44:34 & 64.70  & $0.1^{+0.1}_{-0.1}$\\[3pt]
        4733  & 2004-05-07 13:29:38 & 24.81  & $0.7^{+0.4}_{-0.3}$\\[3pt]
        5323  & 2004-05-09 03:11:31 & 42.62   & $0.3^{+0.2}_{-0.2} $\\[3pt]
        5337  & 2004-07-05 18:42:37 & 9.94   & $<0.5$  \\[3pt]
        5338  & 2004-07-06 13:27:55 & 28.57   & $0.4^{+0.2}_{-0.2} $\\[3pt]
        5339  & 2004-07-07 13:24:00 & 14.32   & $0.2^{+0.3}_{-0.2} $\\[3pt]
        5340  & 2004-07-08 11:07:26 & 54.42   & $0.6^{+0.2}_{-0.1}$\\[3pt]
        4734  & 2004-07-11 01:43:17 & 35.48   & $0.3^{+0.2}_{-0.1} $\\[3pt]
        6114  & 2004-09-05 18:23:48 & 66.20   & $0.9^{+0.2}_{-0.2} $\\[3pt]
        6115  & 2004-09-08 18:16:36 & 35.76   & $0.8^{+0.3}_{-0.3} $\\[3pt]
        6118  & 2004-09-11 15:15:08 & 11.46   & $0.4^{+0.5}_{-0.3} $\\[3pt]
        4735  & 2004-09-12 12:59:30 & 28.78   & $0.4^{+0.3}_{-0.2} $\\[3pt]
        4736  & 2004-11-01 18:43:08 & 77.35   & $0.4^{+0.2}_{-0.1} $\\[3pt]
        6152  & 2004-11-07 07:03:53 & 44.09   & $0.5^{+0.2}_{-0.2} $\\[3pt]
        6170  & 2004-12-22 01:17:57 & 47.95   & $22.3^{+1.2}_{-1.2}$\\[3pt]
        6175  & 2004-12-24 16:38:17 & 40.66   & $43.1^{+1.8}_{-1.9}$\\[3pt]
        6169  & 2004-12-30 02:10:50 & 29.38   & $32.4^{+1.8}_{-1.8}$\\[3pt]
        4737  & 2005-01-01 14:30:53 & 21.85   & $14.5^{+1.4}_{-1.4}$\\[3pt]
        14341 & 2011-08-27 10:37:29 & 49.09   & $0.4^{+0.1}_{-0.2}$\\[3pt]
        19304 & 2017-11-08 20:42:03 & 36.57   & $0.1^{+0.2}_{-0.1}$\\[3pt]
        \hline
    \end{tabular}
    \label{tab:chandra-obs}
    \begin{flushleft}
    $^a$: time at the start of the observation;\\
    $^b$: net ACIS-S3 count rate in the 0.5--7 keV band, computed with {\it srcflux}, with an aperture correction to infinity. It is the observed rate, not corrected for the change in detector sensitivity over the years. 
    \end{flushleft}
\end{table}

\begin{table}
    \caption{Log of the {\it XMM-Newton} observations used in this work}
    \centering
    \begin{tabular}{rccc}
        \hline                                           \\[-5pt]
        ObsID      & Observation Epoch$^a$   & Exp.~Time & Count Rate$^b$\\
                   &                     & (ks) & $\left(10^{-3}\right.$ ct s$\left.^{-1}\right)$         \\[5pt]
        \hline                                           \\[-5pt]
        0104260101 & 2002-06-04 02:06:57 & 42.3 & $90\pm2$ (pn)$^c$\\
        & &   & $42\pm1$ (MOS)$^c$        \\
        0164560701 & 2004-07-23 08:51:10 & 32.5 &  $44\pm1$ (pn)$^d$\\
        & & & $21\pm1$ (MOS)$^d$       \\
        \hline
    \end{tabular}
    \label{tab:xmm-obs}
    \begin{flushleft}
    $^a$: time at the start of the observation;\\
    $^b$: net EPIC pn and MOS1$+$MOS2 count rate in the 0.3--12 keV band, in the source extraction circle (no aperture correction or vignetting correction);\\ 
    $^c$: within a source extraction circle of 16$^{\prime\prime}$ radius;\\ 
    $^d$: within a source extraction circle of 13$^{\prime\prime}$ radius.
    \end{flushleft}
\end{table}

\subsection{{\it Chandra X-Ray Observatory}}
M\,101 has been observed 28 times with \textit{Chandra X-Ray Observatory}'s Advanced CCD Imaging Spectrometer (ACIS). The observation log is listed in Table 1; note that the ObsID numbering bears little relation to the actual time sequence. We downloaded the data from the public archives, then reprocessed and analysed them with the Chandra Interactive Analysis of Observations ({\sc ciao}) software version 4.12 \citep{2006SPIE.6270E..1VF}, Calibration Database 4.9.1. New level-2 event files and aspect solution files were created with the {\it ciao} tasks {\it chandra\_repro}, followed by  {\it reproject\_obs}. We inspected each image with {\sc ds9} \citep{ds9} and used {\it srcflux} to identify those in which J1403 was significantly detected at the 90\% confidence level (using the Bayesian statistics of \citealt{1991ApJ...374..344K}). We found it in 26 observations. We used {\it dmextract} to build background-subtracted light-curves (time resolution of 3.24 s). When there were sufficient counts for spectral analysis, we 
extracted spectra and associated response and ancillary response files with {\it specextract} (with the options ``weight = no'' and ``correctpsf = yes'', for point-like sources). For each {\it srcflux}, {\it dmextract} and {\it specextract} analysis, we defined a source region with suitable size and shape, depending on the properties of the point spread function at that location for that particular observation. For example, for on-axis observations of the source, the extraction circle had a radius of $2\farcs5$. Instead, for epochs in which the source was located several arcmin off-axis, at the edge of the S3 chip, we used a $5^{\prime\prime} \times 7^{\prime\prime}$ source ellipse. Local background regions were at least 4 times the size of the source regions.

\subsection{{\it {XMM-Newton}}}
M\,101 has been observed with \textit{XMM-Newton} four times. Publicly available European Photon Imaging Camera (EPIC) observations were downloaded from the High Energy Astrophysics Science Archive Research Center (HEASARC) data archive. We only used two observations (0104260101 and 0164560701: Table 2), because in the other two observations, the target ULX either fell on a chip gap or was off the field of view. The data were reduced using standard tasks within the Science Analysis System ({\sc {sas}}) version 18.0.0. Intervals of strong background flaring were filtered out. The source was extracted using a circular region with a radius of 16$^{\prime\prime}$ for 0104260101 and 13$^{\prime\prime}$ for 0164560701; the local background was at least three times larger than the source regions. The small size of the source extraction circles was due to the locations of chip gaps and the need to avoid contamination from nearby (variable) point-like sources which are detected in the {\it Chandra} images. We used the standard flagging criteria of \verb|#XMMEA_EP| plus \verb|FLAG = 0| for the pn camera and \verb|#XMMEA_EM| for the MOS1 and MOS2 cameras. Additionally, we selected patterns 0--4 and 0--12 for pn and MOS, respectively. The task \textit{xmmselect} was used to extract individual spectra and corresponding response and ancillary response functions for pn, MOS1 and MOS2. The individual spectra were inspected to verify consistency before being combined into a single EPIC spectrum using the \textit{epicspeccombine} task. 
For our timing analysis, we first performed barycentric corrections using the {\sc {sas}} task \textit{barycen}. Next, pn, MOS1 and MOS2 background-subtracted light curves with the {\sc {sas}} tasks \textit{evselect} and \textit{epiclccorr} before finally combining them into a single EPIC light curve using (the {\sc{ftools}} package \textit{lcmath};  \citealt{1995ASPC...77..367B}). The time resolution of the EPIC light curve is 2.6\,s.

\begin{figure}
    \centering
    \includegraphics[width=0.48\textwidth]{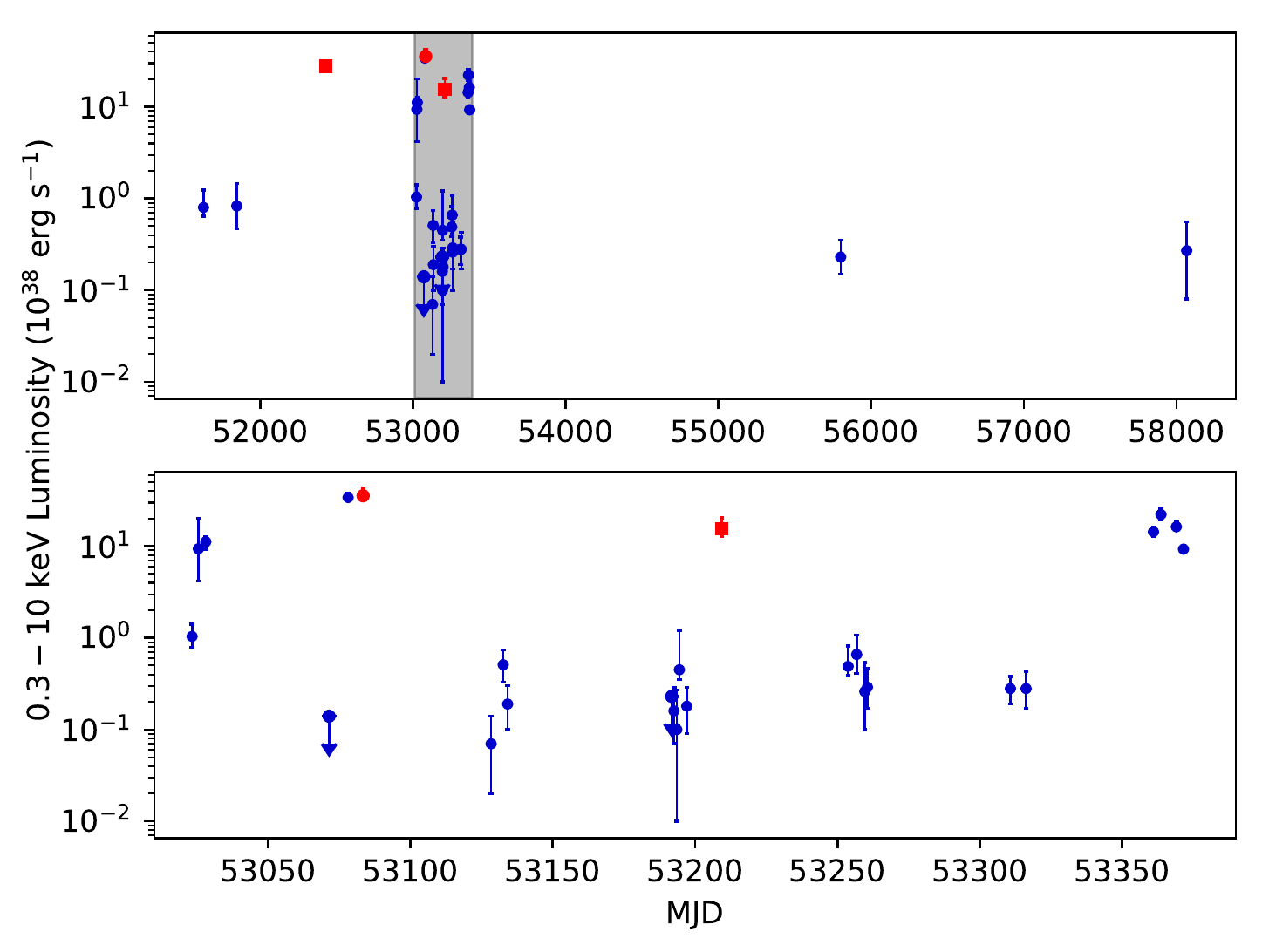}
    \caption{Long term light-curve derived from 28 \textit{Chandra} observations (circles) and 2 \textit{XMM-Newton} observations (squares). The datapoints represent the average unabsorbed 0.3--10 keV luminosity during each observation. Such values were obtained from {\sc xspec} spectral modelling whenever possible (as explained in Section 2.3) or with {\it srcflux} for the observations with few counts. Red data points indicate observations in which strong QPOs are detected (in chronological order: ObsIDs 0104260101, 4732, 0164560701). The grey-shaded time interval in the upper panel is displayed zoomed-in in the lower panel.}
    \label{fig:full_lc}
\end{figure}

\subsection{X-ray spectral analysis}
We used {\sc{xspec}} version 12.11.1 \citep{1996ASPC..101...17A} to perform the spectral fitting, over the 0.3--10 keV range for the {\it XMM-Newton} spectra and the 0.3--7 keV range for {\it Chandra} spectra. The two combined {\it XMM-Newton}/EPIC spectra, and seven {\it Chandra}/ACIS spectra (ObsIDs 4732, 4737, 5297, 5309, 6169, 6170, 6175) with $\gtrsim$300 counts, were grouped to a minimum of 20 counts per bin so that we could use Gaussian statistics. Another seven {\it Chandra}/ACIS spectra (ObsIDs 934, 2065, 4731, 5296, 5340, 6114, 6115) have enough counts for a simple power-law fitting with the Cash statistics \citep{1979ApJ...228..939C}, for which the data were rebinned to 1 count per bin. The absorbed and de-absorbed fluxes of those 16 observations were computed with the {\it cflux} task in {\sc xspec}. The remaining 14 {\it Chandra} observations do not have enough counts for {\sc xspec} fitting: the fluxes of J1403 in those datasets were estimated directly from the filtered event files, with the {\sc ciao} task {\it srcflux} (Table 1).

\begin{figure}
    \centering
    \includegraphics[width=0.48\textwidth]{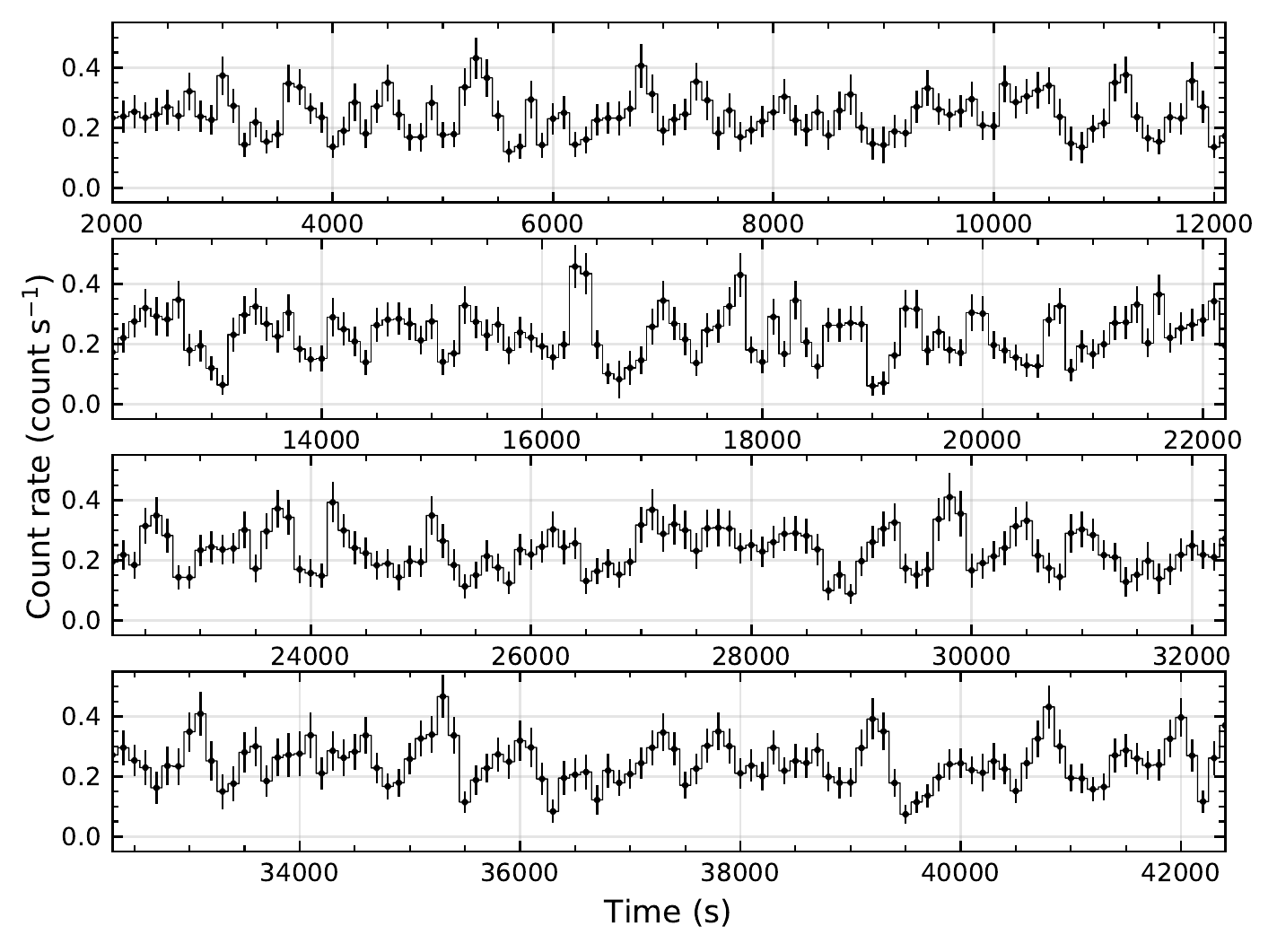}\\
    \includegraphics[width=0.48\textwidth]{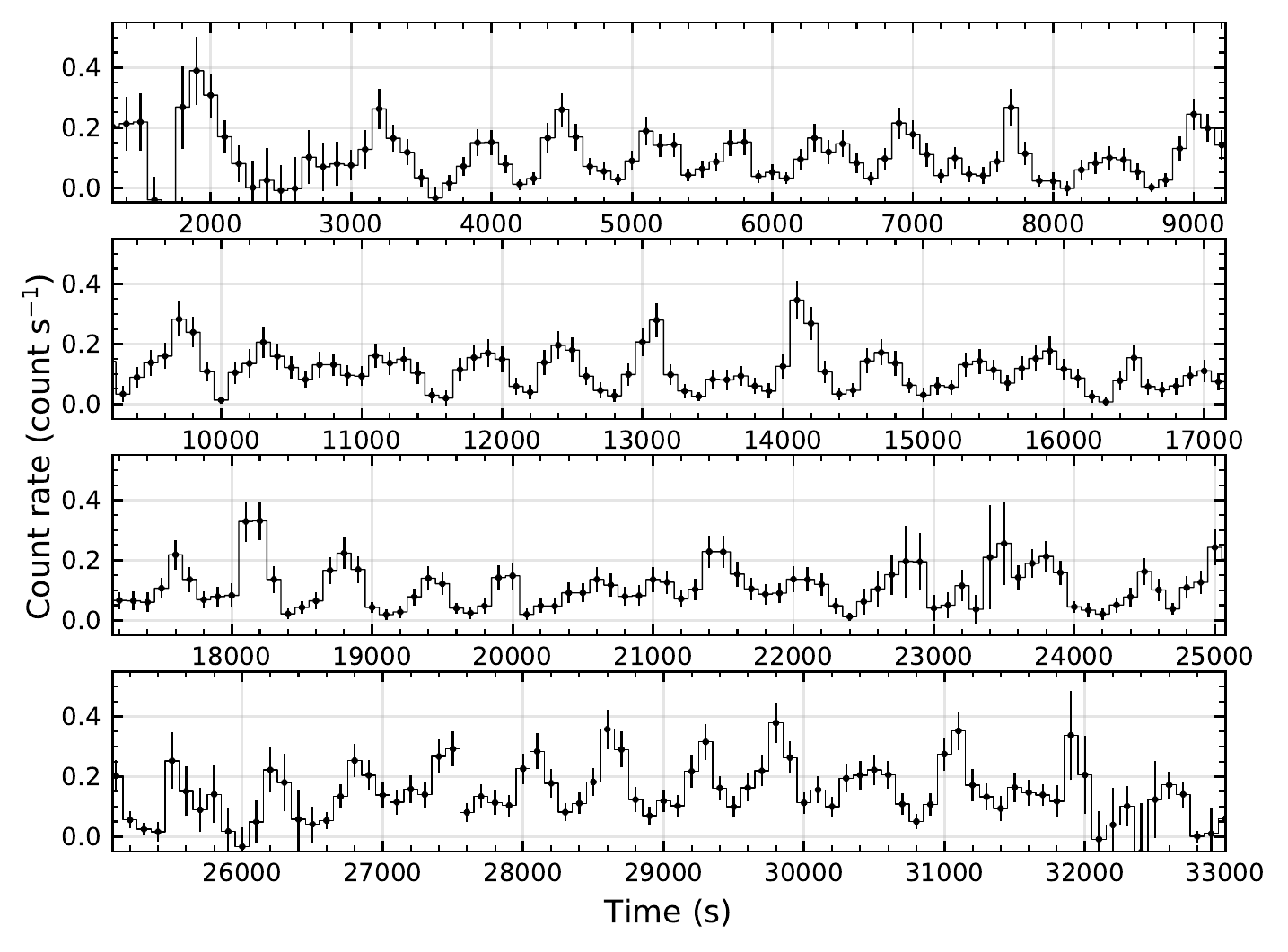}
    \caption{\textit{XMM-Newton}/EPIC background-subtracted light-curves from the observations in 2002 (top; ObsID 010426101) and 2004 (bottom; ObsID 0164560701), binned to 100 s.}
    \label{fig:xmm_lightcurves}
\end{figure}

\subsection{X-ray time-frequency analysis}

First, we did a preliminary timing analysis with basic {\sc{ftools}} tasks such as {\it lcurve} and {\it powspec} to inspect the data. Then, we used the Lomb--Scargle method to identify possible significant QPO peaks. The Lomb--Scargle periodogram \citep[][]{Lomb1976Ap&SS..39..447L,Scargle1982ApJ...263..835S} is a commonly used statistical tool designed to detect periodic signals in unevenly spaced observations. We calculated the Lomb--Scargle periodograms and their corresponding false alarm probabilities based on an implementation of {\sc{Astropy}} \citep[][]{astropy:2013,astropy:2018}. 
We did so for the two {\it XMM-Newton} observations and the seven {\it Chandra} observations with higher signal-to-noise (ObsIDs 4732, 4737, 5297, 5309, 6169, 6170, 6175). 


In order to look for possible variations of a QPO frequency within individual observations, we also computed dynamical power spectra with the weighted wavelet Z-transform (WWZ) technique \citep{Foster1996AJ....112.1709F}. For this, we used {\sc{libwwz}}\footnote{\url{https://github.com/RedVoxInc/libwwz}}, which is a Python package based on the Fortran code of \cite{Foster1996AJ....112.1709F} and \cite{Templeton2004JAVSO..32...41T}, as well as on \citet{m_emre_aydin_2017_375648}'s Python 2.7 WWZ library. WWZs are widely used in many areas ( {\it e.g.} Pulsar timing analysis by \citealt{Lyne2010Sci...329..408L}; supermassive black hole binary search by \citealt{Graham2015MNRAS.453.1562G}; time–frequency analysis for light curve of dwarf nova by \citealt{Ghaderpour2020PASP..132k4504G}). As with other time-frequency analysis methods, there is a trade-off between time resolution and frequency resolution, which is controlled by a decay constant $c$. In our work, we followed  \citet{Foster1996AJ....112.1709F} and chose $ c = 1 / 8 \pi^2 \approx 0.0127$. Thus the e-folding time (defined as the distance when the wavelet power falls to $1/e^2$, and it can be seen as the window width) is $ \sqrt{2} P $. In our case, the QPO period is around 600\,s, the corresponding e-folding time is 848\,s, which is sufficient for period determination. In Fourier space, the Full Width at Half Maximum (FWHM) of the frequency peak at 1/600\,Hz is about $2 \times 10^{-4}\,\mathrm{Hz}$, lower frequency has smaller FWHM. Therefore, both the time resolution and frequency resolution is appropriate for our analysis.

Finally, after determining the most significant QPO frequencies at different times within each observation, we folded the light curves on those frequencies and obtained their phaseograms. We did that with the {\sc{stingray}} \citep[][]{stingrayHuppenkothen2019,stingray2019ApJ...881...39H,stingray_matteo_bachetti_2021_4881255} Python package, designed for spectral-timing analysis of X-ray data. 

\subsection{X-ray/optical astrometry}
Using the Mikulski Archive for Space Telescopes, we selected and  downloaded {\it Hubble Space Telescope} ({\it HST}) observations of M\,101 that included the field of J1403. We found that this region was observed with the Advanced Camera for Surveys/Wide Field Camera (ACS/WFC) on 2002 November 13 (part of Program 9490); and with the Wide Field Camera 3/Ultraviolet and VISible light (WFC3/UVIS) imager on 2014 February 2 (Program 13364) and 2016 September 24 (Program 14166). We used 20 common sources in {\it Gaia} Early Data Release 3 to verify and re-align the {\it HST} astrometry in every filter to better than $\approx$0$\farcs$01. We did not find enough unambiguous {\it Chandra}/{\it HST} point-like associations in the region around J1403 for a direct astrometric alignment between the two sets of images. However, the large number of {\it Chandra} observations enabled us to take an average value of the X-ray position with a smaller uncertainty than the typical astrometric error of individual observations. More specifically, we selected 16 {\it Chandra} observations where J1403 has a high number of counts and is not too far off-axis. In each of them, the position of J1403 was determined with the {\sc {ciao}} task \textit{wavdetect}, and the associated error (statistical uncertainty due to the centroiding algorithm) was computed with Equation (5) from \citet{2005ApJ...635..907H}; this error was then added in quadrature with a 1-$\sigma$ absolute astrometry uncertainty of $\approx$0$\farcs$8. The final X-ray position is a weighted average and its error is the weighted standard deviation. We obtained R.A. $= 14^h03^m14^s.295$, Dec. $= +54^{\circ}18^{\prime}06\farcs26$ with an uncertainty radius of $\approx$0$\farcs$24, which enabled us to pinpoint a likely optical counterpart in the {\it HST} images (Section 3.4).


\section{Results}

\subsection{Transient behaviour}
The first motivation for our study was to characterise the outburst behaviour of this transient ULX. The long-term 0.3--10 keV light curve (Figure 2) was obtained with a combination of measurements. For the sixteen observations (fourteen from {\it Chandra} and two from {\it XMM-Newton}) with a sufficient number of counts, unabsorbed luminosities are the result of our {\sc {xspec}} modelling (Table 4, explained in more details in Section 3.3), with 90\% confidence limits computed with the {\it cflux} pseudo-model component. For the other 14 observations, the luminosities or upper limits plotted in Figure 2 come from {\it {srcflux}}, under the assumption of a line-of-sight Galactic column density $N_{\rm H} = 9 \times 10^{20}$ cm$^{-2}$ \citep{nh16,kalberla05}, a power-law photon index $\Gamma = 1.7$ (which is a representative value for the low state of Galactic black holes: {\it e.g.},  \citealt{chakrabarti95,shaposhnikov06,sobolewska11,yang15}), and a proper calculation of the local auxiliary response function ({\it srcflux} parameter ``point spread function method = arfcorr''). {\it srcflux} was applied over the 0.5--7 keV band, but we extrapolated the flux values to the 0.3--10 keV band, for consistency. We stress that do not have enough counts to determine whether the spectra of the fainter group of observations ($L_{\rm X} \lesssim 10^{38}$ erg s$^{-1}$) is really a power-law or is also dominated by thermal emission in the soft band; a power-law is just the simplest approximation. If the soft X-ray emission is thermal even in the fainter epochs, the estimated unabsorbed 0.3--10 keV luminosity would be about 25\% lower than estimated with a power-law model.

In most cases, J1403 was found at an X-ray luminosity between a few $\times 10^{37}$ erg s$^{-1}$ and $\approx$1 $\times 10^{38}$ erg s$^{-1}$. There were five separate flaring episodes  
to a luminosity $\gtrsim$10$^{39}$ erg s$^{-1}$ recorded over a 17-year timeline ({\it i.e.}, five epochs at ULX level separated by at least one detection at a lower luminosity). There is no evidence of a regular pattern for the outbursts. Transitions from lower to upper luminosities occur sometimes on short timescales: a change from $L_{\rm X} \approx 10^{38}$ erg s$^{-1}$ to $L_{\rm X} \approx 10^{39}$ erg s$^{-1}$ happened over only two days, in 2003 January. By 2003 March, the X-ray source had declined below a detection limit of $\approx$10$^{37}$ erg s$^{-1}$, but a week later it was up again as a ULX at $L_{\rm X} \approx 3 \times 10^{39}$ erg s$^{-1}$. On at least a couple of epochs, namely {\it Chandra} during ObsIDs 4737 (Figure A1) and 6170 (Figure A5), luminosity changes by an order of magnitude occurred on timescales $\lesssim$10$^4$ s. We will discuss the intra-observational variability in more details in Section 3.2. Interestingly, J1403 appears to spend very little time in the luminosity range between $\approx$10$^{38}$ erg s$^{-1}$ and $\approx$10$^{39}$ erg s$^{-1}$, which is where transient stellar-mass BHs spend the majority of their time in outburst \citep{fender16}.

\begin{figure}
\hspace{-0.5cm}
    \includegraphics[height=0.49\textwidth, angle=270]{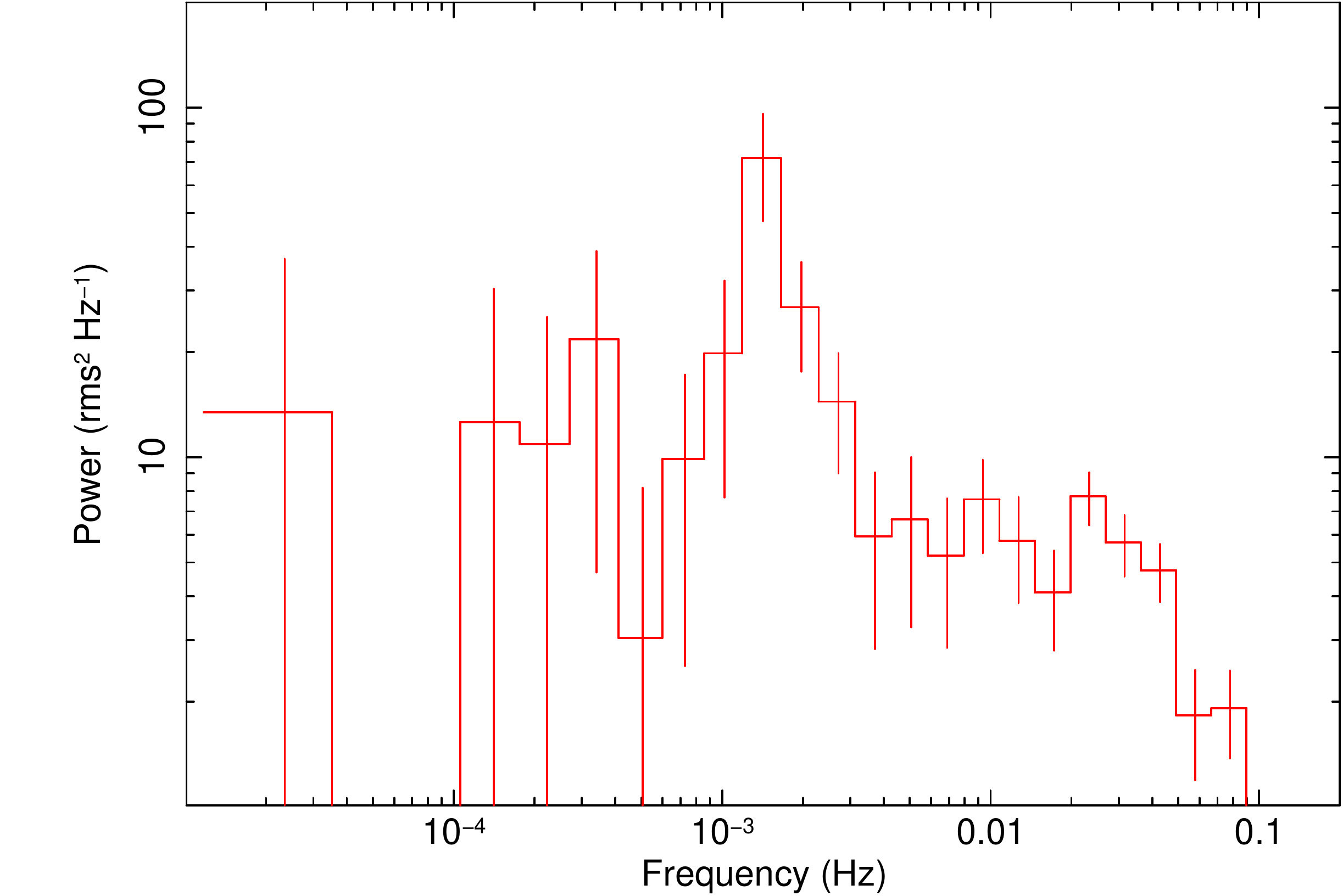}\\[20pt]
    \caption{Power spectral density of J1403 derived from the {\it XMM-Newton}/EPIC light-curve from observation 0104260101. The white noise level has been subtracted (parameter ``norm $= -2$'' in {\it powspec}).}
    \label{fig:xmm101_powspec}
\end{figure}

\begin{figure}
\hspace{-0.5cm}
    \includegraphics[height=0.49\textwidth, angle=270]{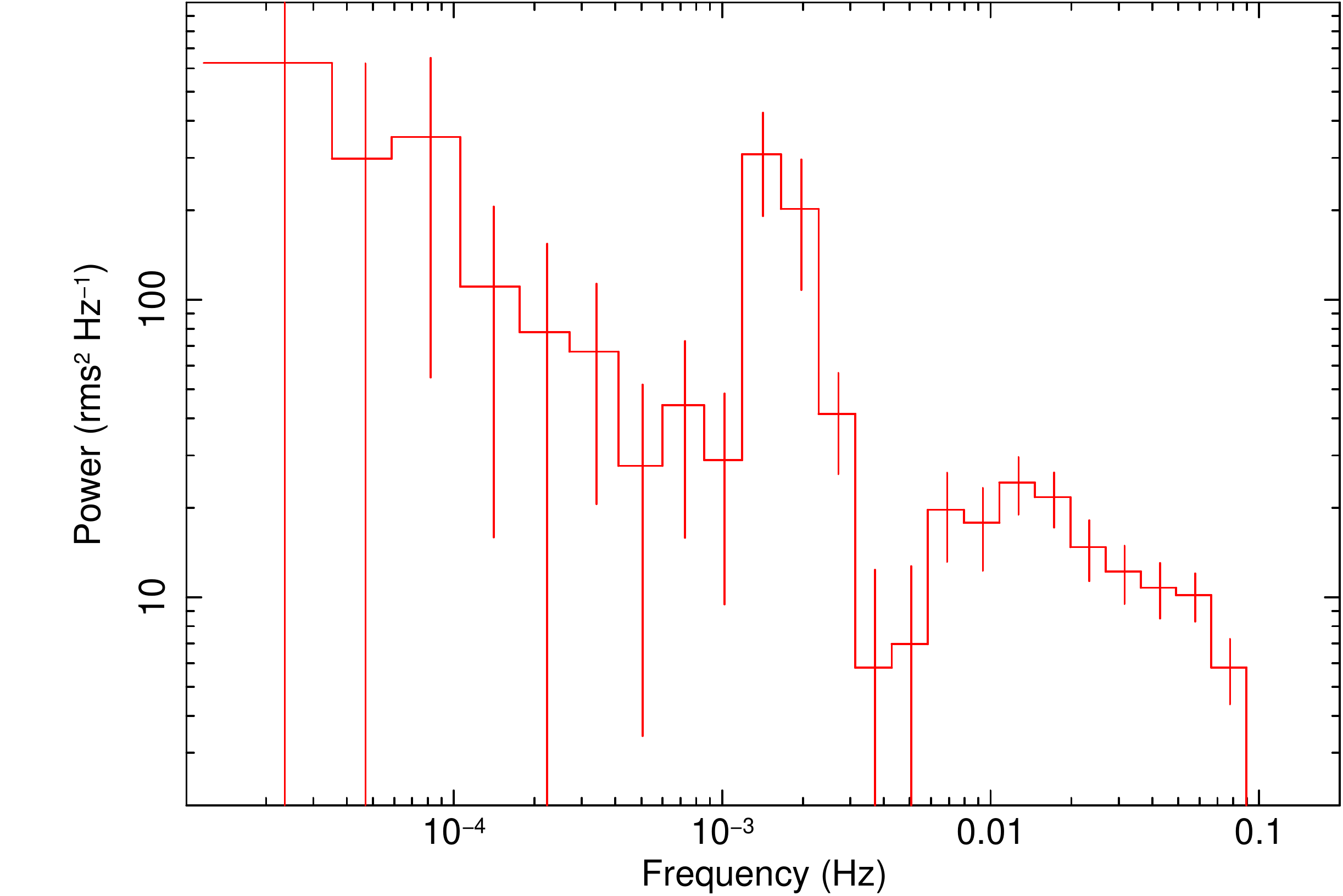}\\[20pt]
    \caption{As in Figure 4, for the {\it XMM-Newton}/EPIC light-curve from observation 164560701.}
    \label{fig:xmm701_powspec}
\end{figure}

\begin{figure}
 \includegraphics[height=0.48\textwidth, angle=270]{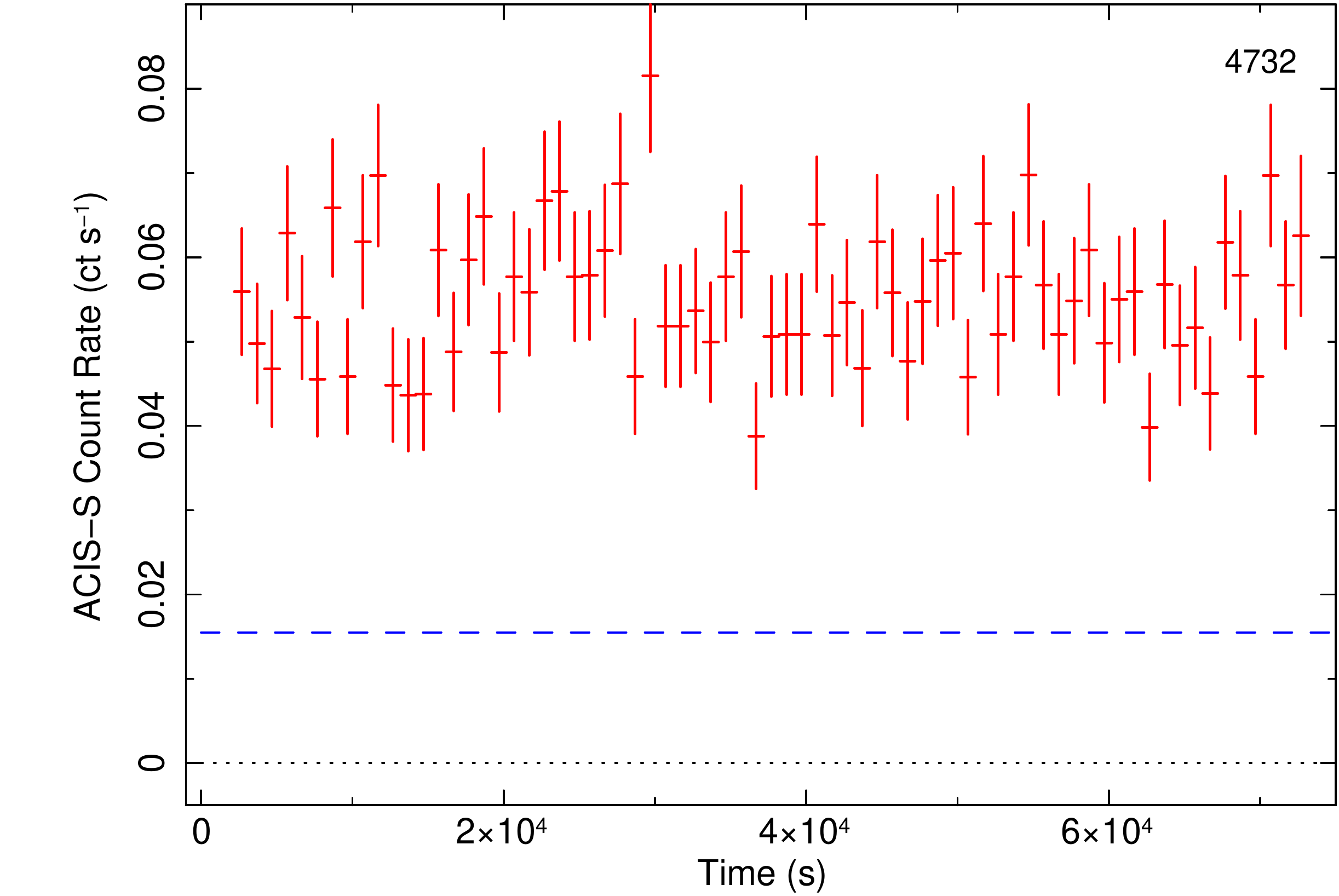}\\
\includegraphics[height=0.48\textwidth, angle=270]{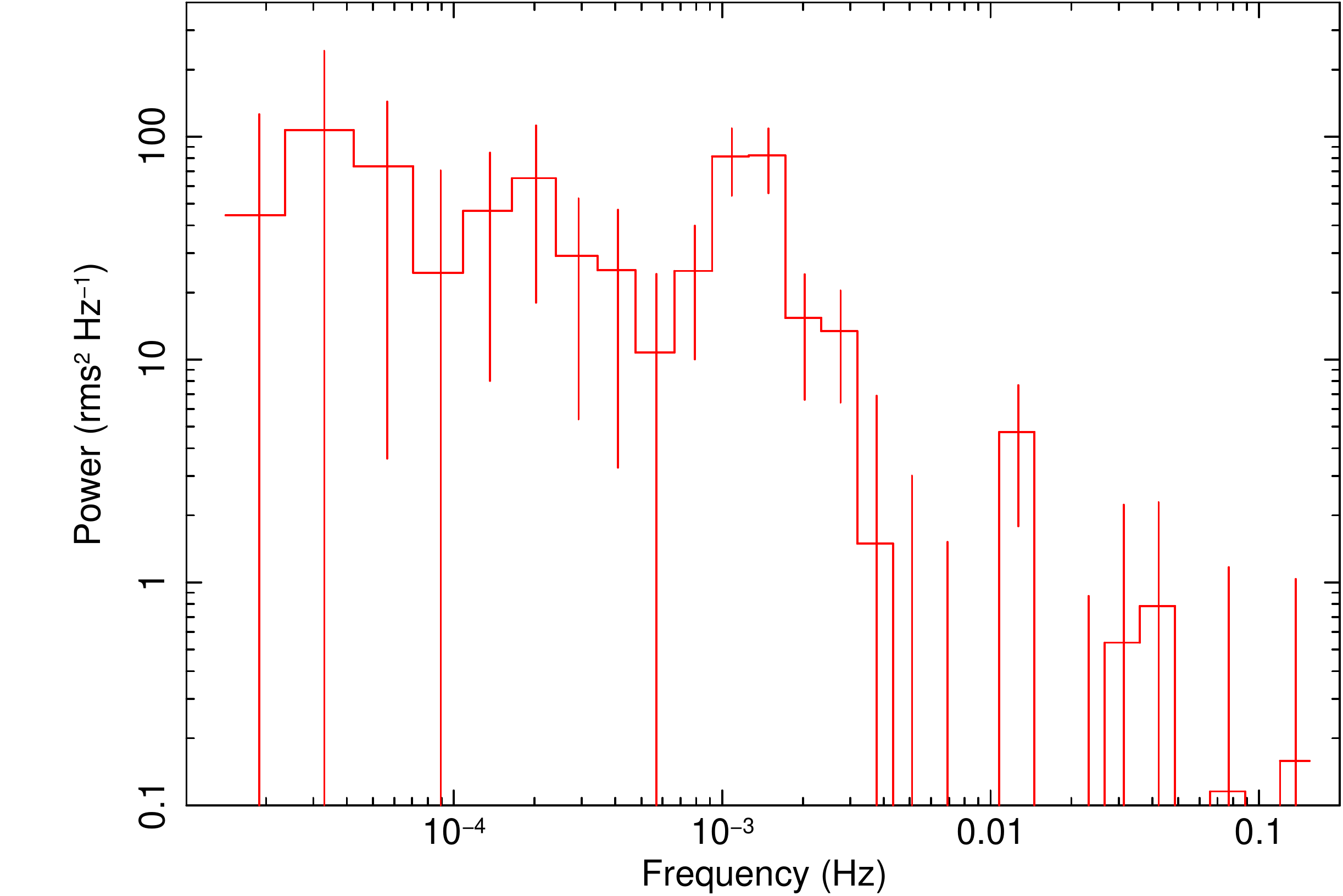}\\[20pt]
    \caption{Top panel: light curve of {\it Chandra} ObsID 4732. The dashed line represents the count rate corresponding to an approximate de-absorbed 0.3--10 keV luminosity of $10^{39}$ erg s$^{-1}$ (based on our detailed spectral modelling). Bottom panel: power spectral density from the same observation, computed with {\it powspec}, after white noise subtraction.}
    \label{fig:4732_powspec}
\end{figure}

\subsection{Short-term variability and QPOs}

The intra-observation variability of J1403 proved to be also unusual. It is immediately clear from visual inspection (Figure~\ref{fig:xmm_lightcurves}) that light curves from the two {\it XMM-Newton} observations contain a markedly regular pattern (particularly ObsID 0164560701). We computed power density spectra of the two light curves with the {\sc ftools} task {\it powspec} (fast Fourier transform algorithm). They show significant peaks at characteristic frequencies just above $\sim$1\,mHz (Figures~\ref{fig:xmm101_powspec}--\ref{fig:xmm701_powspec}). Then, we inspected the  power density spectra of the {\it Chandra} observations extracted using {\it powspec}, and found a significant peak at $\sim$1.632\,mHz in at least one observation, ObsID 4732 (Figure~\ref{fig:4732_powspec}), corresponding to the highest luminosity recorded by {\it Chandra} ($L_{\rm X} \approx 3.6 \times 10^{39}$ erg s$^{-1}$).

We used the Lomb--Scargle analysis to quantify the significant frequencies more precisely. We found significant QPO frequencies (false alarm probability level $<$1\%, which is estimated with the method of \citealt{Baluev2008MNRAS.385.1279B}) in the three observations already mentioned above ({\it XMM-Newton} ObsIDs 0104260101 and 0164560701, {\it Chandra} ObsID 4732); they are summarized in Table \ref{tab:frequencies}. In all three cases, the strongest oscillations have characteristic periods of $\sim$600\,s. In the rest of this paper, we will focus on the analysis of those three observations. Several other {\it Chandra} observations show candidate QPO frequencies (usually also in the range of $\sim$1.5--2.5\,mHz: Table \ref{tab:frequencies}), although they do not reach the 1\% false alarm probability threshold as in ObsID 4732. In most of those cases, the power of frequencies are higher than the short sub-intervals of the observations in which a QPO is present (as we shall illustrate with the dynamical power spectra). A summary of Lomb--Scargle results for a selection of bright {\it Chandra} observations is reported in Appendix A.

For the third step of our timing analysis, we used the WWZ technique, to look for intra-observational changes in the QPO frequencies and intensities. We found that the QPO appears and disappears within individual observations, and their frequencies also have some slightly drifts (Figures \ref{fig:xmm101-WWZ}, \ref{fig:xmm701-WWZ}, and \ref{fig:4732-WWZ}).
We also studied whether the phase of the oscillation was preserved during individual observations (Figures \ref{fig:xmm101-PG}, \ref{fig:xmm701-PG}, and \ref{fig:4732-PG} show three examples of phaseograms). Their phases basically remain constantly, although {\it XMM-Newton} ObsIDs 0104260701 shows a possible slowly trend, we cannot confirm it because our long period of QPO leads to the low time resolution.

Finally, as an independent check, we searched for periodicities in the same three observations ({\it XMM-Newton} ObsIDs 0104260101 and 0164560701, {\it Chandra} ObsID 4732) with the {\sc ftools} task {\it efsearch}, which works by folding the light-curve over a range of test periods, and determining the $\chi^2$ of the folded light-curves for each period. We then used the {\it efold} task to plot the light-curves folded onto the most significant period. The results are consistent with those found from the Lomb--Scargle analysis.
For the first and second {\it XMM-Newton} observations, we find best-fitting folding periods of $(607.5 \pm 1.0)$\,s and $(575.9 \pm 1.0)$\,s, respectively (Figure \ref{fig:xmm101-efold} and \ref{fig:xmm701-efold}). For the {\it Chandra}  ObsID 4732, we find $(612.8 \pm 1.5)$\,s. The folded light-curves of the three observations are shown in Figures \ref{fig:xmm101-efold}--\ref{fig:4732-efold}.

\begin{figure}
    \centering
    \includegraphics[width=0.48\textwidth]{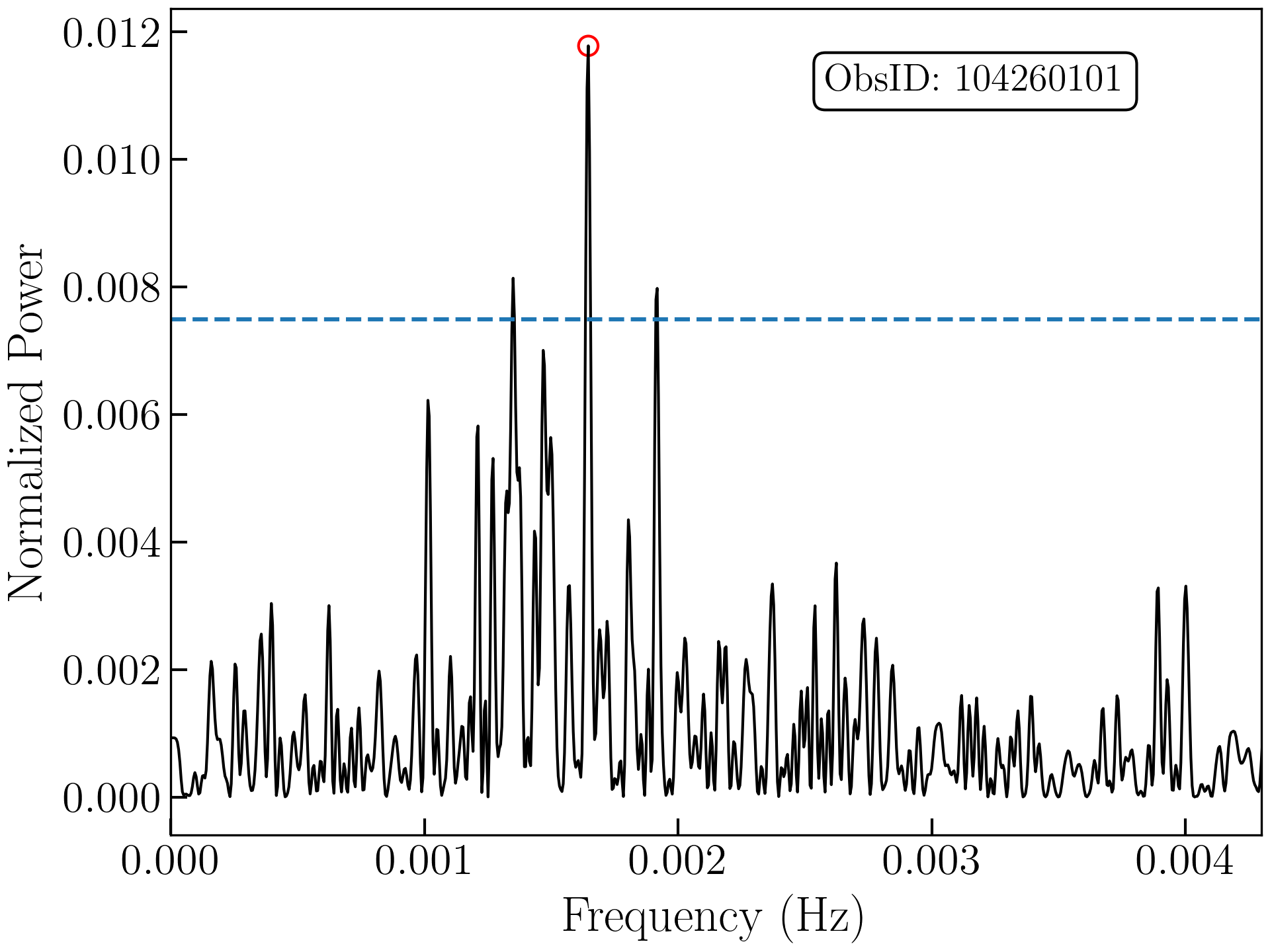}
    \caption{Lomb--Scargle periodogram of {\it XMM-Newton} ObsID 104260101. The open red circle indicates the most significant frequency, and the dashed blue line shows the 1\% false alarm probability level.}
    \label{fig:xmm101-PSD}
\end{figure}

\begin{figure}
    \centering
    \includegraphics[width=0.48\textwidth]{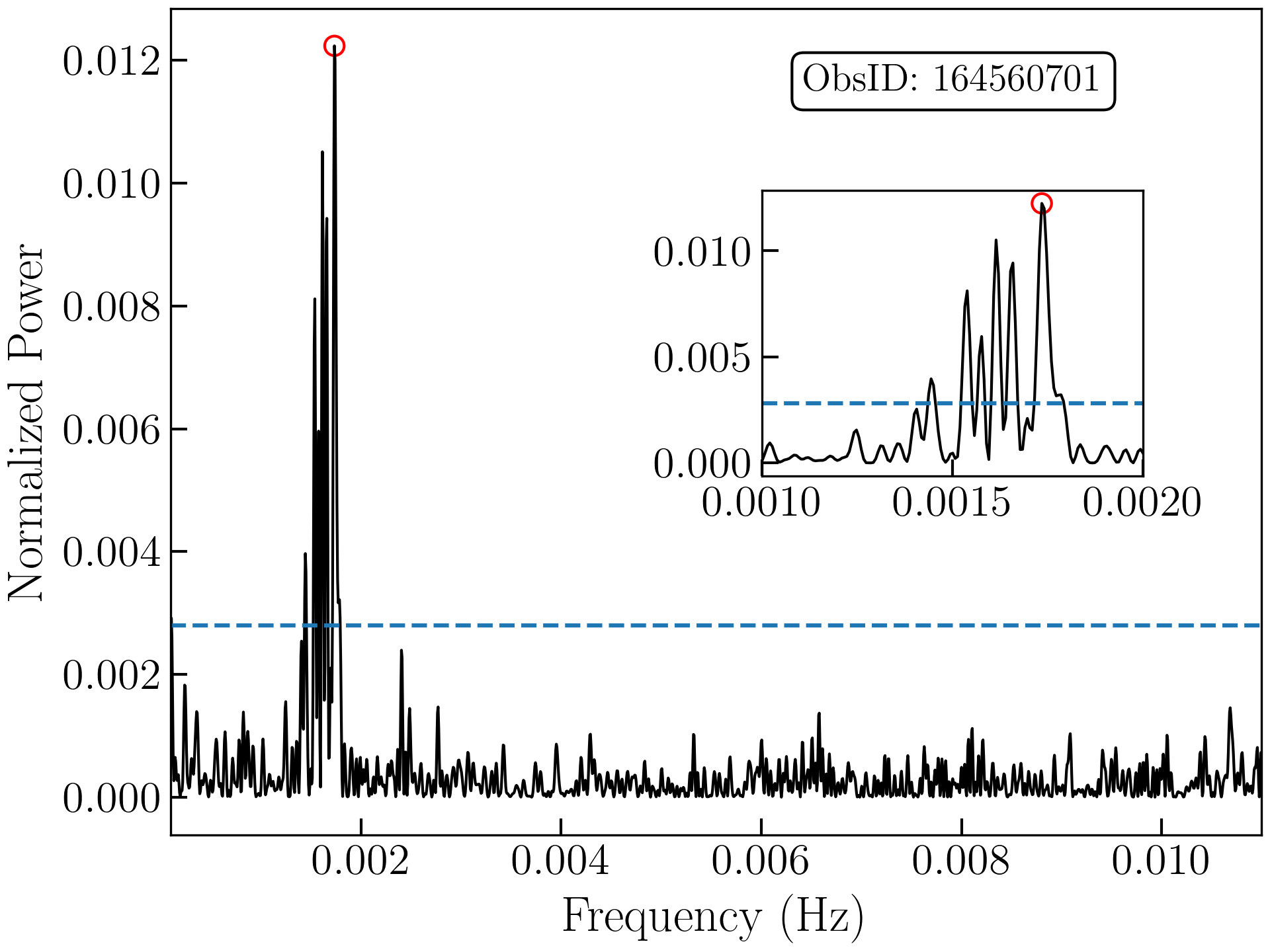}
    \caption{As in Figure 7, for {\it XMM-Newton} ObsID 164560701. 
    A zoomed-in inset of the significant frequencies are shown in the middle right. It shows that there is a series of significant lower-frequency harmonics of the strongest frequency.}
    \label{fig:xmm701-PSD}
 
\end{figure}

\begin{figure}
    \centering
    \includegraphics[width=0.48\textwidth]{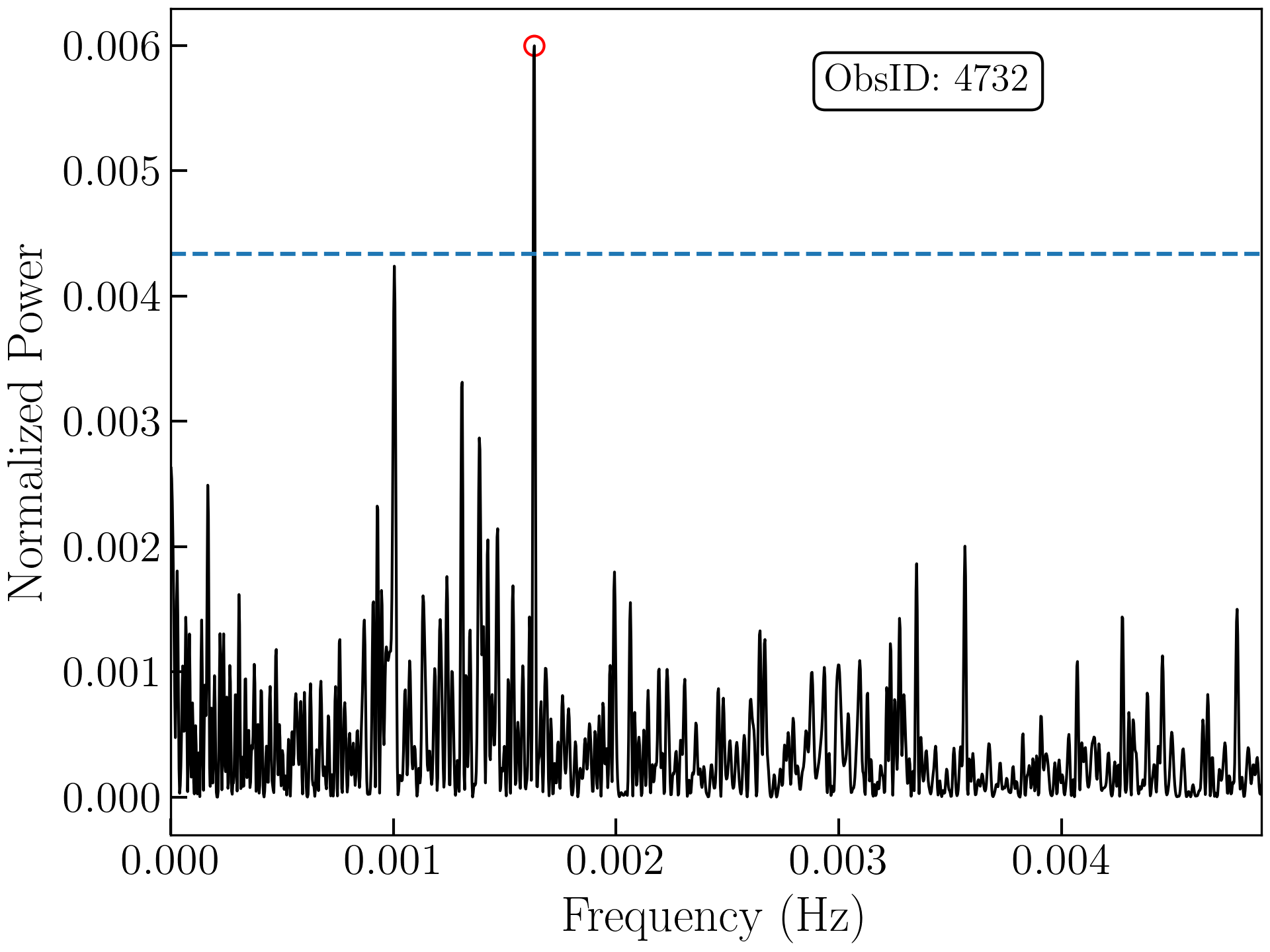}
    \caption{As in Figure 7, for {\it Chandra} ObsID 4732.} 
    \label{fig:4732-PSD}
\end{figure}

\begin{table}
    \caption{Main QPO frequencies and periods found with our WWZ analysis, in both {\it XMM-Newton} observations and in the {\it Chandra} observations with the highest signal-to-noise ratio.}
    \label{tab:frequencies}
    \begin{tabular}{lrr}
        \hline                                                          \\[-5pt]
        ObsID      & Significant QPOs        & Candidate  QPOs          \\ [5pt]
        \hline                                                          \\[-5pt]
        \multicolumn{3}{c}{{\it Chandra}}                               \\[5pt]
        \hline                                                          \\[-5pt]
        4732       & 1.632 mHz $=$ 612.791 s & 1.003 mHz $=$ 997.250 s             \\
                   &                         & 1.308 mHz $=$ 764.700 s  \\
        4737       &                         & 7.682 mHz $=$ 130.183 s  \\
                   &                         & 1.368 mHz $=$ 730.932 s  \\
                   &                         & 3.153 mHz $=$ 317.205 s  \\
        5297       &                         & 6.647 mHz $=$ 150.440 s  \\
                   &                         & 2.688 mHz $=$ 372.054 s  \\
                   &                         & 1.892 mHz $=$ 528.473 s  \\
        5309       &                         & 1.734 mHz $=$ 576.724 s  \\
                   &                         & 1.583 mHz $=$ 631.615 s  \\
        6169       &                         & 1.730 mHz $=$ 578.031 s  \\
                   &                         & 21.099 mHz $=$ 47.396 s  \\
                   &                         & 1.548 mHz $=$ 645.843 s  \\
                   &                         & 2.948 mHz $=$ 339.246 s  \\
                   &                         & 1.912 mHz $=$ 523.106 s  \\
        6170       &                         & 0.182 mHz $=$ 5480.675 s \\
        6175       &                         & 1.595 mHz $=$ 627.021 s  \\
                   &                         & 1.799 mHz $=$ 555.885 s  \\[5pt]
        \hline                                                          \\[-5pt]
        \multicolumn{3}{c}{{\it XMM-Newton}}                            \\[5pt]
        \hline                                                          \\[-5pt]
        0104260101 & 1.645 mHz $=$ 607.739 s &                          \\
                   & 1.349 mHz $=$ 741.208 s &                          \\
                   & 1.917 mHz $=$ 521.635 s &                          \\
        0164560701 & 1.734 mHz $=$ 576.561 s &                          \\
                   & 1.614 mHz $=$ 619.422 s &                          \\
                   & 1.659 mHz $=$ 602.910 s &                          \\
                   & 1.539 mHz $=$ 649.938 s &                          \\
                   & 1.577 mHz $=$ 634.314 s &                          \\
                   & 1.444 mHz $=$ 692.588 s &                          \\
                   & 1.785 mHz $=$ 560.239 s &                          \\[5pt]                 
       \hline
    \end{tabular}
\end{table}

\begin{figure}
    \centering
    \includegraphics[width=0.48\textwidth]{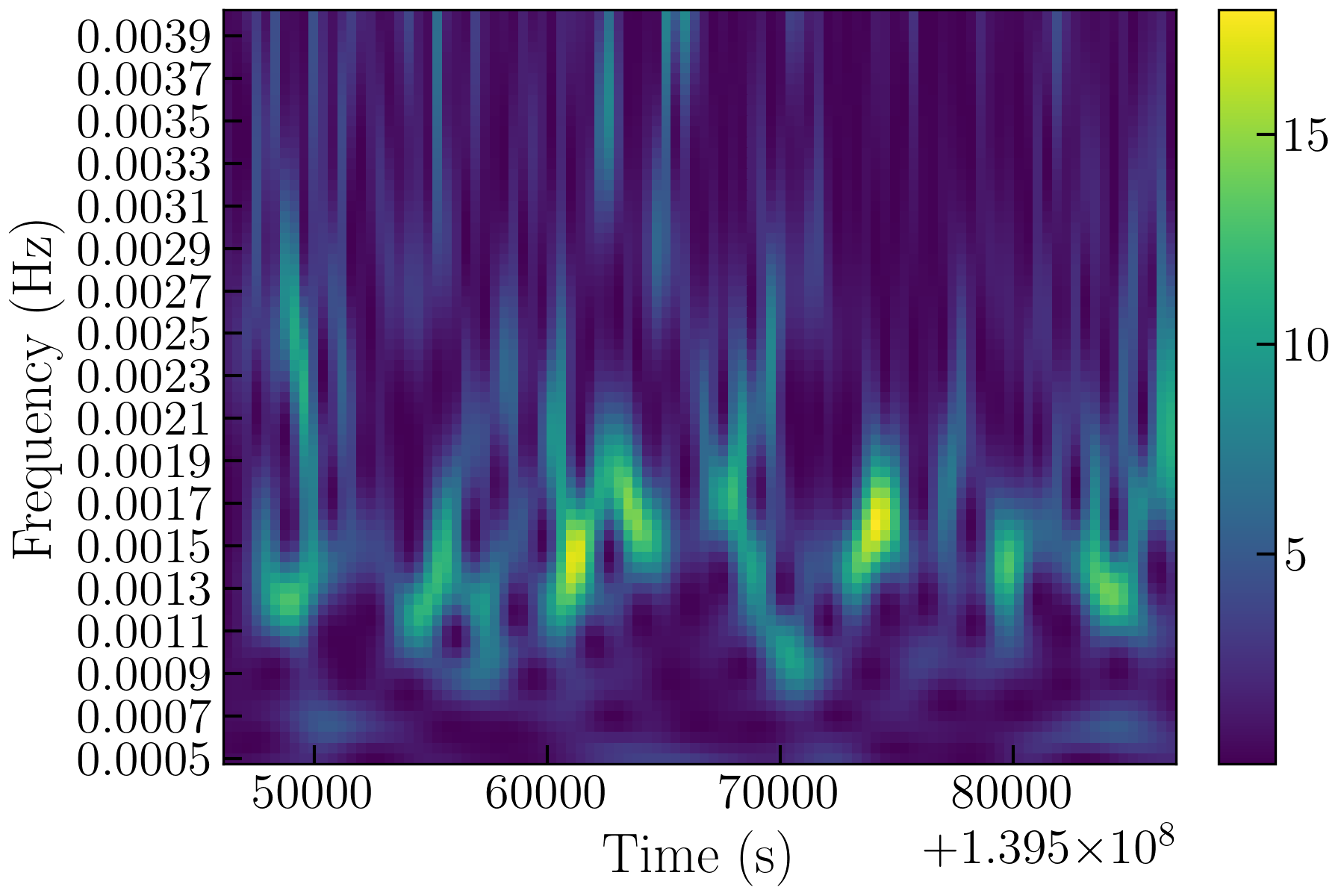}
    \caption{WWZ dynamical power spectrum for {\it XMM-Newton} ObsID 104260101.}
    \label{fig:xmm101-WWZ}
\end{figure}

\begin{figure}
    \centering
    \includegraphics[width=0.48\textwidth]{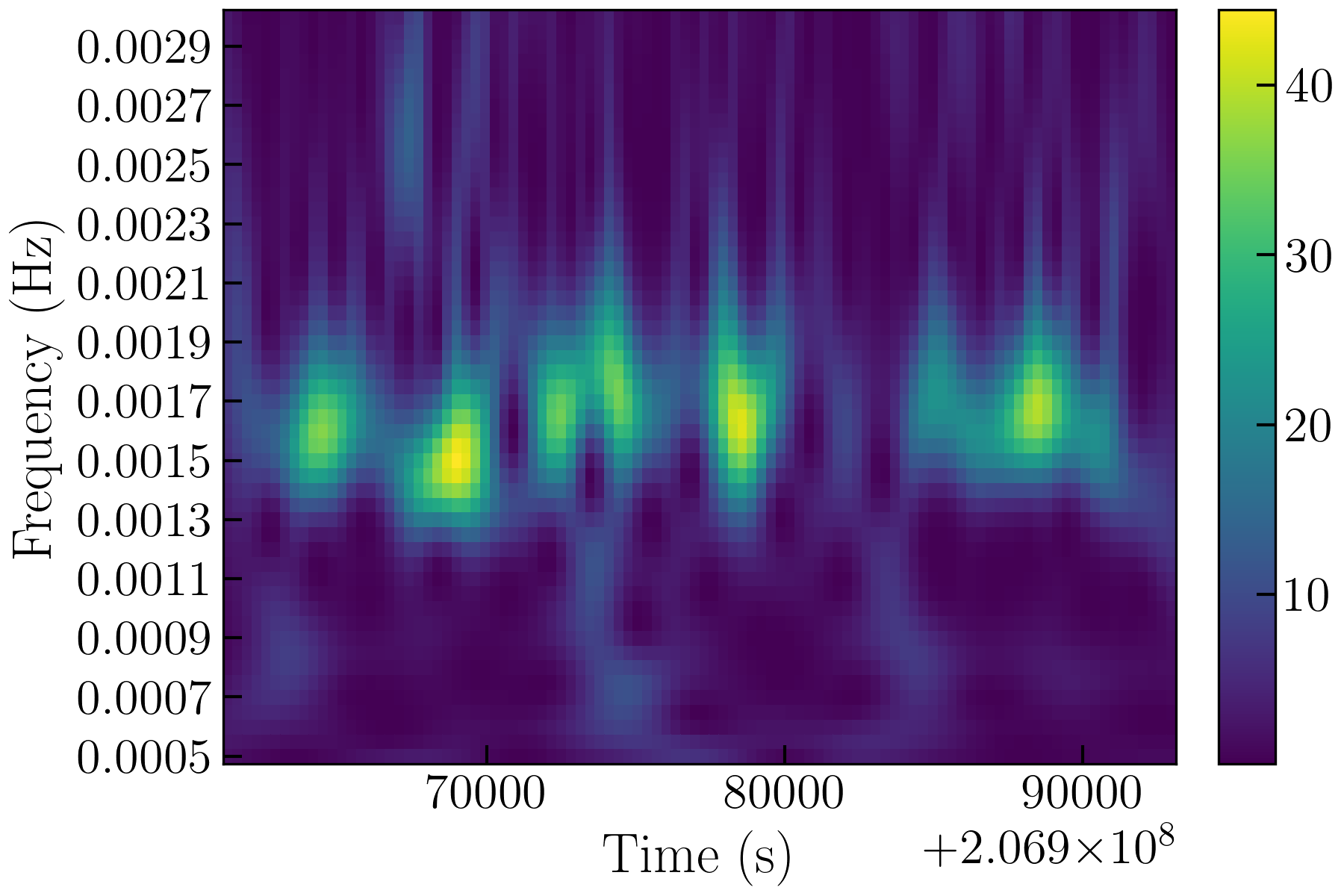}
    \caption{As in Figure 10, for {\it XMM-Newton} ObsID 164560701.}
    \label{fig:xmm701-WWZ}
\end{figure}

\begin{figure}
    \centering
    \includegraphics[width=0.48\textwidth]{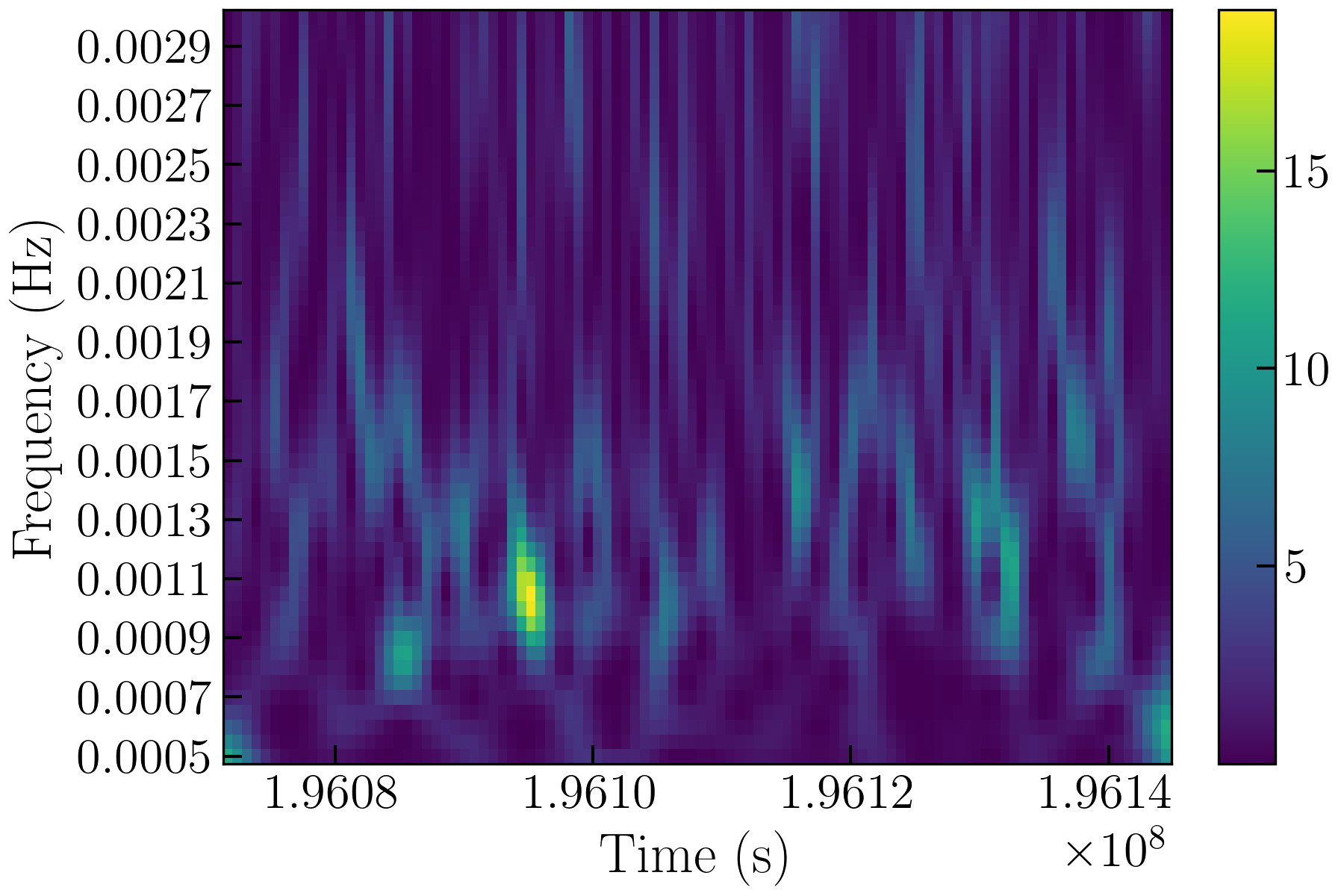}
    \caption{As in Figure 10, for {\it Chandra} ObsID 4732.}
    \label{fig:4732-WWZ}
\end{figure}







\begin{figure}
    \centering
    \includegraphics[width=0.48\textwidth, height=1.1\columnwidth]{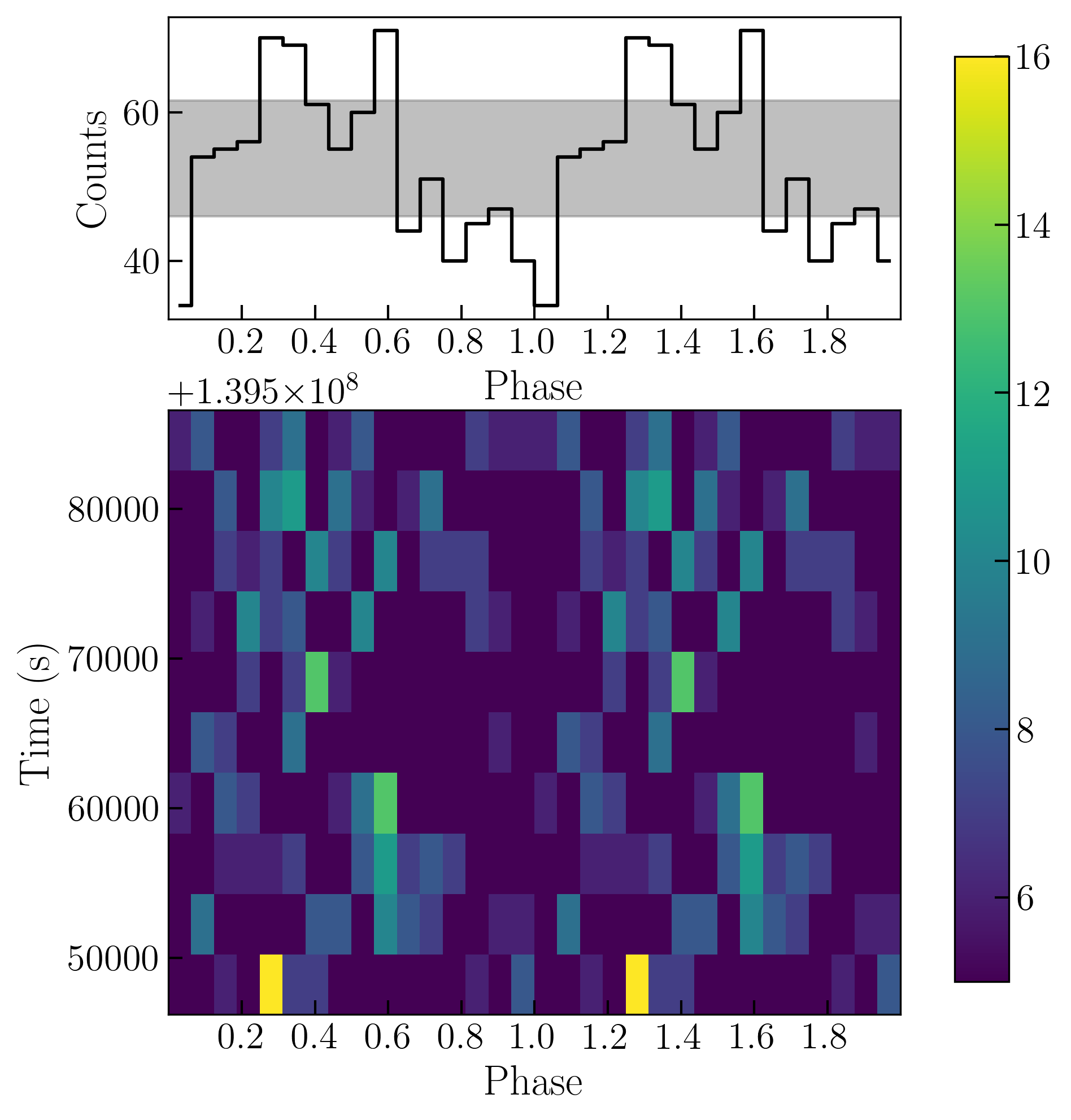}
    \caption{Phaseogram of {\it XMM-Newton} ObsID 104260101. The top panel shows the light curve folded by the 607.739\,s period, which is the most significant frequency indicated in Figure \ref{fig:xmm101-PSD}. The shaded grey area shows the mean Poisson confidence level. The bottom panel shows the phase variations during the observation.}
    \label{fig:xmm101-PG}
\end{figure}

\begin{figure}
    \centering
    \includegraphics[width=0.48\textwidth, height=1.1\columnwidth]{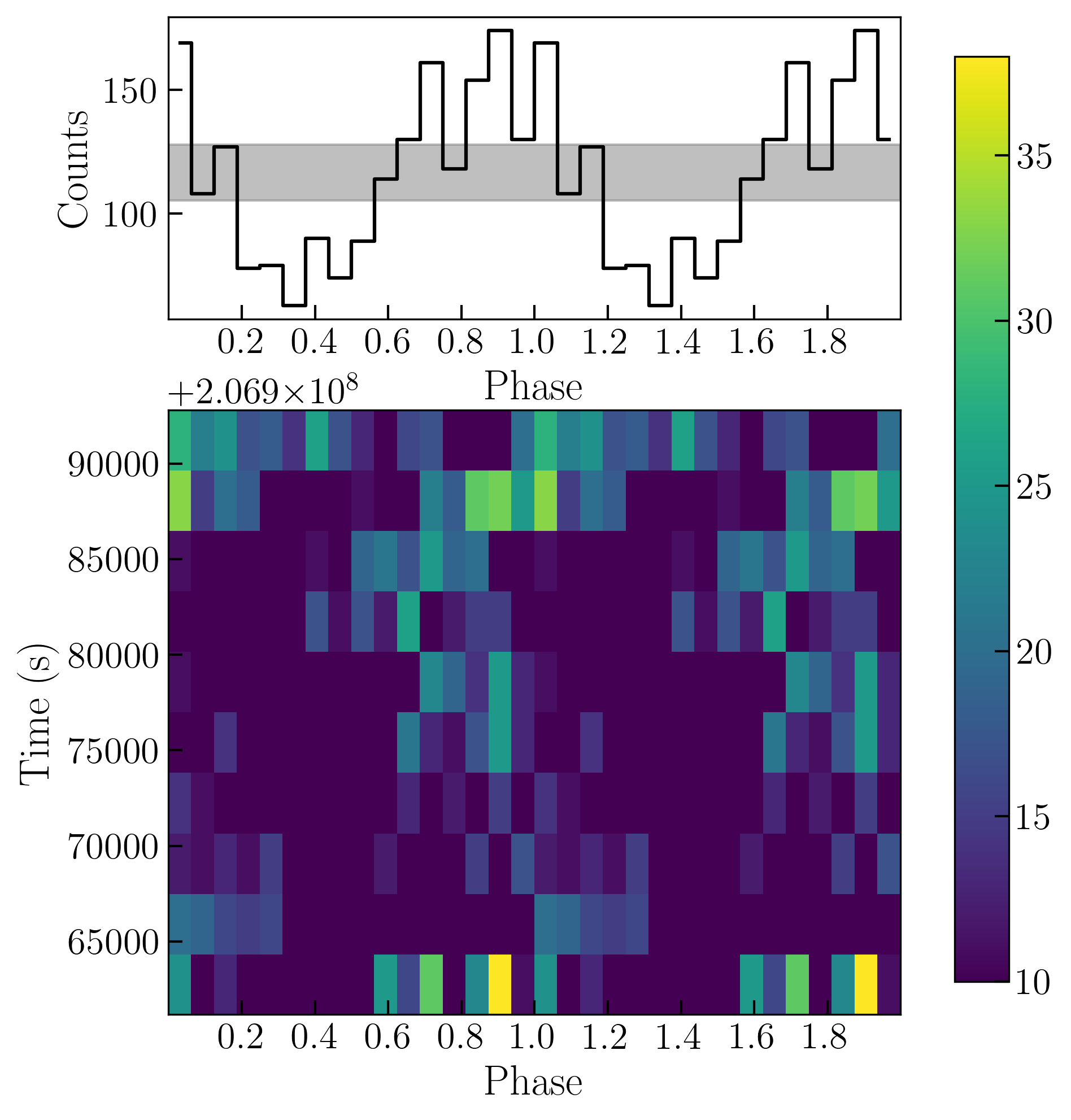}
    \caption{As in Figure 13, for {\it XMM-Newton} ObsID 164560701.}
    \label{fig:xmm701-PG}
\end{figure}

\begin{figure}
    \centering
    \includegraphics[width=0.48\textwidth, height=1.1\columnwidth]{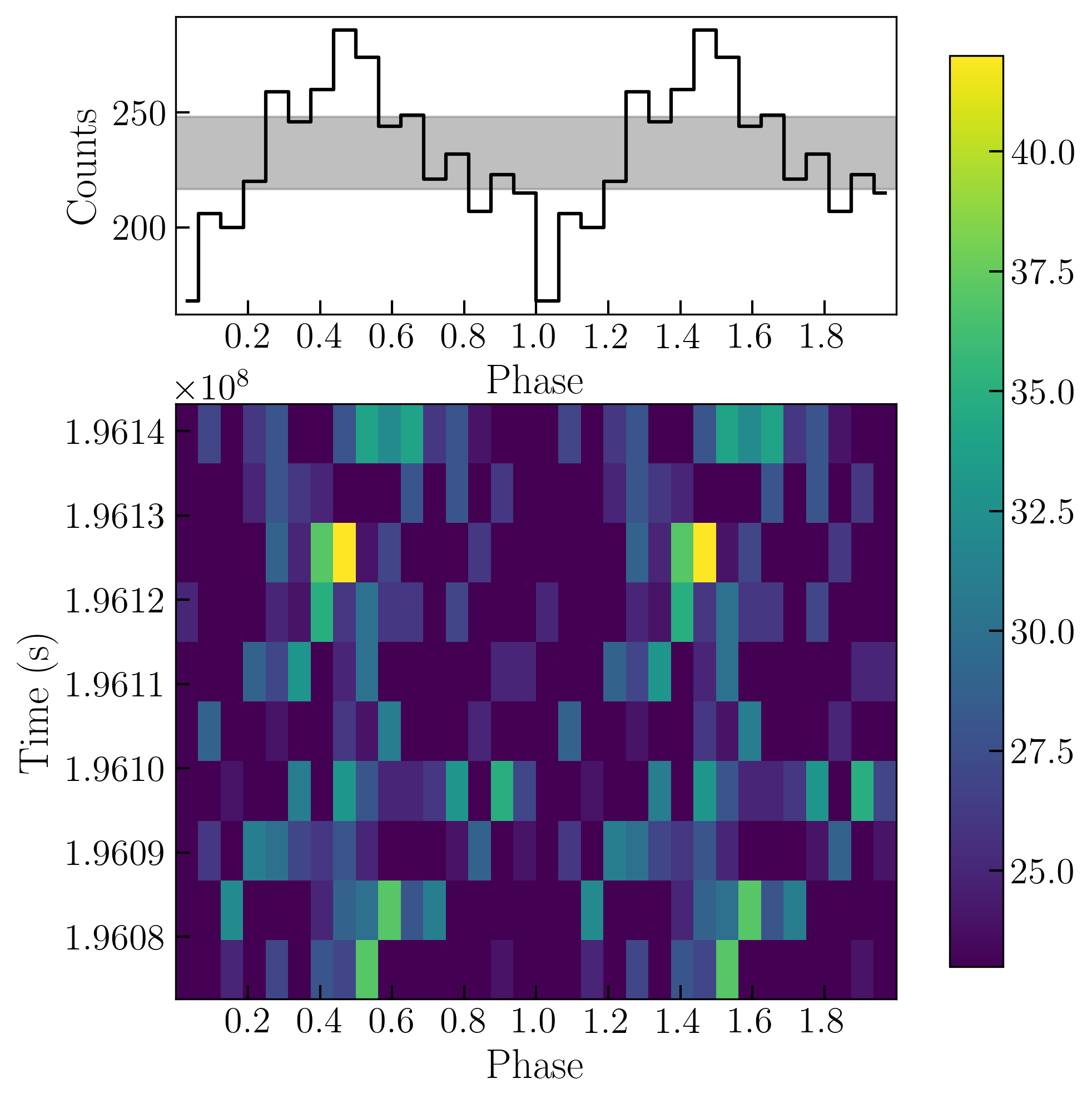}
    \caption{As in Figure 13, for {\it Chandra} ObsID 4732.}
    \label{fig:4732-PG}
\end{figure}

\begin{figure}
\hspace{-0.5cm}
    \includegraphics[width=0.48\textwidth]{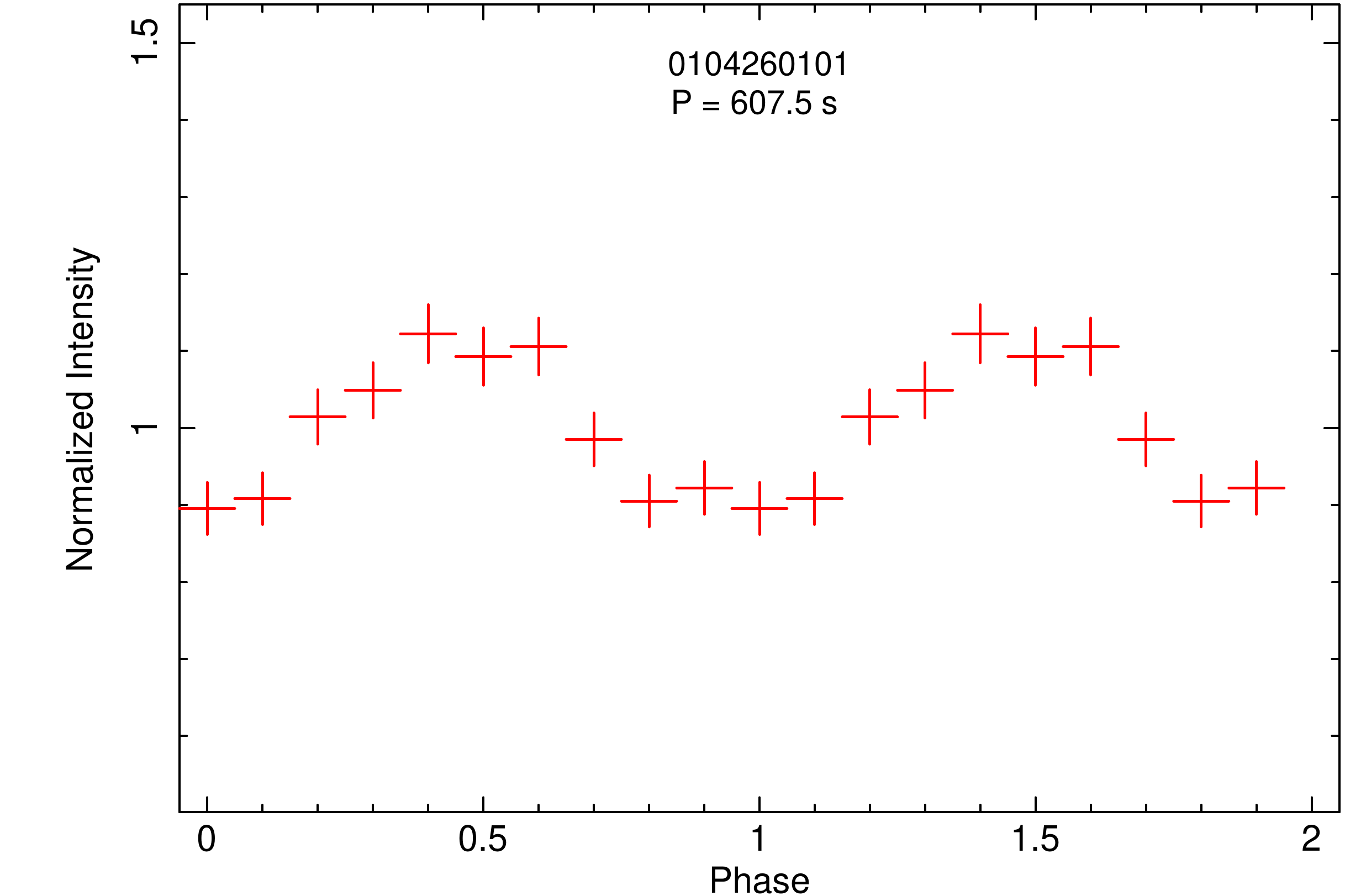}
    \caption{0.3--10 keV background-subtracted light curve of {\it XMM-Newton} ObsID 104260101, folded over a period of 607.5 s. This is the period that provides the best $\chi^2$ in the {\it efsearch} analysis; it is consistent (within the {\it efsearch} uncertainty) with the characteristic period of 607.7 s found with the Lomb--Scargle analysis.}
    \label{fig:xmm101-efold}
\end{figure}

\begin{figure}
\hspace{-0.5cm}
    \includegraphics[width=0.48\textwidth]{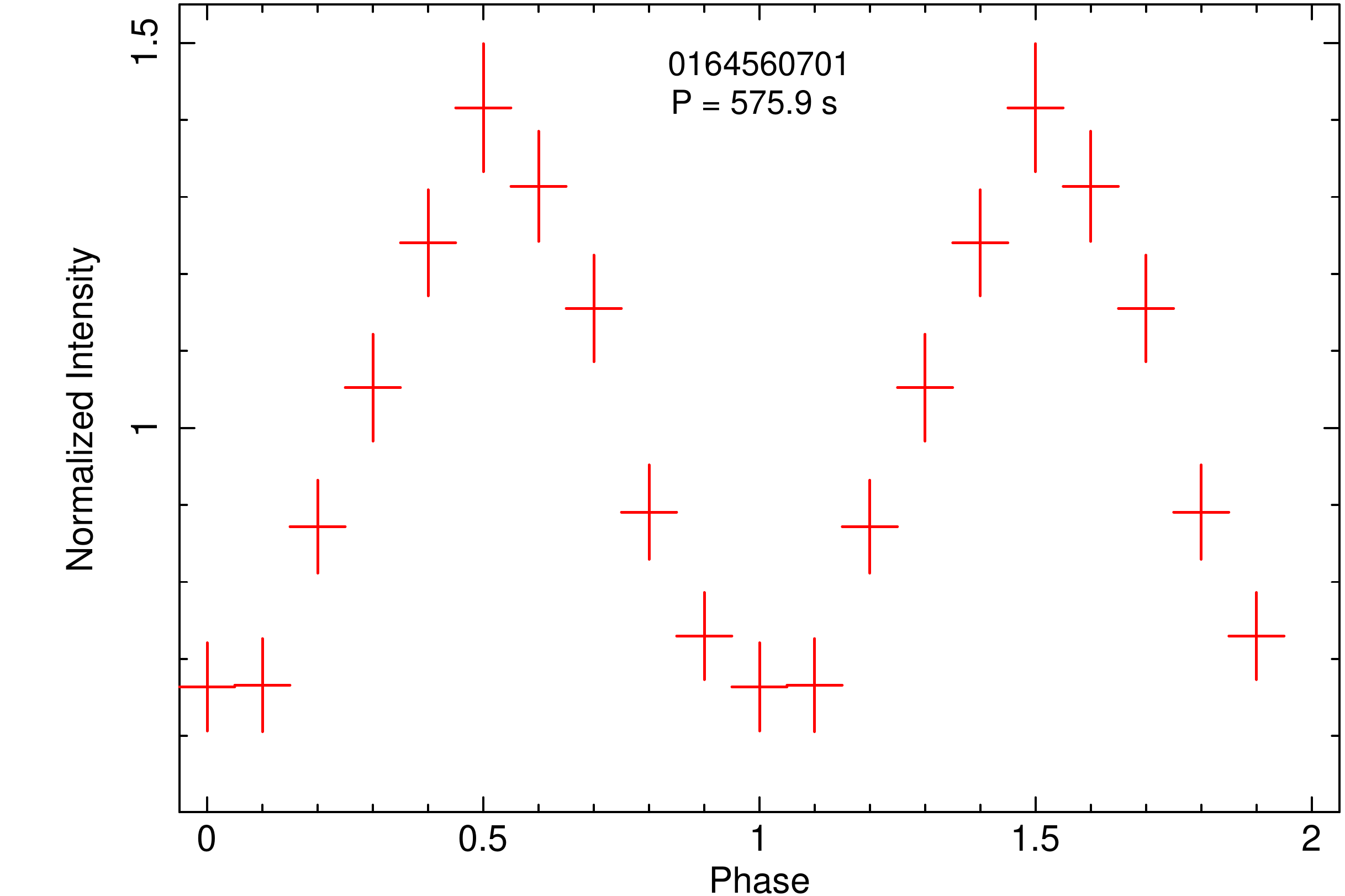}
    \caption{0.3--10 keV background-subtracted light curve of {\it XMM-Newton} ObsID 164560701, folded over a period of 575.9 s (consistent with the strongest period of 576.6 s found with the Lomb--Scargle analysis).}
    \label{fig:xmm701-efold}
\end{figure}

\begin{figure}
\hspace{-0.5cm}
    \includegraphics[width=0.48\textwidth]{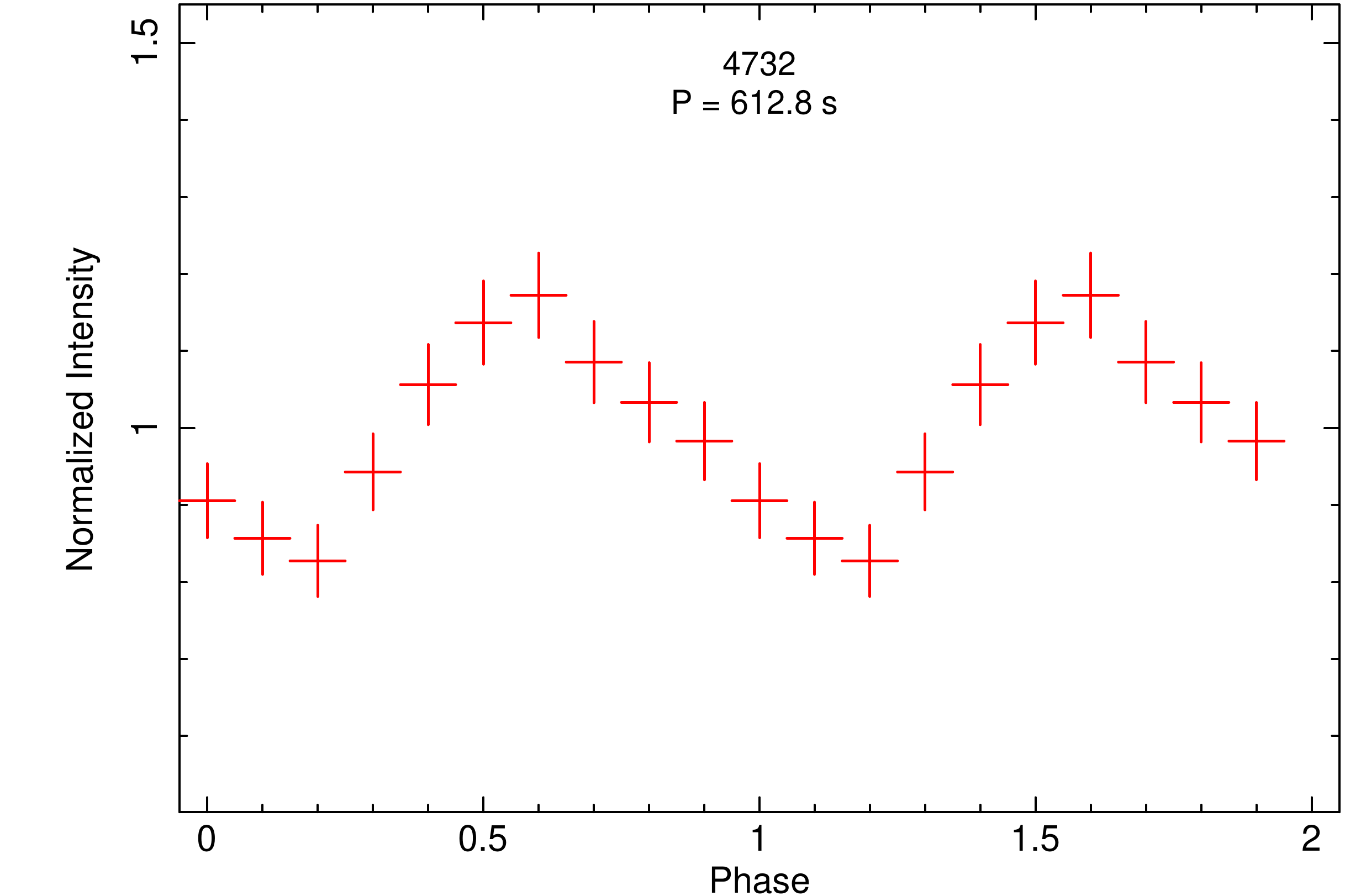}
    \caption{0.3--7 keV background-subtracted light curve of {\it Chandra} ObsID 4732, folded over a period of 612.8 s (identical to the period  found with the Lomb--Scargle analysis).}
    \label{fig:4732-efold}
\end{figure}

\subsection{Spectral properties}
All the X-ray spectra with sufficient signal-to-noise ratio ({\it i.e.}, those corresponding to $L_{\rm X} \gtrsim 10^{39}$ erg s$^{-1}$) share a general property: they are mildly curved over the {\it Chandra}/{\it XMM-Newton} band. Thus, they cannot be fitted with a simple power-law. Single thermal emission components (blackbody, standard disk, $p$-free disk) also do not provide a good fit, because they are more curved than the observed data. This is a common situation in ULXs. By analogy with previous spectral analysis in the literature \citep{walton18}, we fitted the spectra with a double thermal component (blackbody plus $p$-free disk: Tables 4,5).
The double-thermal model provides two characteristic temperatures and two associated radii. The higher temperature is typically $\sim$1.5--2 keV, for a blackbody emitting size $\sim$30--60 km. The lower temperature is $\sim$0.4--0.5 keV and corresponds to inner disk radii $\sim$200--400 km. 

We also tried to fit the spectra with Comptonization models: for example {\it simpl}$\times${\it diskbb}, {\it simpl}$\times${\it diskpbb}, {\it nthcomp}, and {\it diskir}. Such models provide statistically equivalent results to the double thermal models. The two characteristic best-fitting temperatures in the Comptonization models are also around 1.5--2 keV and 0.4--0.5 keV. In this framework, the higher temperature corresponds to the energy of the Comptonizing electrons, and the lower temperature is associated with the seed disk photons. A more detailed comparison of the different spectral models is beyond the scope of this paper, which is more focused on the peculiar time variability properties.

In three of the {\it Chandra} observations (ObsIDs 4732, 5297, 5309), the spectral fits are significantly improved by the addition of a thermal plasma component around 1 keV. Such components are often seen in ULX spectra even at CCD resolution \citep{middleton15} and are resolved into a complex structure of emission and absorption lines at grating resolution \citep{pinto16,pinto17,kosec18}. They are interpreted as signatures of the super-Eddington outflow characteristic of ULXs, especially for accretion disks seen at high inclination (denser outflow along our line of sight). However, for J1403, there is no significant thermal plasma component in the two {\it XMM-Newton} observations, despite their higher signal-to-noise compared with the {\it Chandra} observations at similar luminosity. Thus, based on the available data, we cannot determine whether the outflow signatures are real and whether they are persistent or transient. 

Another caveat is that the $p$ index of the disk model component tends to the lowest limit ($p = 0.5$), both for double-thermal models ({\it bb} plus {\it diskpbb}) and for Comptonization models ({\it simpl}$\ast${\it diskpbb}). This indicates that the spectral curvature of this component is mild, and significantly lower than that expected from a standard disk. Overall, the best-fitting components may be nothing more than phenomenological tools in {\sc xspec} to reproduce a mildly curved spectrum. 

\begin{figure}
\hspace{-0.5cm}
    \includegraphics[width=0.48\textwidth]{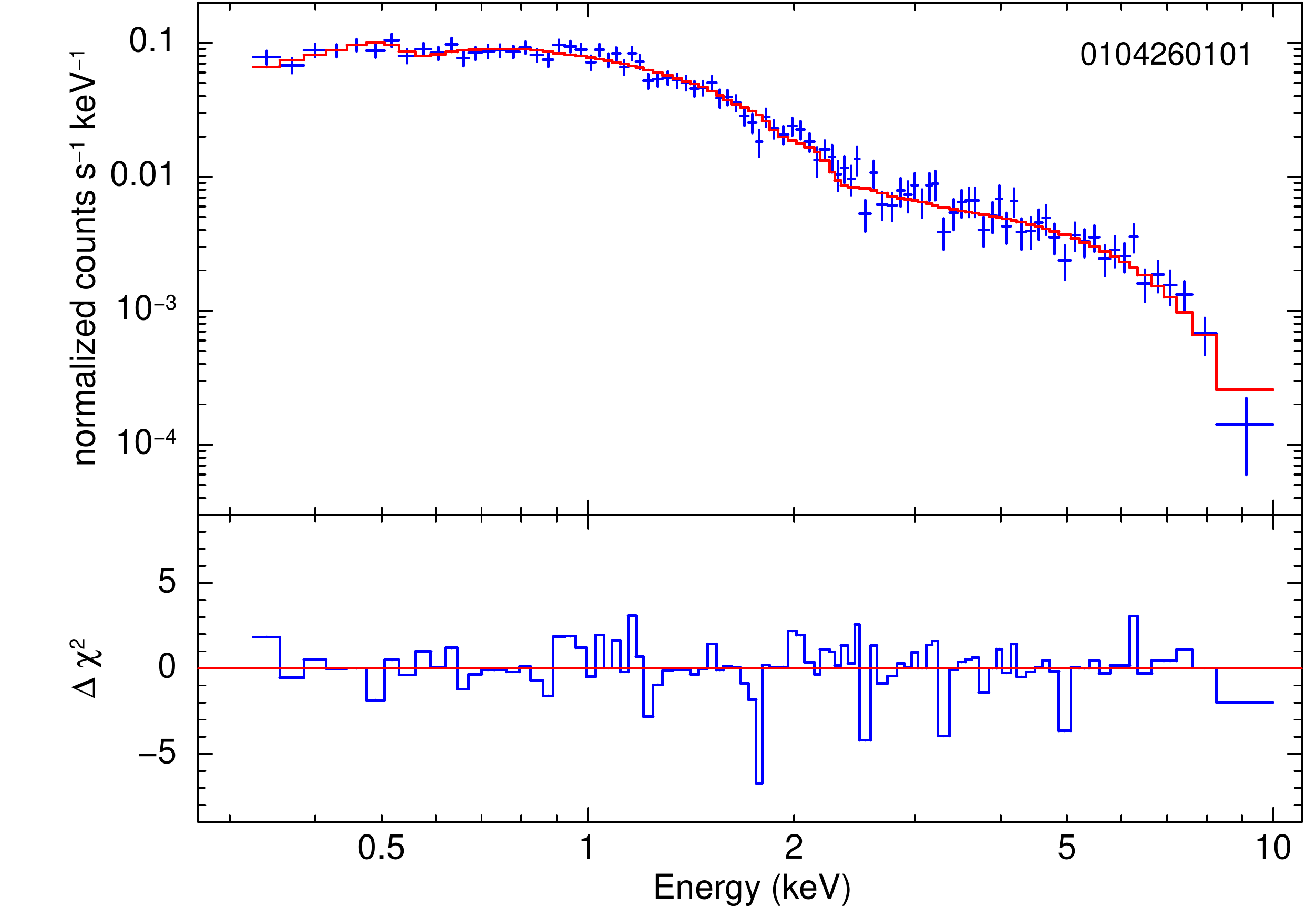}
    \caption{X-ray spectrum and model residuals from {\it XMM-Newton} ObsID 0104260101, fitted with {\it tbabs}$_{\rm Gal} \times$ {\it tbabs} $\times$ ({\it bbodyrad} $+$ {\it diskpbb}). The datasets from pn, MOS1 and MOS2 were combined with the {\sc sas} task {\it epicspeccombine} and grouped to $>$20 counts per bin. The best-fitting parameter values are listed in Table 4. }
    \label{fig:xmm101-spectrum}
\end{figure}

\begin{figure}
\hspace{-0.5cm}
    \includegraphics[width=0.48\textwidth]{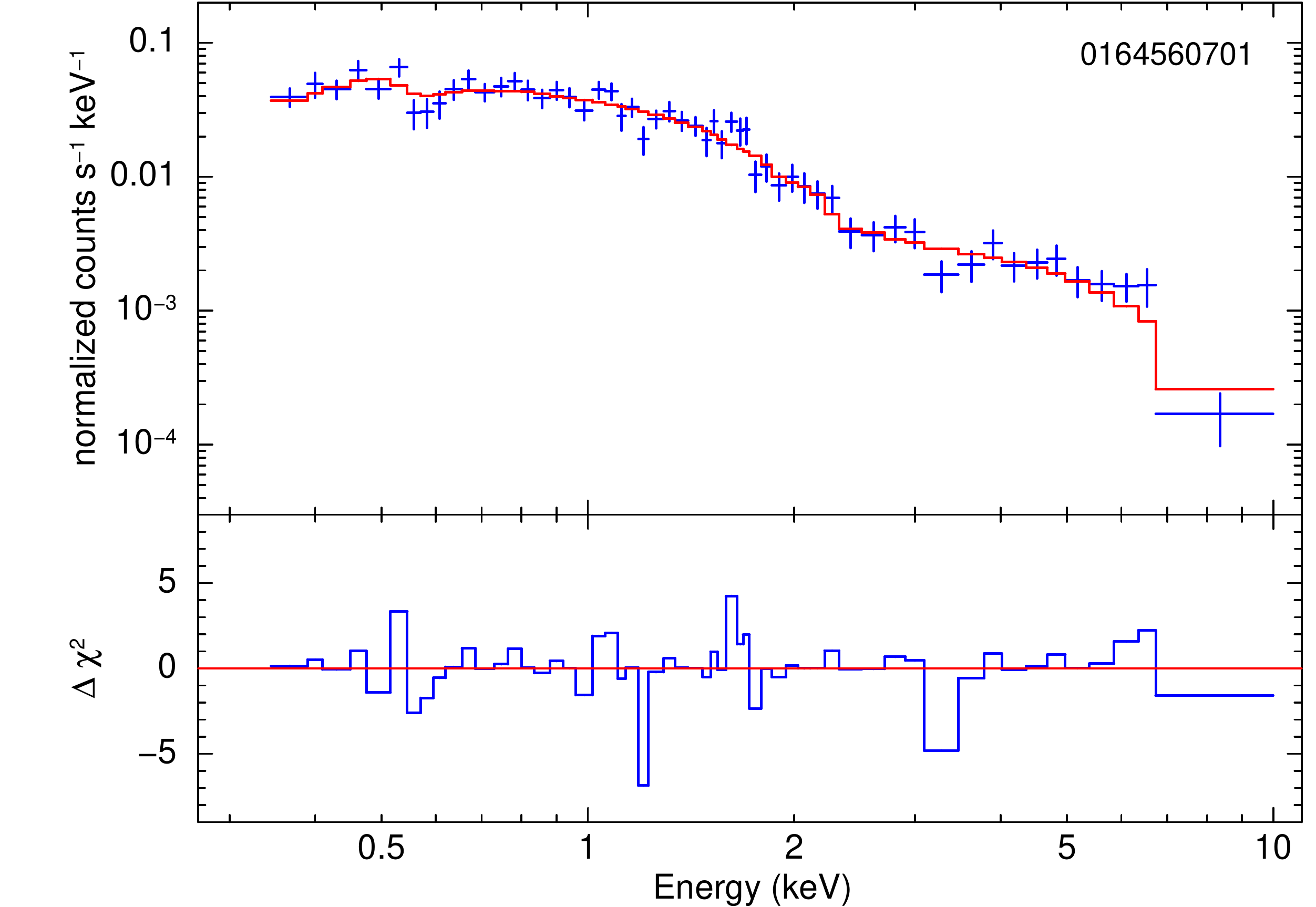}
    \caption{As in Figure 19, for the spectrum from {\it XMM-Newton} ObsID 0164560701.}
    \label{fig:xmm701-spectrum}
\end{figure}

\begin{figure}
    \includegraphics[width=0.48\textwidth]{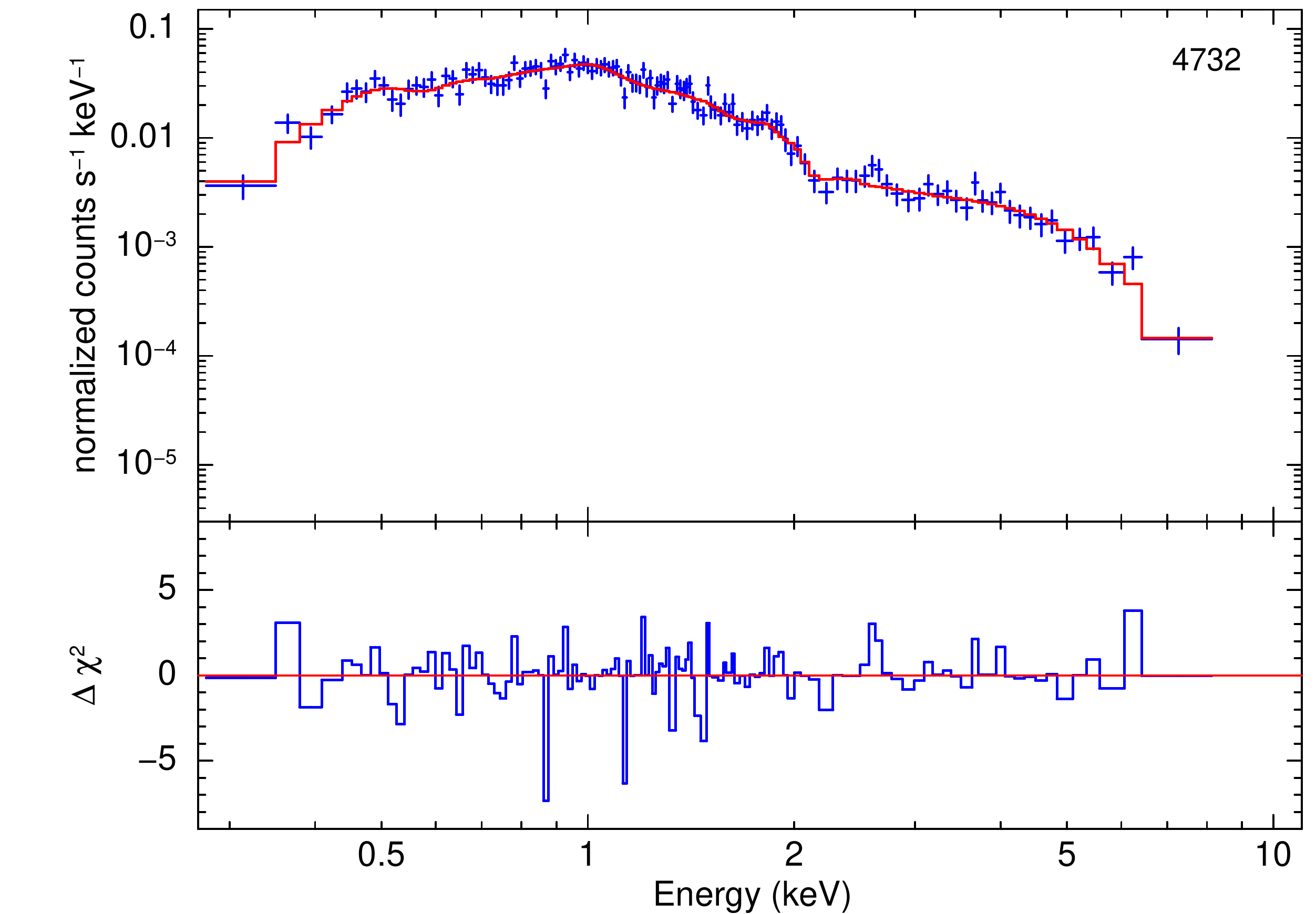}
    \caption{X-ray spectrum and model residuals from {\it Chandra} ObsID 4732, fitted with {\it tbabs}$_{\rm Gal} \times$ {\it tbabs} $\times$ ({\it bbodyrad} $+$ {\it diskpbb}). The data were grouped to $>$20 counts per bin. The best-fitting parameter values are listed in Table 5.}
    \label{fig:4732-spectrum}
\end{figure}

\subsection{Optical counterpart}

We find a single bright optical point source within the 0.24$^{\prime\prime}$ X-ray error circle. The Vega magnitudes are listed in Table \ref{tab:hst}. At first glance, the colours and magnitudes appear consistent with typical optical counterparts of ULXs. We use the latest set of Padova stellar isochrones\footnote{http://stev.oapd.inaf.it/cgi-bin/cmd}, with metallicity $Z=0.019$, to estimate the physical characteristics of the companion star. Both the UVIS and ACS data suggest an age of $\approx$15\,Myr, with a mass of $\approx15$\,M$_{\odot}$, a temperature of $\approx$15,000--25,000\,K and a bolometric luminosity of $\approx$3 $\times10^{38}$\,erg s$^{-1}$.

We also use the combined ACS-WFC and WFC3-UVIS dataset to fit a simple de-reddened blackbody spectrum in {\sc xspec}.  The best-fitting model suggests that the emission is coming from a surface with a radius of $\approx34^{+38}_{-21}$\,R$_{\odot}$ at a temperature of $\approx$16,800$^{+8000}_{-4000}$\,K and a luminosity of $(3.1\pm0.5)\times10^{38}$\,erg s$^{-1}$. This is in agreement with the kind of star expected from the isochrones results. We find a slight (factor of $<$1.2) excess in emission in the F814W band compared with the expected blackbody flux. This is likely to be due to contamination from at least one neighbouring (red) star, partially blended with the much brighter and bluer counterpart. We do not know what the X-ray luminosity was at the time of the optical observations. However, even if the X-ray source was at its peak levels of $\sim$3 $\times 10^{39}$ erg s$^{-1}$, the contribution of an irradiated outer disk to the optical luminosity must be at least an order of magnitude lower than the intrinsic optical/UV luminosity of the donor star. In other words, the brightness of the optical counterpart suggests that J1403 is a high-mass X-ray binary.

Previous studies \citep{2014MNRAS.442.1054H} had suggested a bright infrared source, consistent with a red supergiant, as the counterpart of J1403. In fact, follow-up {\it HST} observations showed four bright stars, potential optical candidates, within a 0.7$^{\prime\prime}$ X-ray positional error circle \citep{2020MNRAS.497..917L}: the red supergiant and three bluer sources. With our improved X-ray astrometric precision (error circle of 0.24$^{\prime\prime}$), we have ruled out the red supergiant and have identified instead as the true optical counterpart the bright blue star labelled `Source D' in \citet{2020MNRAS.497..917L}.





\begin{table}
\caption{Spectral parameters for the {\it XMM-Newton}/EPIC observations, fitted with {\it tbabs}$_{\rm Gal} \times$ {\it tbabs} $\times$ ({\it bbodyrad} $+$ {\it diskpbb}). Uncertainties are the 90\% confidence limits for one interesting parameter.}
\vspace{-0.2cm}
\begin{center}
\begin{tabular}{lcc}  
\hline \hline\\[-5pt]    
Parameter & \multicolumn{2}{c}{Value in Each ObsID}\\
   & 0104260101 & 0164560701 \\
\hline  \\[-5pt]
$N_{\rm {H,Gal}}^a$ &  [0.09]  &  [0.09] \\ [4pt]
$N_{\rm {H,int}}^a$ & $<$0.01 &  $<$0.02 \\[4pt]
%
%
%
%
%
%
$kT_{\rm{bb}}^b$  & $1.69^{+0.26}_{-0.16}$ &  $1.62^{+0.41}_{-0.21}$\\[4pt]
%
%
$N_{\rm{bb}}^c$  & $2.7^{+1.1}_{-0.8} \times 10^{-3}$  & $1.6^{+1.2}_{-0.7} \times 10^{-3}$   \\[4pt]
$kT_{\rm{in}}^d$ & $0.50^{+0.06}_{-0.03}$ & $0.49^{+0.08}_{-0.07}$\\[4pt]
$p$  & $<$0.52 & $<$0.53  \\[4pt]
$N_{\rm{dpbb}}^e$  & $6.1^{+5.5}_{-1.9} \times 10^{-2}$  &  $3.6^{+4.8}_{-1.8} \times 10^{-2}$   \\[4pt]
%
%
%
%
%
\hline\\[-5pt]
$\chi^2_{\nu}$   & 0.98 (87.3/89) & 1.16 (58.2/50)\\[4pt]
%
\hline\\[-5pt]
$F_{0.3-10}^f$  & $38.7^{+2.0}_{-2.0}$  & $18.7^{+1.5}_{-1.4}$ \\[4pt]
$L_{0.3-10}^g$   & $27.8^{+1.2}_{-1.2}$  & $13.5^{+0.9}_{-0.9}$  \\[4pt]
\hline\\[-5pt]
\end{tabular} 
\label{tab:xmm_spec}
\end{center}
\begin{flushleft} 
$^a$: units of $10^{22}$ cm$^{-2}$. \\
$^b$: temperature of the blackbody component ({\it bbodyrad}); units of keV.\\
$^c$: $N_{\rm {bb}} = (R_{\rm{bb}}/d_{10})^2$, where $R_{\rm{bb}}$ is the blackbody radius (units of km), and $d_{10}$ the distance to the galaxy (units of 10 kpc; here, $d_{10} = 680$). \\
$^d$: peak temperature of the slim disk component ({\it diskpbb}); units of keV.\\
$^e$: $N_{\rm {dpbb}} = (r_{\rm{in}}/d_{10})^2 \cos \theta$, where $r_{\rm{in}}$ is the ``apparent'' inner radius of the disk (units of km), $d_{10}$ the distance to the galaxy (units of 10 kpc; here, $d_{10} = 680$), and $\theta$ is our viewing angle. \\
$^f$: observed flux in the 0.3--10 keV band; units of $10^{-14}$ erg cm$^{-2}$ s$^{-1}$.\\
$^g$: de-absorbed luminosity in the 0.3--10 keV band, defined as $4\pi d^2$ times the de-absorbed model flux; units of $10^{38}$ erg s$^{-1}$.\\
\end{flushleft}
\end{table}

\section{Discussion}

\subsection{Comparison with low-frequency QPOs in other systems}


QPOs \citep[see {\it e.g.}][for a recent review]{ingram19} have been studied in stellar-mass BHs and NSs since their first discovery in low-mass X-ray binaries with NS accretors, in the early '80s \citep{Motch1983}.
Although there are open questions about the genesis of QPOs in both BH and NS binaries, 
their study  has the potential to provide clues on the nature of the accretor, and even constrain its mass and spin. 
While being highly statistically significant, the QPOs detected in J1403 do not show any property that allows us to unambiguously classify them. Their most relevant characteristic is that they are detected at a frequency lower than that of the large majority of QPOs observed in Galactic accretors. Several QPO models have been proposed over the years (\citealt{ingram19}), and some of them seem capable of producing QPO behavior with characteristics consistent with those we find in J1403.  

Let us assume first that the 600-s QPOs seen in J1403 are an extreme case of some previously known classes of low-frequency QPOs. Low-frequency QPOs in BH X-ray binaries are observed at frequencies $\gtrsim$0.1~Hz. They are divided into three classes: Type A, Type B, and Type C \citep[][e.g.]{casella05}. Such classes roughly correspond to three QPO classes observed in NS X-ray binaries (flaring branch, normal branch, and horizontal branch oscillations, respectively, see \citealt{motta2017}). Type-C QPOs are the most common class of BH QPOs: they are seen in the low/hard state, hard-intermediate state, and high/soft state of BH transients \citep{belloni05,belloni10}. Their centroid frequency increases with luminosity, starting from as low as $\sim$0.1 Hz in the low/hard state at the beginning and end of an outburst. Type-A and Type-B QPOs generally appear in the soft-intermediate state, at higher frequencies (around 5--7 Hz) and remain currently unexplained.

\begin{figure}
    \centering
    \includegraphics[width=0.48\textwidth]{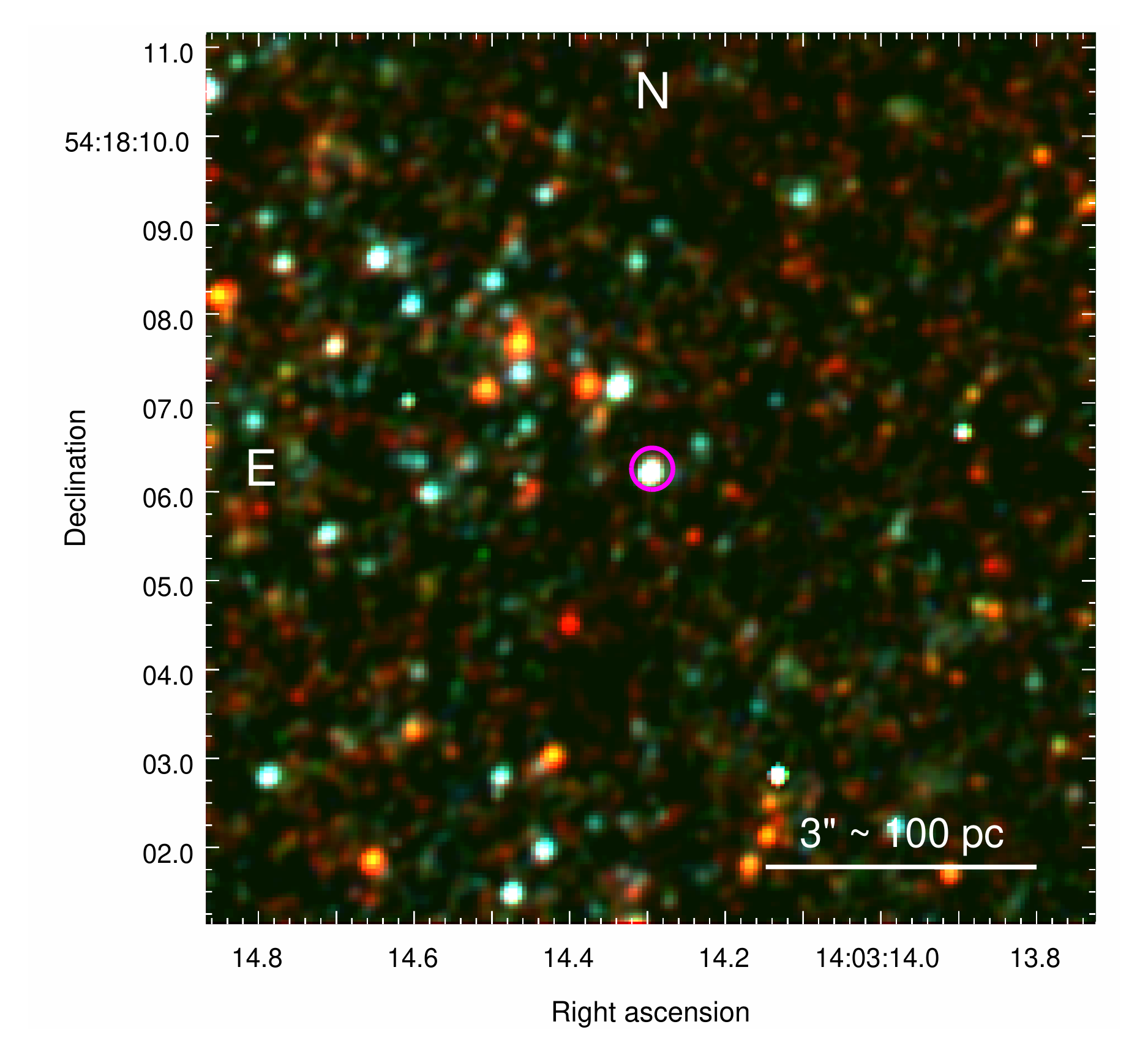}
    \caption{\textit{HST}/ACS-WFC RGB image: blue represents the F435W filter, green is F555W and red is F814W. The 0$\farcs${24} magenta region indicates the X-ray position of J1403. A bright blue optical source is coincident with the X-ray position, likely from a blue giant companion star, optical emission from the outer accretion disk or a combination of the two.}
    \label{fig:hst}
\end{figure}

Lense-Thirring precession \citep{lense18,wilkins72}, a relativistic frame-dragging effect expected to occur in the vicinity of a spinning compact mass, has been invoked as a possible explanation for Type-C QPOs \citep{stella98,ingram09,ingram16,middleton18}. In particular, \cite{ingram09} proposed that Type-C QPOs are associated with a global, rigid precession of a misaligned, geometrically thick inner disk. 
Test-particle models of relativistic precession have also been considered and have been shown to be limiting cases of the global disk precession models \citep{Mottalt2018MNRAS.473..431M}. Although this effect is likely to occur in virtually all accreting NS and BH systems, we cannot directly relate it to the QPOs in J1403 as it would be difficult to explain the very low frequency of the QPO in this system. Such frequencies would require an extremely slowly spinning compact object \citep[Fig. 8 in ][]{Mottalt2018MNRAS.473..431M}, or extreme values of the disc aspect ratio $H/R$, of the viscosity $\alpha$, or of the surface density profile. Much lower Lense-Thirring precession frequencies (consistent with those in J1403) are predicted if the X-ray luminosity oscillations come from the photosphere of a precessing outflow, rather than the precessing inner disk \citep{middleton19}.

The expected precession frequency in Lense-Thirring models is proportional to $a\,r^{-3}$ where $a$ is the spin parameter and $r$ is either the spherization radius of the disk (in disk precession models) or the photosphere of the precessing outflow \citep{middleton19}. Given the observational uncertainties in both parameters, it is easy to find values of $a$ and $r$ that satisfy almost any observed frequency, from minutes to days, including the 600-s range seen in J1403. However, it is precisely the steep dependence on $r$ that makes those models unsuitable to our case. The radius of the geometrically thick part of the precessing flow (spherization radius in super-critical accretion) scales with $\dot{m}$ \citep{poutanen07}; the photospheric radius of the precessing outflow scales with $\dot{m}^{3/2}$ \citep[Eq.~13 in ][]{middleton19}. So, we expect the Lense-Thirring precession frequency to have a strong dependence on $\dot{m}$. On the other hand, the luminosity in the ultraluminous regime is a much weaker function of $\dot{m}$: $L \propto L_{\rm Edd} \, (1+\ln \dot{m})$ for BHs and weakly magnetized NSs \citep{poutanen07}, dominated by massive disk outflows at the spherization radius.
Therefore, in the Lense-Thirring scenario, we expect that changes in the QPO frequency from epoch to epoch should be much larger than the corresponding luminosity changes, for a change of $\dot{m}$. This is not the case in J1403, where the strongest QPOs are detected within a small range of frequencies ($\approx$1.5--1.7 mHz) despite changes in X-ray luminosity by factor of 3 between the corresponding epochs (years apart).

Next, we consider the possibility that the 1.6-mHz oscillations in J1403 are not related to BH Type-C QPOs. Instead, we look for alternative mechanisms of very-low-frequency variability in NSs. Some NS accretors exhibit QPOs at a range of frequencies from about 5 mHz to 15 mHz \citep{Revnivtsev2001}, an order of magnitude lower than Type-C QPOs. 
It has been proposed that such QPOs are associated with the marginally-stable nuclear burning of matter on the surface of the NS in an accretion regime near or an order of magnitude below the Eddington limit \citep{Heger2007,keek14,mancuso19,tse21}. The ignition of matter in different regions of the NS surface produces a quasi-periodic modulation of the emission coming directly from the crust, observable as a QPO with a soft energy spectrum \citep{Heger2007}.  
However, because accretion onto NSs produces more energy per unit mass than does nuclear burning, each burst (which is essentially Eddington-limited) provides only a small fraction of the energy produced by extraction of gravitational potential energy. This would make it difficult to identify the signature of marginally stable burning in a NS accretor located in galaxies as distant as M\,101. Furthermore, theoretical work \citep{Heger2007} predicts that the QPOs arising from marginally stable burning would display frequencies a factor of a few larger than what observed in J1403. Thus, while we cannot exclude such an origin for the QPO in J1403, we lack circumstantial evidence further supporting such an interpretation. 

A comparison with the very-low-frequency QPOs in the high mass X-ray binary LMC X-4 is perhaps more relevant for our system. LMC X-4 is powered by a NS with a spin period of $\approx$13.5 s, a binary period of 1.4 d and a superorbital period of $\approx$30.5 d \citep[{\it e.g.},][]{naik04,falanga15,brumback18}. It is one of the few high-mass X-ray binaries fed via disk accretion rather than wind accretion (a possible similarity with ULX pulsars), from an $\approx$18 $M_{\odot}$ companion filling its Roche lobe. LMC X-4 has a persistent X-ray luminosity of $\sim$10$^{38}$, with occasional aperiodic flares peaking at $\approx$2 $\times 10^{39}$ erg s$^{-1}$ \citep{moon03}. During the super-Eddington flares, very-low-frequency QPOs were detected in two frequency bands ($\approx$0.65--1.35 mHz and $\approx$2--20 mHz; \citealt{moon01}), overlapping with the QPO frequencies in J1403. A possible scenario proposed for the lowest frequency band was the beating between the spin frequency of the NS and the Keplerian frequency of the disk at its inner radius (magnetospheric radius), which would extend a little inside the corotation radius during the flares.

Only a few other systems are known to produce quasi-periodic modulations at frequencies $\sim$1 mHz. These modulations may not always be strictly classified as QPOs, but are nonetheless characterised by a preferred frequency. Particularly noteworthy is the case of the BH X-ray binary GRS\,1915$+$105, which is known to exhibit several different variability classes, some of which highly repetitive and with a period in the range of 10--1000 s \citep{Belloni2000}. Among these classes, the $\rho$ class, sometimes referred to as the ``heartbeats'', is the most regular (and arguably the most studied), and appears with a (mildly variable) period between $\approx$40 s and $\sim$1000 s. This type of variability has been identified in two additional Galactic X-ray binaries: a relatively low-luminosity  BH X-ray binary, IGR\,J17091$-$3624 \citep{Altamirano2011,janiuk15}, and the NS X-ray binary MXB\,1730$-$335 (also called the Rapid Burster, \citealt{Bagnoli2015}). 
A similar variability pattern has been observed in a ULX, 4XMM\,J111816.0$-$324910, with a characteristic timescale of $\approx$3500\,s \citep[][]{Motta2020}, and in a number of Active Galaxies (particularly noteworthy the case of GSN\,069, \citealt{Miniutti2019}).
 
Because of its extreme brightness, the spectral and timing properties of GRS~1915$+$105 have been studied over the last three decades with every major X-ray telescope (the {\it Rossi X-ray Timing Explorer}, {\it ASCA}, {\it Suzaku}, {\it XMM-Newton}, {\it Chandra}, {\it Nicer}, etc.). The inner disk is radiation-dominated, the luminosity is around or slightly above the Eddington limit, and there is evidence of winds from the outer disk. Absorption lines associated with winds have been observed at the time of broadband spectral changes \citep{ueda09}. The heartbeat quasi-periodic variability has been modelled as a limit-cycle behaviour associated with interactions between radiation and accreting matter at super-Eddington accretion rates \citep{done04,2011ApJ...737...69N, 2012MNRAS.421..502N}.  This model could in principle offer a possible explanation for the QPOs in J1403. However, as discussed in \cite{Motta2020} for the case of the ULX in 4XMM\,J111816.0$-$324910, the poor knowledge of the system itself makes it impossible to draw any firm conclusion on the matter.

Finally, among the extragalactic ULX class, M\,74 X-1 (CXOU J013651.1$+$154547) has shown perhaps the most impressive example of intra-observational flaring, on timescales of $\approx$5000 s $\approx$0.2 mHz \citep{krauss05} in at least one {\it Chandra} observations (but not in others). There is currently no information on the nature of this compact object and of its donor star, and the nature of the above modulation remains unclear. Other examples of ULXs flaring on timescales of $\sim$10,000 s are NGC\,7456 ULX-1 \citep{pintore20} and NGC\,4559 X7 \citep{pintore21}. However, the flaring variability in those sources appears stochastic (possibly related to inhomogeneities in the super-critical disk outflow), without the quasi-periodic pattern so strongly detected in J1403.
\begin{sidewaystable*}
\vspace{15cm}
\caption{Spectral properties in selected {\it Chandra}/ACIS observations. The seven spectra with $>$300 counts were grouped to a minimum of 20 counts per bin, and fitted with {\it tbabs}$_{\rm Gal} \times$ {\it tbabs} $\times$ ({\it apec} $+$ {\it bbodyrad} $+$ {\it diskpbb}), with the $\chi^2$ statistics. The other seven spectra were grouped to 1 count per bin, and fitted with {\it tbabs}$_{\rm Gal} \times$ {\it tbabs} $\times$ {\it power-law}, with the Cash statistics. Uncertainties are the 90\% confidence limits for one interesting parameter.}
\scriptsize{
\begin{center}
\begin{tabular}{lcccccccccccccccc}  
\hline \hline\\[-5pt]    
Parameter & \multicolumn{13}{c}{Value in Each ObsID}\\
  & 934 & 2065 & 4731 & 4732 & 4737 & 5296 & 5297 & 5309 & 5340 & 6114 & 6115 & 6169 & 6170 & 6175 \\
\hline  \\[-5pt]
$N_{\rm {H,Gal}}^a$ &  [0.09]  &  [0.09] &  [0.09] & [0.09] & [0.09] &  [0.09] &  [0.09] &  [0.09] &  [0.09] &  [0.09] &  [0.09] &  [0.09] & [0.09] &  [0.09] \\ [4pt]
$N_{\rm {H,int}}^a$ & $<$0.13 & $<$0.37  &  $<$0.12 & $<$0.02 & $<$0.09 & $<$0.12 & $<$0.09 & $<$0.01 & $<$0.23 & $<$0.09 & $<$0.27 & $<$0.07   & $<$0.04 & $<$0.07 \\[4pt]
$kT_{\rm{apec}}^b$ & -- & -- & -- & $1.19^{+0.19}_{-0.16}$ & -- & -- & $0.68^{+0.20}_{-0.36}$ & $1.75^{+2.50}_{-0.54}$ & -- & -- & -- & -- & -- & -- \\[4pt]
$N_{\rm{apec}}^c$  & -- & -- & -- & $1.3^{+0.7}_{-0.6}$ & -- & -- & $0.49^{+0.28}_{-0.29}$ & $2.2^{+7.8}_{-1.5}$& -- & -- & -- & -- & -- & -- \\[4pt]
%
%
%
%
%
$kT_{\rm{bb}}^d$  & -- & -- & -- & $1.45^{+0.26}_{-0.16}$   & $0.73^{+0.27}_{-0.14}$ & -- & $1.54^{+1.65}_{-0.54}$ & $1.65^{+0.47}_{-0.21}$ & -- & -- & -- & $1.00^{+0.32}_{-0.15}$ & $1.80^{+1.00}_{-0.40}$ & $1.43^{+0.69}_{-0.33}$ \\[4pt]
$N_{\rm{bb}}^e$ $\left(10^{-3}\right)$ & -- & -- & -- & $6.2^{+3.4}_{-2.4}$ & $25^{+34}_{-20}$ & -- & $2.0^{+5.8}_{-1.6}$ & $3.8^{+2.5}_{-2.0}$ & -- & -- & -- & $11.5^{+8.4}_{-8.2}$ & $1.7^{+2.7}_{-1.1}$ &   $3.7^{+5.2}_{-2.5}$\\[4pt]
$kT_{\rm{in}}^f$  & -- & -- & -- & $0.45^{+0.07}_{-0.07}$ & $0.23^{+0.14}_{-0.07}$ & -- & $0.36^{+0.27}_{-0.13}$ & $0.49^{+0.10}_{-0.10}$ & -- & -- & -- & $0.34^{+0.09}_{-0.10}$ &$0.48^{+0.16}_{-0.12}$ & $0.41^{+0.11}_{-0.07}$  \\[4pt]
$p$  & -- & -- & -- & $<$0.54 & [0.5] & -- & [0.5] & $<$0.52 & -- & -- & -- & $<$0.8 & $<$0.59 & $<$0.64  \\[4pt]
$N_{\rm{dpbb}}^g$  & -- & -- & -- & $0.10^{+0.19}_{-0.04}$ & $0.9^{+7.5}_{-0.8}$ & -- & $0.08^{+0.44}_{-0.07}$ & $0.07^{+0.11}_{-0.04}$ & -- & -- & -- & $0.24^{+3.15}_{-0.16}$ & $0.03^{+0.16}_{-0.02}$ & $0.16^{+0.78}_{-0.11}$   \\[4pt]
%
%
%
$\Gamma_{\rm{pl}}^h$  &  $2.49^{+0.58}_{-0.36}$ &  $1.9^{+2.4}_{-0.7}$ &  $1.40^{+0.35}_{-0.34}$ & --  &  $2.23^{+0.25}_{-0.23}$ & $0.85^{+0.83}_{-0.80}$ & -- & -- & $1.64^{+1.18}_{-0.47}$ & $2.65^{+0.77}_{-0.50}$ & $1.29^{+0.73}_{-0.53}$ & -- & -- & --  \\[4pt]
$N_{\rm{pl}}^i$  & $3.0^{+1.3}_{-0.6}$  & $2.5^{+6.5}_{-1.0}$  & $2.0^{+0.6}_{-0.5}$ & --
& $33.1^{+4.0}_{-3.7}$ & $8.7^{+7.5}_{-4.9}$ & -- & -- &$1.6^{+1.9}_{-0.4}$ & $1.8^{+0.7}_{-0.4}$ & $1.1^{+1.2}_{-0.4}$ & -- & -- & -- \\[4pt]
\hline\\[-5pt]
$\chi^2_{\nu}$  & -- & -- & -- & 0.95 (112.6/118) & 1.06 (10.6/10) & -- & 1.04 (12.5/12) & 0.77 (94.5/123) & -- & -- & -- & 1.04 (36.4/35) & 0.65 (25.3/39) & 1.15 (63.0/55) \\[4pt]
C-stat/dof   & 0.88 (87.0/99) & 1.43(17.11/12) & 0.71 (49.1/69) & -- & -- & 2.1 (22.9/11) 
& -- & -- & 0.84 (38.8/46) & 0.92 (54.9/50) & 0.83 (23.3/28) & -- & -- & -- \\[1pt]
\hline\\[-5pt]
$F_{0.3-10}^j$  & $0.89^{+0.22}_{-0.18}$  & $1.2^{+1.2}_{-0.7}$ & $1.71^{+0.70}_{-0.49}$ & $47.4^{+2.6}_{-3.1}$
& $11.8^{+1.9}_{-1.8}$  &  $15.2^{+20.8}_{-8.5}$ & $16.5^{+7.4}_{-4.4}$ & $48.3^{+2.1}_{-3.3}$ & $0.58^{+0.21}_{-0.20}$ & $0.56^{+0.20}_{-0.14}$ & $1.10^{+0.76}_{-0.46}$ & $22.3^{+2.6}_{-2.2}$ & $23.2^{+5.5}_{-4.9}$ & $29.5^{+5.0}_{-3.3}$\\[4pt]
$L_{0.3-10}^k$   & $0.80^{+0.44}_{-0.16}$  & $0.83^{+0.63}_{-0.36}$ & $1.04^{+0.38}_{-0.26}$ &  $33.5^{+2.0}_{-1.9}$
& $9.5^{+4.7}_{-1.3}$ & $9.4^{+10.7}_{-5.2}$ & $11.2^{+4.0}_{-2.4}$ & $34.0^{+2.5}_{-2.1}$ & $0.45^{+0.76}_{-0.10}$ & $0.49^{+0.33}_{-0.10}$ & $0.66^{+0.41}_{-0.25}$ & $17.1^{+1.6}_{-0.9}$  & $15.6^{+2.9}_{-2.2}$ & $24.5^{+3.2}_{-4.1}$ \\[4pt]
\hline\\[-5pt]
\end{tabular} 
\label{tab:chandra_spec}
\end{center}
\begin{flushleft} 
$^a$: units of $10^{22}$ cm$^{-2}$. \\
$^b$: units of keV. \\
$^c$: units of $10^{-19}/\{4\pi d^2\}\,\int n_en_{\rm H} \, dV$, where $d$ is the luminosity distance in cm, and $n_e$ and $n_H$ are the electron and H densities in cm$^{-3}$.\\
$^d$: temperature of the blackbody component ({\it bbodyrad}); units of keV.\\
$^e$: $N_{\rm {bb}} = (R_{\rm{bb}}/d_{10})^2$, where $R_{\rm{bb}}$ is the blackbody radius (units of km), and $d_{10}$ the distance to the galaxy (units of 10 kpc; here, $d_{10} = 680$). \\
$^f$: peak temperature of the slim disk component ({\it diskpbb}); units of keV.\\
$^g$: $N_{\rm {dpbb}} = (r_{\rm{in}}/d_{10})^2 \cos \theta$, where $r_{\rm{in}}$ is the ``apparent'' inner radius of the disk (units of km), $d_{10}$ the distance to the galaxy (units of 10 kpc; here, $d_{10} = 680$), and $\theta$ is our viewing angle. \\
$^h$: power-law photon index (for spectra with a low number of counts).\\
$^i$: power-law normalization; units of $10^{-6}$ photons keV$^{-1}$ cm$^{-2}$ s$^{-1}$ at 1 keV.\\
$^j$: observed flux in the 0.3--10 keV band; units of $10^{-14}$ erg cm$^{-2}$ s$^{-1}$.\\
$^k$: de-absorbed luminosity in the 0.3--10 keV band, defined as $4\pi d^2$ times the de-absorbed model flux; units of $10^{38}$ erg s$^{-1}$.\\
\end{flushleft}
}
\end{sidewaystable*}  

\begin{table}
    \centering
    \scriptsize{
    \caption{Apparent brightness of the optical counterpart (Vegamag system), measured from the \textit{HST} observations.}
    \label{tab:hst}
    \begin{tabular}{lcccccc} 
        \hline\hline\\[-5pt]
        Obs.~Date & \multicolumn{5}{c}{Filter}               \\
         & F275 & F336 & F435/438   & F555  & F814               \\
        \hline\\[-5pt]
        2002-11-13$^a$ &&& $22.9 \pm 0.1$ & $22.9 \pm 0.1$ & $22.7 \pm 0.1 $ \\
        2014-02-15$^b$ & $21.3\pm0.1$ & $21.5 \pm 0.1$ &&& \\
        2016-09-24$^b$ &&& $22.8 \pm 0.1$ & $22.8 \pm 0.1$ &                   \\
        \hline
        
    \end{tabular}}
\begin{flushleft}
$^a$ ACS/WFC\\
$^b$ WFC3/UVIS
\end{flushleft}
\end{table}


\begin{figure}
    \centering
    \vspace{-0.4cm}
    \includegraphics[width=0.33\textwidth, angle=270]{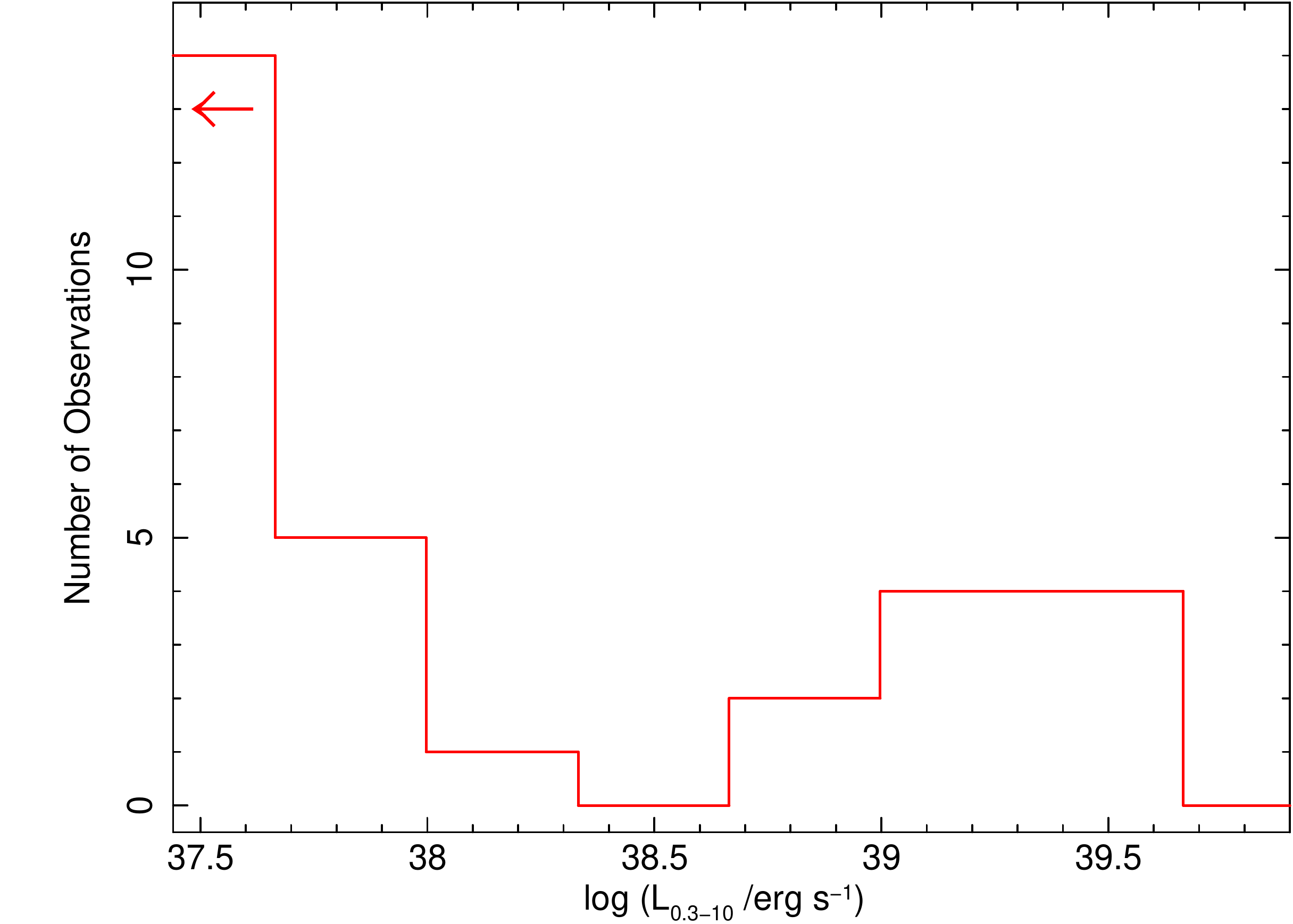}\\\vspace{0.5cm}
    \includegraphics[width=0.33\textwidth, angle=270]{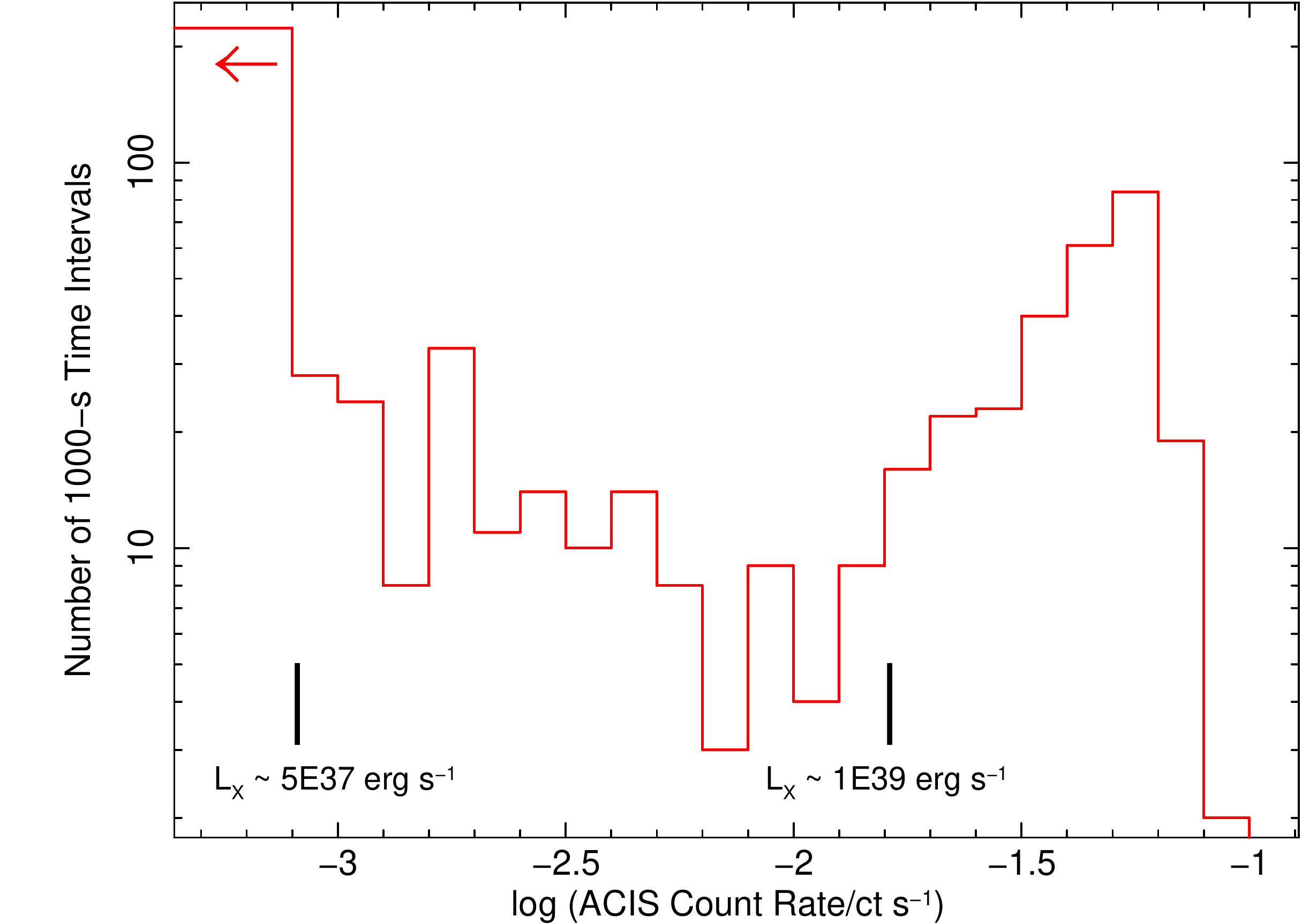}\\\vspace{1cm}
    \caption{Top panel: distribution of the average 0.3--10 keV de-absorbed luminosities of all the 28 {\it Chandra} observations and two 
    {\it XMM-Newton} observations. (The first bin on the left corresponds to observations with $L_{\rm X} < 4.6 \times 10^{37}$ erg s$^{-1}$.)
    Bottom panel: distribution of 0.3--7 keV background-subtracted ACIS count rates in 1000-s intervals, for the 14 {\it Chandra} observations with higher luminosity. Two black markers identify the approximate count rates corresponding to 0.3--10 keV de-absorbed luminosities of $5 \times 10^{37}$ erg s$^{-1}$ and $1 \times 10^{39}$ erg s$^{-1}$. To account for the decline in sensitivity of the ACIS detector, the observed count rates were rescaled to ``equivalent'' Cycle-5 count rates, with {\sc pimms}. (The first histogram bin on the left corresponds to the number of 1000-s intervals with no detected net counts.) }
    \label{fig:histograms}
\end{figure}

\subsection{Nature of the compact object}

The X-ray spectral properties of J1403 do not give substantial clues about the nature of the accretor. A slightly curved 0.3--10 keV spectrum, consistent with a double thermal model or with Comptonization, is typical of ULXs \citep{walton18} but does not discriminate between a NS and a BH accretor. Both classes of compact objects would be in the ultraluminous regime in the epochs when $L_{\rm X} > 10^{39}$ erg s$^{-1}$. As for the epochs in which $L_{\rm X} < 10^{38}$ erg s$^{-1}$, the lack of counts does not permit us to discriminate between NS and BH models. The general evolution of the flares, with the rapid rise from $L_{\rm X} < 10^{38}$ erg s$^{-1}$ to $L_{\rm X} > 10^{39}$ erg s$^{-1}$, and equally rapid declines, does not resemble the behaviour of BH transients \citep{tetarenko18} in Galactic low-mass X-ray binaries, and is also very different from the few confirmed BHs in high-mass systems (Cyg X-1, LMC X-1, LMC X-3). Giant flares with peak X-ray luminosities sometimes in excess of $10^{39}$ erg s$^{-1}$ are a feature of NSs in Be X-ray binaries \citep{martin14}, which are associated with moderately young stellar populations. Even in those cases, though, a smooth, gradual decline from peak luminosity is observed over several weeks, consistent with the viscous timescale of the (transient) accretion disk \citep{vasilopoulos20,wilson-hodge18,tsigankov17,lutovinov17}.

Perhaps the best clue to the nature of the compact object comes from the luminosity distribution of the various epochs. We have already noted (Figure 2) that J1403 tends to be found at $L_{\rm X} \gtrsim 10^{39}$ erg s$^{-1}$ or at $L_{\rm X} \lesssim 10^{38}$ erg s$^{-1}$, but more rarely in between. In the top panel of Figure \ref{fig:histograms}, we replotted the same average de-absorbed luminosities of all 30 observations as a histogram over the luminosity range. From that, we confirm the presence of a scarcely populated gap of about an order of magnitude in luminosity. If J1403 is an ordinary stellar-mass black hole, the luminosity gap corresponds to the typical high/soft state range, a state in which other black hole transients typically spend several weeks during each complete outburst \citep{remillard06,fender16}. A small caveat for the estimate of the gap size is that the luminosities below the gap have been estimated with a power-law model, while those above the gap with double-thermal model. As we noted in Section 3.3, if the spectra below the gap are also thermal, their luminosity would be about 25\% lower and the gap correspondingly larger.

To obtain an alternative estimate of the gap, we took the {\it Chandra} background-subtracted light curves from the 14 observations with higher signal-to-noise ratio (the same ones listed in Table 4), and rebinned them to 1000 s.  The histogram of the count rate distribution (in log scale) is plotted in the bottom panel of Figure \ref{fig:histograms}. This time, the approximate luminosity conversion is based on the best-fitting spectral model for ObsID 4732 for all bins. The lowest portion of the distribution (bins with $<$20 occurrences) is between net ACIS-S rates (converted to equivalent Cycle 5 rates with {\sc pimms}) of 
$\approx$1.6 $\times 10^{-3}$ ct s$^{-1}$ and $\approx$1.6 $\times 10^{-2}$ ct s$^{-1}$.



A possible interpretation of this bimodal distribution is that J1403 is a NS that switches between accretor and propeller states \citep{illarionov75,ghosh79} between observations--or even within individual observations, as suggested by the rapid count rate evolution for example during {\it Chandra} ObsIDs 4737 (Figure A1) and 6170  (Figure A5). In the rest of Section 4.2, we will assume this is the case. We will explore the consequences of this assumption for other system parameters, to determine whether it is at least a physical scenario (without claiming it should be the only possible one).

The luminosity gap between accretor and propeller state is the ratio between the gravitational energy released from infinity to the surface of the neutron star (of radius $R$), and the energy released only down to the corotation radius $R_{\rm {cor}}$ \citep{corbet96,campana98}. Assuming radiatively efficient accretion in both cases, 
\begin{equation}
    \frac{L_{\rm acc}}{L_{\rm prop}} \approx 2\,R_{\rm {cor}}/R =
    \left( \frac{2GMP^2}{\pi^2R^3}  \right)^{1/3}
    \approx 340 \, P^{2/3} \, M_{1.4}^{1/3} \, R_6^{-1}.
\end{equation}

The assumption of radiative efficiency in the accretor state is because of the hard surface of the neutron star; in the propeller state, because the luminosity is just below Eddington, at a level usually characterized by a standard disk. The factor of 2 in the ratio of the radii (often forgotten in the literature) is because surface accretion releases an amount of energy $\propto M\dot{M}/R$ while an accretion disk radiates only an energy $\propto M\dot{M}/2R$.

For the observed ratio $L_{\rm acc}/L_{\rm prop} \approx 10$,
Equation (1) suggests a spin period $P \approx 5\,M_{1.4}^{-1/2}$ ms. Even if we relax the definition of luminosity gap by a factor of 2, the period is still $\lesssim$15 ms.

The magnetic field can then be estimated from the observed luminosity at the lower limit of the accretor state, just before the propeller transition \citep{stella86,campana02}:
\begin{equation}
L_{\rm acc,m} \approx \frac{GM\dot{M}}{R} 
\approx 3.9 \times 10^{37} \, \xi^{7/2} \, B_{12}^2 \, P^{-7/3} \, M_{1.4}^{-2/3} \, R_6^5  \ \ {\mathrm{erg~s}}^{-1}.
\end{equation}

The empirical factor $\xi$ parameterizes the difference between the classical Alfven radius in spherical acccretion and in a disk geometry. Adopting the best-fitting coefficients of \cite{campana18}, we have
\begin{equation}
L_{\rm acc,m} \approx 1.32^{+0.75}_{-0.48} \times 10^{37}  \, B_{12}^2 \, P^{-7/3} \, M_{1.4}^{-2/3} \, R_6^5  \ \ {\mathrm{erg~s}}^{-1}
\end{equation}
From our {\it Chandra} spectral fitting, the transition luminosity is $L_{0.3-10} \approx 1 \times 10^{39}$ erg s$^{-1}$. Allowing for a bolometric correction (based on the model parameters of Table 1), we take
\begin{equation}
\left(L_{\rm acc,m}\right)_{\rm obs} \approx 1.5 \times 10^{39}  \ \ {\mathrm{erg~s}}^{-1}.
\end{equation}
Hence, from Eq.(3), we obtain a surface field $B \approx 2.3 \times 10^{10}\,P_{\rm 5ms}$ G (where $P_{\rm 5ms}$ is the period in units of 5 ms), or, equivalently, $B \approx 2.3 \times 10^{10}\,M^{1/2}_{1.4}$ G. If we had assumed instead a luminosity ratio between accretor and propeller as large as 20 (taking into account uncertainties in the fitting models and in the exact definition of the luminosity gap), we would have obtained a spin period of $\approx$14 ms and a surface field $\approx$7 $\times 10^{12}$ G.

Typical magnetic fields in NS high-mass X-ray binaries are between $10^{12}$--$10^{13}$ G  \citep{2018MNRAS.480..692P,vandeneijnden21}. There is at least one well-known example of a luminous (at times, super-Eddington) high-mass X-ray binary, SMC X-1, with $B \approx 10^{11}$ G, estimated from its spin-up rate \citep{1997A&A...321L..25L}, although a much higher field was recently proposed ($B \approx 4.2 \times 10^{12}$ G) if an X-ray feature at $\approx 55$ keV is interpreted as a cyclotron line \citep{2020ApJ...895...10P}. However, there are no known examples of high-mass X-ray binaries with a spin frequency as high as $\sim$200 Hz and a surface field as low as $\approx$2 $\times 10^{10}$ G. The high spin frequency is more typical of low-mass X-ray binary, while the inferred field is intermediate between those measured in high- and low-mass systems \citep{mukherjee15,vandeneijnden21}. We will comment on this issue further in Section 4.3.

Using again the parameters of \citet{campana18}, we can estimate the magnetospheric radius in the accretor state:
\begin{equation}
R_{\rm m} = (480 \pm 70) \, B_{12}^{4/7} \, R_6^{10/7} \, L_{39}^{-2/7} \, M_{1.4}^{1/7} \ \  {\mathrm{km}},
\end{equation}
which corresponds to $R_{\rm m} \approx 35$--50 km during the epochs in the ultraluminous regime. It is plausible that at those epochs, the inner disk was in the super-critical, advective regime (slim disk). An alternative expression for the magnetospheric radius in the slim disk model is:
\begin{equation}
R_{\rm m} \approx 560 (\alpha/0.1)^{2/7} \, \dot{m}^{-2/7} \, B_{12}^{4/7} \, R_6^{12/7} \, M_{1.4}^{-3/7} \ \ {\mathrm{km}}  
\end{equation}
\citep{takahashi17}, where $\alpha$ is the viscosity parameter ($\alpha \sim 0.1$--0.3) and $\dot{m} \ge 1$ represents super-critical accretion ($\dot{m} \equiv \dot{M}/\dot{M}_{\rm{Edd}}$, and $\dot{M}_{\rm{Edd}}$ for a NS is defined such that $L_{\rm{Edd}} = 0.2 \dot{M}_{\rm{Edd}}\,c^2$). For the ultraluminous epochs ($\dot{m} \approx 10$--30), Equation(6) gives similar estimate of the magnetospheric radius, $R_{\rm m} \approx 25$--$35 \, (\alpha/0.1)^{2/7}$ km.  

An alternative derivation of the accretor/propeller transition threshold and magnetospheric radius in the case of radiation pressure dominated disks was presented by \citet{2017MNRAS.470.2799C}. Inserting our estimated value of $P \approx 5 \times 10^{-3}$ s into their Equation (67), 
we obtain a magnetic field 
\begin{equation}
B_{12} \approx 5.6 \times 10^{-3} \, M_{1.4}^{7/6} \, R_6^{-3} \, 
\left[ \frac{16.0 M_{1.4}^{-1/3}}{(\alpha/0.1)} - \frac{L}{L_{\rm{Edd}}}  \right]^{1/2} \ \ {\mathrm{G}},   
\end{equation}
where $L/L_{\rm{Edd}}$ is the peak Eddington ratio in the propeller state. The $L/L_{\rm{Edd}}$ term in brackets is negligible in our case, and  Equation (7) gives $B \approx 2.2 \times 10^{10} \,  (\alpha/0.1)^{-1/2}\, M_{1.4}^{-1} \, R_6^{-3} \,$ G, consistent with our previous estimate. From Equation (61) of \citet{2017MNRAS.470.2799C}, we then obtain a magnetospheric radius $R_{\rm m} \approx 50\,M_{1.4}^{1/3}$ km, independent of accretion rate. This is again roughly consistent (within a factor of 2) with our previous derivations.

Finally, we compare the magnetospheric radius (inner radius of the disk) with the spherization radius $R_{\rm {sph}}$, to determine whether we expect strong outflows from the truncated disk, similar to what we expect from a super-critical disk around a black hole. Using again the model of radiation-pressure-dominated disk in \citet{2017MNRAS.470.2799C}, the condition $H/R \approx 1$ (standard definition of spherization radius) in the inner disk is reached for $\dot{m} \approx 10$, corresponding to $L \approx 2 \times 10^{39}$ erg s$^{-1}$ (note that the definition of $\dot{m}$ in \citet{2017MNRAS.470.2799C} is different from our definition because we have included the efficiency factor $\eta \approx 0.2$). We compare this estimate with the well-known derivation of $R_{\rm {sph}}/R_{\rm in} \approx (5/3)\,\dot{m}$, from \cite{poutanen07}. The latter expression was derived for a black hole disk, and requires some adaptation for the case of a disk truncated at the magnetospheric radius. However, we can follow \cite{poutanen07} and adopt their definition of $R_{\rm {sph}}$ as the radius where the integrated disk luminosity $L(R>R_{\rm {sph}}) = L_{\rm Edd}$ (independent of the magnetospheric radius). For J1403, the bolometric disk luminosity in the propeller state is always below Eddington (up to $L \approx 1.5 \times 10^{38}$ erg s$^{-1}$ at the transition to accretor state). Thus, there is no spherization radius and probably no strong outflow in the propeller state. In the accretor state, we expect the disk component of the luminosity to become super-Eddington for $L \approx 2 \times 10^{39}$ erg s$^{-1}$ ($\dot{m} \approx 10$), in agreement with the estimate from \citet{2017MNRAS.470.2799C}. 

Recapping our argument, we have suggested that the two characteristic radii in J1403, when observed in the ultraluminous regime, are a magnetospheric radius at $\sim$30--50 km (inside the corotation radius and quite close to the NS surface, because of the low magnetic field) and a spherization radius located an order of magnitude further out (corresponding to $\dot{m} \sim 10$). It is an interesting coincidence that these two values are similar to the  characteristic radii of the two thermal components used for our spectral modelling (Section 3.3). A common interpretation of the higher temperature component in pulsar ULXs \citep{walton18} is that it comes from the geometrically thick portion of the (non-standard) disk between the spherization radius and the magnetospheric radius. An alternative interpretation \citep{mushtukov17,koliopanos17} is that the hotter thermal emission comes mostly from accretion curtains at the magnetospheric radius, where the inflow gets deflected along the poloidal field lines out of the disk plane.

\subsection{A NS spun up by accretion?}

We noted earlier than the inferred fast spin and moderately low magnetic field are more typical of pulsars in low-mass rather than high-mass X-ray binaries, even though J1403 appears to belong to a young stellar population. Here we use a simple model to assess whether the pulsar evolution towards fast spin and low magnetic field could have been caused by the huge amount of accretion, on a time-scale consistent with the life time of a high-mass donor star.

For a radiation-pressure-dominated disk near the Eddington limit, we use again the model of \citet{2017MNRAS.470.2799C} (their Eq.~69) to derive the period derivative:
\begin{equation}
           \dot{P} \approx - 1.3 \times 10^{-11} \, \dot{m}\, B_{12}^{2/9} \, P^2 \ \ {\mathrm{s~s}}^{-1}. 
\end{equation}
where we have assumed for simplicity that the moment of inertia is $10^{45}$ g cm$^{-2}$, $R = 10$km, and we have used the definition of $\dot{m} = \left(L/L_{\rm{Edd}}\right)$ rather than the definition $\dot{m} = \left(L/\eta L_{\rm{Edd}}\right)$ adopted by \citet{2017MNRAS.470.2799C}.

Magnetic field decay caused by accretion is usually invoked to explain the low field of NSs in low-mass X-ray binaries \citep{shibazaki89,romani90,konar97,mukherjee17}. Following the review of \cite{igoshev21}, we adopt a simple model of magnetic field decay:
\begin{equation}
    B(t) = \frac{B_0}{1+\dot{M}t/m_{\rm B}}
\end{equation}
where $m_{\rm B} \sim 10^{-4} M_{\odot}$, and $B_0$ is the initial field. We define an accretion timescale $t_{\rm A} \equiv m_{\rm B}/\dot{M}$, and recall that the Eddington accretion rate is $\dot{M}_{\rm {Edd}} \approx 1.0 \times 10^{18}$ g s$^{-1}$. Thus, $t_{\rm A} \approx 2 \times 10^{11}\, \dot{m}^{-1}$ s.

After scaling the magnetic field to units of $10^{12}$ G, we re-write Equation (9) as $B_{12}(t) = B_{0,12}/(1+t/t_{\rm A})$. Substituting into Equation (8), we obtain a simple first-order differential equation
\begin{equation}
    \frac{dP}{P^2} \approx - 1.3 \times 10^{-11} \, \dot{m}\, B_{0,12}^{2/9}\, \left(1+t/t_{\rm A}\right)^{-2/9} \,dt.
\end{equation}
Equation (10) can be easily integrated analytically as 
\begin{equation}
    1/P(t) - 1/P_0 \approx 3.3 \, B_{0,12}^{2/9}\, \left(1+t/t_{\rm A}\right)^{7/9}.
\end{equation}
where $P_0$ is the birth spin period in s. Assuming then that $P_0 \gg 5$ ms and $B_0 \gg 10^{10}$ G ({\it i.e.}, $t/t_{\rm A} \gg 1$), we can approximate Equation (11) as 
\begin{equation}
    1/P(t) \approx 3.3 \, B_{0,12}^{2/9}\, \left(t/t_{\rm A}\right)^{7/9},
\end{equation}
and Equation (9) as 
\begin{equation}
   B_{12}(t) \approx B_{0,12}\, \left(t/t_{\rm A}\right)^{-1}. 
\end{equation}
Finally, we use the values of the spin period and magnetic field today ($t = t_0)$, estimated in Section 4.2, to solve the system of Equations (12--13) for the accretion age $t_0$ and the initial field $B_{0,12}$. We obtain $B_0 \approx 3.2 \times 10^{12}$ G and $t_0 \approx 140 \, t_{\rm A} \approx 9 \times 10^5$\,$\dot{m}^{-1}$ yr.

In summary, if J1403 was constantly accreting in the super-Eddington regime, at a luminosity $\sim$10$^{39}$ erg s$^{-1}$, it may take only $\sim$10$^5$ yr to get spun up to a period of 5 ms and to have its magnetic field reduced from a ``canonical'' initial value $\sim$10$^{12}$ G to the current inferred value $\sim$10$^{10}$ G. The NS would accrete a mass $\sim$0.1 $M_{\odot}$ during this evolution. In fact, J1403 is only in the ULX state for $\sim$10\% of the time, and we may assume that the spin and surface field evolution in the propeller state is small compared with that in the accretor state. Thus, the age of the source is more likely $\sim$10$^6$ yr. In any case, it is a short time scale, perfectly consistent with the life time of a massive donor star. If this scenario is correct, J1403's evolution is very similar to the evolution proposed by \cite{kluzniak15} for the pulsar ULX M\,82 X-2 (also a high-mass X-ray binary). The latter source has a spin period of 1.37 s but is spinning up (during its super-Eddington regime) at a fast rate $\dot{P} \approx - 2 \times 10^{-10}$ s s$^{-1}$ \citep{bachetti14} and may become a millisecond pulsar in $\sim$10$^5$ yr \citep{kluzniak15}. J1403 may be a similar system, simply at a later stage of evolution, when the NS has already been spun up to a millisecond period.

\section{Conclusions}
The time variability behaviour of the high mass X-ray binary J1403 makes it difficult to classify it into previously known classes of accreting compact objects. Its optical counterpart is consistent with a young, massive donor star, with an age of $\approx$15\,Myr and a mass of $\approx15$\,M$_{\odot}$. In most of the observations over a baseline of 17 years, J1403 is detected at X-ray luminosities between a few $\times 10^{37}$ erg s$^{-1}$ and $\approx$10$^{38}$ erg s${-1}$, typical for a wind-fed high mass X-ray binary. However, it has flared occasionally to luminosities of about 1--4$\times 10^{39}$ erg s$^{-1}$. The flaring episodes have short rise and decay times, much shorter than the viscous timescale of a temporary accretion disk (as is the case for example for Type II bursts in a Be X-ray pulsar). In some observations, the transition between luminosities $<$10$^{38}$ erg ${-1}$ and $>$10$^{39}$ erg ${-1}$ occurred in less than 10$^4$ s. A magnetospheric gate provides a mechanism for sharp changes of luminosity. However, if the observed luminosity jumps are due to the accretor/propeller switch, the classical magnetospheric conditions (valid for a Keplerian disk coupled with a dipole field) imply a very fast spin ($P \sim 5$--15 ms) and a low magnetic field ($B \sim 2$--7 $\times 10^{10}$ G), unusual for a young pulsar. We speculated that the supercritical accretion rate is reducing the surface field, and, at the same time,  spinning up the NS. Alternatively, the magnetic field has strong non-dipolar components, so that our simple disk truncation model is not valid. This is the case proposed for example for the transient super-Eddington pulsar SMC X-3 \citep{tsigankov17} and for the ULX NGC\,5907 X-1 \citep{israel17a}. Exploring this non-dipolar scenario is beyond the scope of this work. 

The observational class of confirmed (from X-ray pulsations) pulsar ULXs does not contain millisecond pulsars yet. The shortest spin period in this class is $\approx$0.42 s for NGC\,7793 P13 \citep{furst16,israel17b}. All other pulsar ULXs have spin periods longer than 1 s (see a review table in \citealt{chen21}). However, the fast spin-up rate in some of those ULXs suggests that they may evolve to become millisecond pulsars on timescales shorter than the life time of the super-Eddington accretion phase \citep{kluzniak15}. We propose that J1403 is one example of this evolution process. So, why have we not detected any millisecond pulsar ULXs yet? We suggest that it is an observational bias. X-ray pulsations have been detected so far only in the brightest sources, with $\gtrsim$10,000 detected counts per observation. In order to detect pulsations on a millisecond timescales from ULXs in external galaxies, we need at the same time high count rate, good spatial resolution (or an isolated source), and millisecond time resolution. Only {\it XMM-Newton}/EPIC-pn in timing mode can provide enough sensitivity and time resolution, but at the expense of spatial resolution. Only two bright, isolated ULXs (Holmberg II X-1 and Holmberg IX X-1) have been significantly tested for millisecond pulsations, without a detection (G.L. Israel, private communications). 


Regardless of the nature of the compact object, what makes J1403 truly remarkable is the presence of strong QPOs, at frequencies variable from epoch to epoch, but generally in the range of $\approx$1.3--1.8 mHz. Very little is known about the physical origin of such very-low-frequency variability in X-ray binaries in general, let alone in ULXs. We reviewed several alternative scenarios proposed to explain low-frequency variability in other luminous X-ray binaries: Lense-Thirring precession of the inner disk or of a disk outflow; nuclear burning on the surface of the NS; ``heartbeat'' limit-cycle instability; beat frequency between material rotating at the inner edge of the Keplerian disk and the NS surface. There is no strong evidence in favour of any one of those scenarios. Exploration of new theoretical scenarios is left to further work.

\section*{Acknowledgements}
We thank Manfred Pakull for his comments and suggestions.
This research is based on data obtained from the Chandra Data Archive and has made use of software provided by the Chandra X-ray Center (CXC) in the application package {\sc ciao}. We also used data obtained from the ESA’s XMM-Newton Science Archive (XSA), and data from the NASA/ESA {\it Hubble Space Telescope}. The latter are associated with programs 9490, 13364, and 14166, and were obtained from the Space Telescope Science Institute, which is operated by the Association of Universities for Research in Astronomy, Inc., under NASA contract NAS 5–26555.  
To analyze some of the data, we used the {\it galaXy} (\url{https://github.com/sammarth-k/galaXy}) software and its data products.
We also used the VizieR catalogue access tool, CDS, Strasbourg, France (DOI : 10.26093/cds/vizier). The original description of the VizieR service was published in 2000, A\&AS 143, 23.

RTU acknowledges support from the Packard Foundation. RS acknowledges grant number 12073029 from the National Science Foundation of China. PE and GLI acknowledge financial support from the Italian Ministry for Education and Research through the PRIN grant 2017LJ39LM.
SEM acknowledges financial contribution from the agreement ASI-INAF n.2017-14-H.0, the INAF mainstream grant, and the PRIN-INAF 2019 n.15.

\section*{DATA AVAILABILITY}
All data are public and retrievable from the {\it XMM-Newton}, {\it Chandra} data archives, and from the Mikulski Archive for Space Telescopes (for {\it HST}).



\bibliographystyle{mnras}
\bibliography{references} 

\begin{thebibliography}{}
\makeatletter
\relax
\def\mn@urlcharsother{\let\do\@makeother \do\$\do\&\do\#\do\^\do\_\do\%\do\~}
\def\mn@doi{\begingroup\mn@urlcharsother \@ifnextchar [ {\mn@doi@}
  {\mn@doi@[]}}
\def\mn@doi@[#1]#2{\def\@tempa{#1}\ifx\@tempa\@empty \href
  {http://dx.doi.org/#2} {doi:#2}\else \href {http://dx.doi.org/#2} {#1}\fi
  \endgroup}
\def\mn@eprint#1#2{\mn@eprint@#1:#2::\@nil}
\def\mn@eprint@arXiv#1{\href {http://arxiv.org/abs/#1} {{\tt arXiv:#1}}}
\def\mn@eprint@dblp#1{\href {http://dblp.uni-trier.de/rec/bibtex/#1.xml}
  {dblp:#1}}
\def\mn@eprint@#1:#2:#3:#4\@nil{\def\@tempa {#1}\def\@tempb {#2}\def\@tempc
  {#3}\ifx \@tempc \@empty \let \@tempc \@tempb \let \@tempb \@tempa \fi \ifx
  \@tempb \@empty \def\@tempb {arXiv}\fi \@ifundefined
  {mn@eprint@\@tempb}{\@tempb:\@tempc}{\expandafter \expandafter \csname
  mn@eprint@\@tempb\endcsname \expandafter{\@tempc}}}

\bibitem[\protect\citeauthoryear{{Altamirano} et~al.,}{{Altamirano}
  et~al.}{2011}]{Altamirano2011}
{Altamirano} D.,  et~al., 2011, \mn@doi [\apjl] {10.1088/2041-8205/742/2/L17},
  \href {http://adsabs.harvard.edu/abs/2011ApJ...742L..17A} {742, L17}

\bibitem[\protect\citeauthoryear{{Arnaud}}{{Arnaud}}{1996}]{1996ASPC..101...17A}
{Arnaud} K.~A.,  1996, in {Jacoby} G.~H.,  {Barnes} J.,  eds,  Astronomical
  Society of the Pacific Conference Series Vol. 101, Astronomical Data Analysis
  Software and Systems V. p.~17

\bibitem[\protect\citeauthoryear{{Astropy Collaboration} et~al.,}{{Astropy
  Collaboration} et~al.}{2013}]{astropy:2013}
{Astropy Collaboration} et~al., 2013, \mn@doi [\aap]
  {10.1051/0004-6361/201322068}, \href
  {http://adsabs.harvard.edu/abs/2013A%26A...558A..33A} {558, A33}

\bibitem[\protect\citeauthoryear{{Astropy Collaboration} et~al.,}{{Astropy
  Collaboration} et~al.}{2018}]{astropy:2018}
{Astropy Collaboration} et~al., 2018, \mn@doi [\aj] {10.3847/1538-3881/aabc4f},
  \href {https://ui.adsabs.harvard.edu/abs/2018AJ....156..123A} {156, 123}

\bibitem[\protect\citeauthoryear{Aydin}{Aydin}{2017}]{m_emre_aydin_2017_375648}
Aydin M.~E.,  2017, eaydin/WWZ: First release, \mn@doi{10.5281/zenodo.375648},
  \url {https://doi.org/10.5281/zenodo.375648}

\bibitem[\protect\citeauthoryear{{Bachetti} et~al.,}{{Bachetti}
  et~al.}{2014}]{bachetti14}
{Bachetti} M.,  et~al., 2014, \mn@doi [\nat] {10.1038/nature13791}, \href
  {https://ui.adsabs.harvard.edu/abs/2014Natur.514..202B} {514, 202}

\bibitem[\protect\citeauthoryear{Bachetti et~al.,}{Bachetti
  et~al.}{2021}]{stingray_matteo_bachetti_2021_4881255}
Bachetti M.,  et~al., 2021, StingraySoftware/stingray: Version 0.3,
  \mn@doi{10.5281/zenodo.4881255}, \url
  {https://doi.org/10.5281/zenodo.4881255}

\bibitem[\protect\citeauthoryear{{Bagnoli} \& {in't Zand}}{{Bagnoli} \& {in't
  Zand}}{2015}]{Bagnoli2015}
{Bagnoli} T.,  {in't Zand} J.~J.~M.,  2015, \mn@doi [\mnras]
  {10.1093/mnrasl/slv045}, \href
  {https://ui.adsabs.harvard.edu/abs/2015MNRAS.450L..52B} {450, L52}

\bibitem[\protect\citeauthoryear{{Baluev}}{{Baluev}}{2008}]{Baluev2008MNRAS.385.1279B}
{Baluev} R.~V.,  2008, \mn@doi [\mnras] {10.1111/j.1365-2966.2008.12689.x},
  \href {https://ui.adsabs.harvard.edu/abs/2008MNRAS.385.1279B} {385, 1279}

\bibitem[\protect\citeauthoryear{{Belloni}}{{Belloni}}{2010}]{belloni10}
{Belloni} T.~M.,  2010, {States and Transitions in Black Hole Binaries}.
p.~53, \mn@doi{10.1007/978-3-540-76937-8\_3}

\bibitem[\protect\citeauthoryear{{Belloni}, {Klein-Wolt}, {M{\'e}ndez}, {van
  der Klis}  \& {van Paradijs}}{{Belloni} et~al.}{2000}]{Belloni2000}
{Belloni} T.,  {Klein-Wolt} M.,  {M{\'e}ndez} M.,  {van der Klis} M.,   {van
  Paradijs} J.,  2000, \aap, \href
  {http://adsabs.harvard.edu/abs/2000A%26A...355..271B} {355, 271}

\bibitem[\protect\citeauthoryear{{Belloni}, {Homan}, {Casella}, {van der Klis},
  {Nespoli}, {Lewin}, {Miller}  \& {M{\'e}ndez}}{{Belloni}
  et~al.}{2005}]{belloni05}
{Belloni} T.,  {Homan} J.,  {Casella} P.,  {van der Klis} M.,  {Nespoli} E.,
  {Lewin} W.~H.~G.,  {Miller} J.~M.,   {M{\'e}ndez} M.,  2005, \mn@doi [\aap]
  {10.1051/0004-6361:20042457}, \href
  {https://ui.adsabs.harvard.edu/abs/2005A&A...440..207B} {440, 207}

\bibitem[\protect\citeauthoryear{{Blackburn}}{{Blackburn}}{1995}]{1995ASPC...77..367B}
{Blackburn} J.~K.,  1995, in {Shaw} R.~A.,  {Payne} H.~E.,   {Hayes} J.~J.~E.,
  eds,  Astronomical Society of the Pacific Conference Series Vol. 77,
  Astronomical Data Analysis Software and Systems IV. p.~367

\bibitem[\protect\citeauthoryear{{Brumback} et~al.,}{{Brumback}
  et~al.}{2018}]{brumback18}
{Brumback} M.~C.,  et~al., 2018, \mn@doi [\apjl] {10.3847/2041-8213/aacd13},
  \href {https://ui.adsabs.harvard.edu/abs/2018ApJ...861L...7B} {861, L7}

\bibitem[\protect\citeauthoryear{{Campana}, {Colpi}, {Mereghetti}, {Stella}  \&
  {Tavani}}{{Campana} et~al.}{1998}]{campana98}
{Campana} S.,  {Colpi} M.,  {Mereghetti} S.,  {Stella} L.,   {Tavani} M.,
  1998, \mn@doi [\aapr] {10.1007/s001590050012}, \href
  {https://ui.adsabs.harvard.edu/abs/1998A&ARv...8..279C} {8, 279}

\bibitem[\protect\citeauthoryear{{Campana}, {Stella}, {Israel}, {Moretti},
  {Parmar}  \& {Orlandini}}{{Campana} et~al.}{2002}]{campana02}
{Campana} S.,  {Stella} L.,  {Israel} G.~L.,  {Moretti} A.,  {Parmar} A.~N.,
  {Orlandini} M.,  2002, \mn@doi [\apj] {10.1086/343074}, \href
  {https://ui.adsabs.harvard.edu/abs/2002ApJ...580..389C} {580, 389}

\bibitem[\protect\citeauthoryear{{Campana}, {Stella}, {Mereghetti}  \& {de
  Martino}}{{Campana} et~al.}{2018}]{campana18}
{Campana} S.,  {Stella} L.,  {Mereghetti} S.,   {de Martino} D.,  2018, \mn@doi
  [\aap] {10.1051/0004-6361/201730769}, \href
  {https://ui.adsabs.harvard.edu/abs/2018A&A...610A..46C} {610, A46}

\bibitem[\protect\citeauthoryear{{Casella}, {Belloni}  \& {Stella}}{{Casella}
  et~al.}{2005}]{casella05}
{Casella} P.,  {Belloni} T.,   {Stella} L.,  2005, \mn@doi [\apj]
  {10.1086/431174}, \href
  {https://ui.adsabs.harvard.edu/abs/2005ApJ...629..403C} {629, 403}

\bibitem[\protect\citeauthoryear{{Cash}}{{Cash}}{1979}]{1979ApJ...228..939C}
{Cash} W.,  1979, \mn@doi [\apj] {10.1086/156922}, \href
  {http://adsabs.harvard.edu/abs/1979ApJ...228..939C} {228, 939}

\bibitem[\protect\citeauthoryear{{Chakrabarti} \& {Titarchuk}}{{Chakrabarti} \&
  {Titarchuk}}{1995}]{chakrabarti95}
{Chakrabarti} S.,  {Titarchuk} L.~G.,  1995, \mn@doi [\apj] {10.1086/176610},
  \href {https://ui.adsabs.harvard.edu/abs/1995ApJ...455..623C} {455, 623}

\bibitem[\protect\citeauthoryear{{Chashkina}, {Abolmasov}  \&
  {Poutanen}}{{Chashkina} et~al.}{2017}]{2017MNRAS.470.2799C}
{Chashkina} A.,  {Abolmasov} P.,   {Poutanen} J.,  2017, \mn@doi [\mnras]
  {10.1093/mnras/stx1372}, \href
  {https://ui.adsabs.harvard.edu/abs/2017MNRAS.470.2799C} {470, 2799}

\bibitem[\protect\citeauthoryear{{Chen}, {Wang}  \& {Tong}}{{Chen}
  et~al.}{2021}]{chen21}
{Chen} X.,  {Wang} W.,   {Tong} H.,  2021, \mn@doi [Journal of High Energy
  Astrophysics] {10.1016/j.jheap.2021.04.002}, \href
  {https://ui.adsabs.harvard.edu/abs/2021JHEAp..31....1C} {31, 1}

\bibitem[\protect\citeauthoryear{{Corbet}}{{Corbet}}{1996}]{corbet96}
{Corbet} R. H.~D.,  1996, \mn@doi [\apjl] {10.1086/309890}, \href
  {https://ui.adsabs.harvard.edu/abs/1996ApJ...457L..31C} {457, L31}

\bibitem[\protect\citeauthoryear{{Done}, {Wardzi{\'n}ski}  \&
  {Gierli{\'n}ski}}{{Done} et~al.}{2004}]{done04}
{Done} C.,  {Wardzi{\'n}ski} G.,   {Gierli{\'n}ski} M.,  2004, \mn@doi [\mnras]
  {10.1111/j.1365-2966.2004.07545.x}, \href
  {https://ui.adsabs.harvard.edu/abs/2004MNRAS.349..393D} {349, 393}

\bibitem[\protect\citeauthoryear{{Evans} et~al.,}{{Evans}
  et~al.}{2010}]{evans10}
{Evans} I.~N.,  et~al., 2010, \mn@doi [\apjs] {10.1088/0067-0049/189/1/37},
  \href {https://ui.adsabs.harvard.edu/abs/2010ApJS..189...37E} {189, 37}

\bibitem[\protect\citeauthoryear{{Evans} et~al.,}{{Evans}
  et~al.}{2019}]{vizier:IX/57}
{Evans} I.,  et~al., 2019, {VizieR Online Data Catalog: The Chandra Source
  Catalog (CSC), Release 2.0}

\bibitem[\protect\citeauthoryear{{Falanga}, {Bozzo}, {Lutovinov},
  {Bonnet-Bidaud}, {Fetisova}  \& {Puls}}{{Falanga} et~al.}{2015}]{falanga15}
{Falanga} M.,  {Bozzo} E.,  {Lutovinov} A.,  {Bonnet-Bidaud} J.~M.,  {Fetisova}
  Y.,   {Puls} J.,  2015, \mn@doi [\aap] {10.1051/0004-6361/201425191}, \href
  {https://ui.adsabs.harvard.edu/abs/2015A&A...577A.130F} {577, A130}

\bibitem[\protect\citeauthoryear{{Fender} \& {Mu{\~n}oz-Darias}}{{Fender} \&
  {Mu{\~n}oz-Darias}}{2016}]{fender16}
{Fender} R.,  {Mu{\~n}oz-Darias} T.,  2016, {The Balance of Power: Accretion
  and Feedback in Stellar Mass Black Holes}.
p.~65, \mn@doi{10.1007/978-3-319-19416-5\_3}

\bibitem[\protect\citeauthoryear{{Fender}, {Belloni}  \& {Gallo}}{{Fender}
  et~al.}{2004}]{fender04}
{Fender} R.~P.,  {Belloni} T.~M.,   {Gallo} E.,  2004, \mn@doi [\mnras]
  {10.1111/j.1365-2966.2004.08384.x}, \href
  {https://ui.adsabs.harvard.edu/abs/2004MNRAS.355.1105F} {355, 1105}

\bibitem[\protect\citeauthoryear{{Foster}}{{Foster}}{1996}]{Foster1996AJ....112.1709F}
{Foster} G.,  1996, \mn@doi [\aj] {10.1086/118137}, \href
  {https://ui.adsabs.harvard.edu/abs/1996AJ....112.1709F} {112, 1709}

\bibitem[\protect\citeauthoryear{{Fruscione} et~al.,}{{Fruscione}
  et~al.}{2006}]{2006SPIE.6270E..1VF}
{Fruscione} A.,  et~al., 2006, in Society of Photo-Optical Instrumentation
  Engineers (SPIE) Conference Series. p.~1, \mn@doi{10.1117/12.671760}

\bibitem[\protect\citeauthoryear{{F{\"u}rst} et~al.,}{{F{\"u}rst}
  et~al.}{2016}]{furst16}
{F{\"u}rst} F.,  et~al., 2016, \mn@doi [\apjl] {10.3847/2041-8205/831/2/L14},
  \href {https://ui.adsabs.harvard.edu/abs/2016ApJ...831L..14F} {831, L14}

\bibitem[\protect\citeauthoryear{{Ghaderpour} \& {Ghaderpour}}{{Ghaderpour} \&
  {Ghaderpour}}{2020}]{Ghaderpour2020PASP..132k4504G}
{Ghaderpour} E.,  {Ghaderpour} S.,  2020, \mn@doi [\pasp]
  {10.1088/1538-3873/abaf04}, \href
  {https://ui.adsabs.harvard.edu/abs/2020PASP..132k4504G} {132, 114504}

\bibitem[\protect\citeauthoryear{{Ghosh} \& {Lamb}}{{Ghosh} \&
  {Lamb}}{1979}]{ghosh79}
{Ghosh} P.,  {Lamb} F.~K.,  1979, \mn@doi [\apj] {10.1086/157498}, \href
  {https://ui.adsabs.harvard.edu/abs/1979ApJ...234..296G} {234, 296}

\bibitem[\protect\citeauthoryear{{Graham} et~al.,}{{Graham}
  et~al.}{2015}]{Graham2015MNRAS.453.1562G}
{Graham} M.~J.,  et~al., 2015, \mn@doi [\mnras] {10.1093/mnras/stv1726}, \href
  {https://ui.adsabs.harvard.edu/abs/2015MNRAS.453.1562G} {453, 1562}

\bibitem[\protect\citeauthoryear{{HI4PI Collaboration} et~al.,}{{HI4PI
  Collaboration} et~al.}{2016}]{nh16}
{HI4PI Collaboration} et~al., 2016, \mn@doi [\aap]
  {10.1051/0004-6361/201629178}, \href
  {https://ui.adsabs.harvard.edu/abs/2016A&A...594A.116H} {594, A116}

\bibitem[\protect\citeauthoryear{{Heger}, {Cumming}  \& {Woosley}}{{Heger}
  et~al.}{2007a}]{heger07}
{Heger} A.,  {Cumming} A.,   {Woosley} S.~E.,  2007a, \mn@doi [\apj]
  {10.1086/517491}, \href
  {https://ui.adsabs.harvard.edu/abs/2007ApJ...665.1311H} {665, 1311}

\bibitem[\protect\citeauthoryear{{Heger}, {Cumming}  \& {Woosley}}{{Heger}
  et~al.}{2007b}]{Heger2007}
{Heger} A.,  {Cumming} A.,   {Woosley} S.~E.,  2007b, \mn@doi [\apj]
  {10.1086/517491}, \href
  {https://ui.adsabs.harvard.edu/abs/2007ApJ...665.1311H} {665, 1311}

\bibitem[\protect\citeauthoryear{{Heida} et~al.,}{{Heida}
  et~al.}{2014}]{2014MNRAS.442.1054H}
{Heida} M.,  et~al., 2014, \mn@doi [\mnras] {10.1093/mnras/stu928}, \href
  {https://ui.adsabs.harvard.edu/abs/2014MNRAS.442.1054H} {442, 1054}

\bibitem[\protect\citeauthoryear{{Hong}, {van den Berg}, {Schlegel},
  {Grindlay}, {Koenig}, {Laycock}  \& {Zhao}}{{Hong}
  et~al.}{2005}]{2005ApJ...635..907H}
{Hong} J.,  {van den Berg} M.,  {Schlegel} E.~M.,  {Grindlay} J.~E.,  {Koenig}
  X.,  {Laycock} S.,   {Zhao} P.,  2005, \mn@doi [\apj] {10.1086/496966}, \href
  {https://ui.adsabs.harvard.edu/abs/2005ApJ...635..907H} {635, 907}

\bibitem[\protect\citeauthoryear{Huppenkothen et~al.,}{Huppenkothen
  et~al.}{2019a}]{stingrayHuppenkothen2019}
Huppenkothen D.,  et~al., 2019a, \mn@doi [Journal of Open Source Software]
  {10.21105/joss.01393}, 4, 1393

\bibitem[\protect\citeauthoryear{{Huppenkothen} et~al.,}{{Huppenkothen}
  et~al.}{2019b}]{stingray2019ApJ...881...39H}
{Huppenkothen} D.,  et~al., 2019b, \mn@doi [apj] {10.3847/1538-4357/ab258d},
  \href {https://ui.adsabs.harvard.edu/abs/2019ApJ...881...39H} {881, 39}

\bibitem[\protect\citeauthoryear{{Igoshev}, {Popov}  \& {Hollerbach}}{{Igoshev}
  et~al.}{2021}]{igoshev21}
{Igoshev} A.~P.,  {Popov} S.~B.,   {Hollerbach} R.,  2021, \mn@doi [Universe]
  {10.3390/universe7090351}, \href
  {https://ui.adsabs.harvard.edu/abs/2021Univ....7..351I} {7, 351}

\bibitem[\protect\citeauthoryear{{Illarionov} \& {Sunyaev}}{{Illarionov} \&
  {Sunyaev}}{1975}]{illarionov75}
{Illarionov} A.~F.,  {Sunyaev} R.~A.,  1975, \aap, \href
  {https://ui.adsabs.harvard.edu/abs/1975A&A....39..185I} {39, 185}

\bibitem[\protect\citeauthoryear{{Ingram} \& {Motta}}{{Ingram} \&
  {Motta}}{2019}]{ingram19}
{Ingram} A.~R.,  {Motta} S.~E.,  2019, \mn@doi [\nar]
  {10.1016/j.newar.2020.101524}, \href
  {https://ui.adsabs.harvard.edu/abs/2019NewAR..8501524I} {85, 101524}

\bibitem[\protect\citeauthoryear{{Ingram}, {Done}  \& {Fragile}}{{Ingram}
  et~al.}{2009}]{ingram09}
{Ingram} A.,  {Done} C.,   {Fragile} P.~C.,  2009, \mn@doi [\mnras]
  {10.1111/j.1745-3933.2009.00693.x}, \href
  {https://ui.adsabs.harvard.edu/abs/2009MNRAS.397L.101I} {397, L101}

\bibitem[\protect\citeauthoryear{{Ingram}, {van der Klis}, {Middleton}, {Done},
  {Altamirano}, {Heil}, {Uttley}  \& {Axelsson}}{{Ingram}
  et~al.}{2016}]{ingram16}
{Ingram} A.,  {van der Klis} M.,  {Middleton} M.,  {Done} C.,  {Altamirano} D.,
   {Heil} L.,  {Uttley} P.,   {Axelsson} M.,  2016, \mn@doi [\mnras]
  {10.1093/mnras/stw1245}, \href
  {https://ui.adsabs.harvard.edu/abs/2016MNRAS.461.1967I} {461, 1967}

\bibitem[\protect\citeauthoryear{{Israel} et~al.,}{{Israel}
  et~al.}{2017a}]{israel17b}
{Israel} G.~L.,  et~al., 2017a, \mn@doi [Science] {10.1126/science.aai8635},
  \href {https://ui.adsabs.harvard.edu/abs/2017Sci...355..817I} {355, 817}

\bibitem[\protect\citeauthoryear{{Israel} et~al.,}{{Israel}
  et~al.}{2017b}]{israel17a}
{Israel} G.~L.,  et~al., 2017b, \mn@doi [\mnras] {10.1093/mnrasl/slw218}, \href
  {https://ui.adsabs.harvard.edu/abs/2017MNRAS.466L..48I} {466, L48}

\bibitem[\protect\citeauthoryear{{James}, {Paul}, {Devasia}  \&
  {Indulekha}}{{James} et~al.}{2010}]{james10}
{James} M.,  {Paul} B.,  {Devasia} J.,   {Indulekha} K.,  2010, \mn@doi
  [\mnras] {10.1111/j.1365-2966.2010.16880.x}, \href
  {https://ui.adsabs.harvard.edu/abs/2010MNRAS.407..285J} {407, 285}

\bibitem[\protect\citeauthoryear{{Janiuk}, {Grzedzielski}, {Capitanio}  \&
  {Bianchi}}{{Janiuk} et~al.}{2015}]{janiuk15}
{Janiuk} A.,  {Grzedzielski} M.,  {Capitanio} F.,   {Bianchi} S.,  2015,
  \mn@doi [\aap] {10.1051/0004-6361/201425003}, \href
  {https://ui.adsabs.harvard.edu/abs/2015A&A...574A..92J} {574, A92}

\bibitem[\protect\citeauthoryear{{Joye}}{{Joye}}{2019}]{ds9}
{Joye} W.,  2019, {SAOImageDS9/SAOImageDS9 v8.0.1},
  \mn@doi{10.5281/zenodo.2530958}

\bibitem[\protect\citeauthoryear{{Kalberla}, {Burton}, {Hartmann}, {Arnal},
  {Bajaja}, {Morras}  \& {P{\"o}ppel}}{{Kalberla} et~al.}{2005}]{kalberla05}
{Kalberla} P.~M.~W.,  {Burton} W.~B.,  {Hartmann} D.,  {Arnal} E.~M.,  {Bajaja}
  E.,  {Morras} R.,   {P{\"o}ppel} W.~G.~L.,  2005, \mn@doi [\aap]
  {10.1051/0004-6361:20041864}, \href
  {https://ui.adsabs.harvard.edu/abs/2005A&A...440..775K} {440, 775}

\bibitem[\protect\citeauthoryear{{Keek}, {Cyburt}  \& {Heger}}{{Keek}
  et~al.}{2014}]{keek14}
{Keek} L.,  {Cyburt} R.~H.,   {Heger} A.,  2014, \mn@doi [\apj]
  {10.1088/0004-637X/787/2/101}, \href
  {https://ui.adsabs.harvard.edu/abs/2014ApJ...787..101K} {787, 101}

\bibitem[\protect\citeauthoryear{{Kluzniak} \& {Lasota}}{{Kluzniak} \&
  {Lasota}}{2015}]{kluzniak15}
{Kluzniak} W.,  {Lasota} J.~P.,  2015, \mn@doi [\mnras]
  {10.1093/mnrasl/slu200}, \href
  {https://ui.adsabs.harvard.edu/abs/2015MNRAS.448L..43K} {448, L43}

\bibitem[\protect\citeauthoryear{{Koliopanos}, {Vasilopoulos}, {Godet},
  {Bachetti}, {Webb}  \& {Barret}}{{Koliopanos} et~al.}{2017}]{koliopanos17}
{Koliopanos} F.,  {Vasilopoulos} G.,  {Godet} O.,  {Bachetti} M.,  {Webb}
  N.~A.,   {Barret} D.,  2017, \mn@doi [\aap] {10.1051/0004-6361/201730922},
  \href {https://ui.adsabs.harvard.edu/abs/2017A&A...608A..47K} {608, A47}

\bibitem[\protect\citeauthoryear{{Konar} \& {Bhattacharya}}{{Konar} \&
  {Bhattacharya}}{1997}]{konar97}
{Konar} S.,  {Bhattacharya} D.,  1997, \mn@doi [\mnras]
  {10.1093/mnras/284.2.311}, \href
  {https://ui.adsabs.harvard.edu/abs/1997MNRAS.284..311K} {284, 311}

\bibitem[\protect\citeauthoryear{{Kosec}, {Pinto}, {Fabian}  \&
  {Walton}}{{Kosec} et~al.}{2018}]{kosec18}
{Kosec} P.,  {Pinto} C.,  {Fabian} A.~C.,   {Walton} D.~J.,  2018, \mn@doi
  [\mnras] {10.1093/mnras/stx2695}, \href
  {https://ui.adsabs.harvard.edu/abs/2018MNRAS.473.5680K} {473, 5680}

\bibitem[\protect\citeauthoryear{{Kraft}, {Burrows}  \& {Nousek}}{{Kraft}
  et~al.}{1991}]{1991ApJ...374..344K}
{Kraft} R.~P.,  {Burrows} D.~N.,   {Nousek} J.~A.,  1991, \mn@doi [\apj]
  {10.1086/170124}, \href {http://adsabs.harvard.edu/abs/1991ApJ...374..344K}
  {374, 344}

\bibitem[\protect\citeauthoryear{{Krauss}, {Kilgard}, {Garcia}, {Roberts}  \&
  {Prestwich}}{{Krauss} et~al.}{2005}]{krauss05}
{Krauss} M.~I.,  {Kilgard} R.~E.,  {Garcia} M.~R.,  {Roberts} T.~P.,
  {Prestwich} A.~H.,  2005, \mn@doi [\apj] {10.1086/431784}, \href
  {https://ui.adsabs.harvard.edu/abs/2005ApJ...630..228K} {630, 228}

\bibitem[\protect\citeauthoryear{{Lense} \& {Thirring}}{{Lense} \&
  {Thirring}}{1918}]{lense18}
{Lense} J.,  {Thirring} H.,  1918, Physikalische Zeitschrift, \href
  {https://ui.adsabs.harvard.edu/abs/1918PhyZ...19..156L} {19, 156}

\bibitem[\protect\citeauthoryear{{Li} \& {van den Heuvel}}{{Li} \& {van den
  Heuvel}}{1997}]{1997A&A...321L..25L}
{Li} X.~D.,  {van den Heuvel} E.~P.~J.,  1997, \aap, \href
  {https://ui.adsabs.harvard.edu/abs/1997A&A...321L..25L} {321, L25}

\bibitem[\protect\citeauthoryear{{Lomb}}{{Lomb}}{1976}]{Lomb1976Ap&SS..39..447L}
{Lomb} N.~R.,  1976, \mn@doi [\apss] {10.1007/BF00648343}, \href
  {https://ui.adsabs.harvard.edu/abs/1976Ap&SS..39..447L} {39, 447}

\bibitem[\protect\citeauthoryear{{L{\'o}pez}, {Heida}, {Jonker}, {Torres},
  {Roberts}, {Walton}, {Moon}  \& {Harrison}}{{L{\'o}pez}
  et~al.}{2020}]{2020MNRAS.497..917L}
{L{\'o}pez} K.~M.,  {Heida} M.,  {Jonker} P.~G.,  {Torres} M.~A.~P.,  {Roberts}
  T.~P.,  {Walton} D.~J.,  {Moon} D.~S.,   {Harrison} F.~A.,  2020, \mn@doi
  [\mnras] {10.1093/mnras/staa1920}, \href
  {https://ui.adsabs.harvard.edu/abs/2020MNRAS.497..917L} {497, 917}

\bibitem[\protect\citeauthoryear{{Lutovinov}, {Tsygankov}, {Krivonos}, {Molkov}
   \& {Poutanen}}{{Lutovinov} et~al.}{2017}]{lutovinov17}
{Lutovinov} A.~A.,  {Tsygankov} S.~S.,  {Krivonos} R.~A.,  {Molkov} S.~V.,
  {Poutanen} J.,  2017, \mn@doi [\apj] {10.3847/1538-4357/834/2/209}, \href
  {https://ui.adsabs.harvard.edu/abs/2017ApJ...834..209L} {834, 209}

\bibitem[\protect\citeauthoryear{{Lyne}, {Hobbs}, {Kramer}, {Stairs}  \&
  {Stappers}}{{Lyne} et~al.}{2010}]{Lyne2010Sci...329..408L}
{Lyne} A.,  {Hobbs} G.,  {Kramer} M.,  {Stairs} I.,   {Stappers} B.,  2010,
  \mn@doi [Science] {10.1126/science.1186683}, \href
  {https://ui.adsabs.harvard.edu/abs/2010Sci...329..408L} {329, 408}

\bibitem[\protect\citeauthoryear{{Mancuso}, {Altamirano}, {Garc{\'\i}a}, {Lyu},
  {M{\'e}ndez}, {Combi}, {D{\'\i}az-Trigo}  \& {in't Zand}}{{Mancuso}
  et~al.}{2019}]{mancuso19}
{Mancuso} G.~C.,  {Altamirano} D.,  {Garc{\'\i}a} F.,  {Lyu} M.,  {M{\'e}ndez}
  M.,  {Combi} J.~A.,  {D{\'\i}az-Trigo} M.,   {in't Zand} J.~J.~M.,  2019,
  \mn@doi [\mnras] {10.1093/mnrasl/slz057}, \href
  {https://ui.adsabs.harvard.edu/abs/2019MNRAS.486L..74M} {486, L74}

\bibitem[\protect\citeauthoryear{{Martin}, {Nixon}, {Armitage}, {Lubow}  \&
  {Price}}{{Martin} et~al.}{2014}]{martin14}
{Martin} R.~G.,  {Nixon} C.,  {Armitage} P.~J.,  {Lubow} S.~H.,   {Price}
  D.~J.,  2014, \mn@doi [\apjl] {10.1088/2041-8205/790/2/L34}, \href
  {https://ui.adsabs.harvard.edu/abs/2014ApJ...790L..34M} {790, L34}

\bibitem[\protect\citeauthoryear{{Middleton}, {Walton}, {Fabian}, {Roberts},
  {Heil}, {Pinto}, {Anderson}  \& {Sutton}}{{Middleton}
  et~al.}{2015}]{middleton15}
{Middleton} M.~J.,  {Walton} D.~J.,  {Fabian} A.,  {Roberts} T.~P.,  {Heil} L.,
   {Pinto} C.,  {Anderson} G.,   {Sutton} A.,  2015, \mn@doi [\mnras]
  {10.1093/mnras/stv2214}, \href
  {https://ui.adsabs.harvard.edu/abs/2015MNRAS.454.3134M} {454, 3134}

\bibitem[\protect\citeauthoryear{{Middleton} et~al.,}{{Middleton}
  et~al.}{2018}]{middleton18}
{Middleton} M.~J.,  et~al., 2018, \mn@doi [\mnras] {10.1093/mnras/stx2986},
  \href {https://ui.adsabs.harvard.edu/abs/2018MNRAS.475..154M} {475, 154}

\bibitem[\protect\citeauthoryear{{Middleton}, {Fragile}, {Ingram}  \&
  {Roberts}}{{Middleton} et~al.}{2019}]{middleton19}
{Middleton} M.~J.,  {Fragile} P.~C.,  {Ingram} A.,   {Roberts} T.~P.,  2019,
  \mn@doi [\mnras] {10.1093/mnras/stz2005}, \href
  {https://ui.adsabs.harvard.edu/abs/2019MNRAS.489..282M} {489, 282}

\bibitem[\protect\citeauthoryear{{Miniutti} et~al.,}{{Miniutti}
  et~al.}{2019}]{Miniutti2019}
{Miniutti} G.,  et~al., 2019, \mn@doi [\nat] {10.1038/s41586-019-1556-x}, \href
  {https://ui.adsabs.harvard.edu/abs/2019Natur.573..381M} {573, 381}

\bibitem[\protect\citeauthoryear{{Moon} \& {Eikenberry}}{{Moon} \&
  {Eikenberry}}{2001}]{moon01}
{Moon} D.-S.,  {Eikenberry} S.~S.,  2001, \mn@doi [\apjl] {10.1086/319160},
  \href {https://ui.adsabs.harvard.edu/abs/2001ApJ...549L.225M} {549, L225}

\bibitem[\protect\citeauthoryear{{Moon}, {Eikenberry}  \& {Wasserman}}{{Moon}
  et~al.}{2003}]{moon03}
{Moon} D.-S.,  {Eikenberry} S.~S.,   {Wasserman} I.~M.,  2003, \mn@doi [\apj]
  {10.1086/367826}, \href
  {https://ui.adsabs.harvard.edu/abs/2003ApJ...586.1280M} {586, 1280}

\bibitem[\protect\citeauthoryear{{Motch}, {Ricketts}, {Page}, {Ilovaisky}  \&
  {Chevalier}}{{Motch} et~al.}{1983}]{Motch1983}
{Motch} C.,  {Ricketts} M.~J.,  {Page} C.~G.,  {Ilovaisky} S.~A.,   {Chevalier}
  C.,  1983, \aap, \href {http://adsabs.harvard.edu/abs/1983A%26A...119..171M}
  {119, 171}

\bibitem[\protect\citeauthoryear{{Motta}, {Rouco Escorial}, {Kuulkers},
  {Mu{\~n}oz-Darias}  \& {Sanna}}{{Motta} et~al.}{2017}]{motta2017}
{Motta} S.~E.,  {Rouco Escorial} A.,  {Kuulkers} E.,  {Mu{\~n}oz-Darias} T.,
  {Sanna} A.,  2017, \mn@doi [\mnras] {10.1093/mnras/stx570}, \href
  {https://ui.adsabs.harvard.edu/abs/2017MNRAS.468.2311M} {468, 2311}

\bibitem[\protect\citeauthoryear{{Motta}, {Franchini}, {Lodato}  \&
  {Mastroserio}}{{Motta} et~al.}{2018}]{Mottalt2018MNRAS.473..431M}
{Motta} S.~E.,  {Franchini} A.,  {Lodato} G.,   {Mastroserio} G.,  2018,
  \mn@doi [\mnras] {10.1093/mnras/stx2358}, \href
  {https://ui.adsabs.harvard.edu/abs/2018MNRAS.473..431M} {473, 431}

\bibitem[\protect\citeauthoryear{{Motta} et~al.,}{{Motta}
  et~al.}{2020}]{Motta2020}
{Motta} S.~E.,  et~al., 2020, \mn@doi [\apj] {10.3847/1538-4357/ab9b81}, \href
  {https://ui.adsabs.harvard.edu/abs/2020ApJ...898..174M} {898, 174}

\bibitem[\protect\citeauthoryear{{Mukherjee}}{{Mukherjee}}{2017}]{mukherjee17}
{Mukherjee} D.,  2017, \mn@doi [Journal of Astrophysics and Astronomy]
  {10.1007/s12036-017-9465-6}, \href
  {https://ui.adsabs.harvard.edu/abs/2017JApA...38...48M} {38, 48}

\bibitem[\protect\citeauthoryear{{Mukherjee}, {Bult}, {van der Klis}  \&
  {Bhattacharya}}{{Mukherjee} et~al.}{2015}]{mukherjee15}
{Mukherjee} D.,  {Bult} P.,  {van der Klis} M.,   {Bhattacharya} D.,  2015,
  \mn@doi [\mnras] {10.1093/mnras/stv1542}, \href
  {https://ui.adsabs.harvard.edu/abs/2015MNRAS.452.3994M} {452, 3994}

\bibitem[\protect\citeauthoryear{{Mushtukov}, {Suleimanov}, {Tsygankov}  \&
  {Ingram}}{{Mushtukov} et~al.}{2017}]{mushtukov17}
{Mushtukov} A.~A.,  {Suleimanov} V.~F.,  {Tsygankov} S.~S.,   {Ingram} A.,
  2017, \mn@doi [\mnras] {10.1093/mnras/stx141}, \href
  {https://ui.adsabs.harvard.edu/abs/2017MNRAS.467.1202M} {467, 1202}

\bibitem[\protect\citeauthoryear{{Naik} \& {Paul}}{{Naik} \&
  {Paul}}{2004}]{naik04}
{Naik} S.,  {Paul} B.,  2004, \mn@doi [\apj] {10.1086/379803}, \href
  {https://ui.adsabs.harvard.edu/abs/2004ApJ...600..351N} {600, 351}

\bibitem[\protect\citeauthoryear{{Neilsen}, {Remillard}  \& {Lee}}{{Neilsen}
  et~al.}{2011}]{2011ApJ...737...69N}
{Neilsen} J.,  {Remillard} R.~A.,   {Lee} J.~C.,  2011, \mn@doi [\apj]
  {10.1088/0004-637X/737/2/69}, \href
  {https://ui.adsabs.harvard.edu/abs/2011ApJ...737...69N} {737, 69}

\bibitem[\protect\citeauthoryear{{Neilsen}, {Petschek}  \& {Lee}}{{Neilsen}
  et~al.}{2012}]{2012MNRAS.421..502N}
{Neilsen} J.,  {Petschek} A.~J.,   {Lee} J.~C.,  2012, \mn@doi [\mnras]
  {10.1111/j.1365-2966.2011.20329.x}, \href
  {https://ui.adsabs.harvard.edu/abs/2012MNRAS.421..502N} {421, 502}

\bibitem[\protect\citeauthoryear{{Pan}, {Zhang}, {Song}, {Wang}, {Li}  \&
  {Yang}}{{Pan} et~al.}{2018}]{2018MNRAS.480..692P}
{Pan} Y.~Y.,  {Zhang} C.~M.,  {Song} L.~M.,  {Wang} N.,  {Li} D.,   {Yang}
  Y.~Y.,  2018, \mn@doi [\mnras] {10.1093/mnras/sty1851}, \href
  {https://ui.adsabs.harvard.edu/abs/2018MNRAS.480..692P} {480, 692}

\bibitem[\protect\citeauthoryear{{Pinto}, {Middleton}  \& {Fabian}}{{Pinto}
  et~al.}{2016}]{pinto16}
{Pinto} C.,  {Middleton} M.~J.,   {Fabian} A.~C.,  2016, \mn@doi [\nat]
  {10.1038/nature17417}, \href
  {https://ui.adsabs.harvard.edu/abs/2016Natur.533...64P} {533, 64}

\bibitem[\protect\citeauthoryear{{Pinto} et~al.,}{{Pinto}
  et~al.}{2017}]{pinto17}
{Pinto} C.,  et~al., 2017, \mn@doi [\mnras] {10.1093/mnras/stx641}, \href
  {https://ui.adsabs.harvard.edu/abs/2017MNRAS.468.2865P} {468, 2865}

\bibitem[\protect\citeauthoryear{{Pintore} et~al.,}{{Pintore}
  et~al.}{2020}]{pintore20}
{Pintore} F.,  et~al., 2020, \mn@doi [\apj] {10.3847/1538-4357/ab6ffd}, \href
  {https://ui.adsabs.harvard.edu/abs/2020ApJ...890..166P} {890, 166}

\bibitem[\protect\citeauthoryear{{Pintore} et~al.,}{{Pintore}
  et~al.}{2021}]{pintore21}
{Pintore} F.,  et~al., 2021, \mn@doi [\mnras] {10.1093/mnras/stab913}, \href
  {https://ui.adsabs.harvard.edu/abs/2021MNRAS.504..551P} {504, 551}

\bibitem[\protect\citeauthoryear{{Poutanen}, {Lipunova}, {Fabrika}, {Butkevich}
   \& {Abolmasov}}{{Poutanen} et~al.}{2007}]{poutanen07}
{Poutanen} J.,  {Lipunova} G.,  {Fabrika} S.,  {Butkevich} A.~G.,   {Abolmasov}
  P.,  2007, \mn@doi [\mnras] {10.1111/j.1365-2966.2007.11668.x}, \href
  {https://ui.adsabs.harvard.edu/abs/2007MNRAS.377.1187P} {377, 1187}

\bibitem[\protect\citeauthoryear{{Pradhan}, {Maitra}  \& {Paul}}{{Pradhan}
  et~al.}{2020}]{2020ApJ...895...10P}
{Pradhan} P.,  {Maitra} C.,   {Paul} B.,  2020, \mn@doi [\apj]
  {10.3847/1538-4357/ab8224}, \href
  {https://ui.adsabs.harvard.edu/abs/2020ApJ...895...10P} {895, 10}

\bibitem[\protect\citeauthoryear{{Remillard} \& {McClintock}}{{Remillard} \&
  {McClintock}}{2006}]{remillard06}
{Remillard} R.~A.,  {McClintock} J.~E.,  2006, \mn@doi [\araa]
  {10.1146/annurev.astro.44.051905.092532}, \href
  {https://ui.adsabs.harvard.edu/abs/2006ARA&A..44...49R} {44, 49}

\bibitem[\protect\citeauthoryear{{Revnivtsev}, {Churazov}, {Gilfanov}  \&
  {Sunyaev}}{{Revnivtsev} et~al.}{2001}]{Revnivtsev2001}
{Revnivtsev} M.,  {Churazov} E.,  {Gilfanov} M.,   {Sunyaev} R.,  2001, \mn@doi
  [\aap] {10.1051/0004-6361:20010434}, \href
  {https://ui.adsabs.harvard.edu/abs/2001A&A...372..138R} {372, 138}

\bibitem[\protect\citeauthoryear{{Romani}}{{Romani}}{1990}]{romani90}
{Romani} R.~W.,  1990, \mn@doi [\nat] {10.1038/347741a0}, \href
  {https://ui.adsabs.harvard.edu/abs/1990Natur.347..741R} {347, 741}

\bibitem[\protect\citeauthoryear{{Sathyaprakash} et~al.,}{{Sathyaprakash}
  et~al.}{2019}]{sathyaprakash19}
{Sathyaprakash} R.,  et~al., 2019, \mn@doi [\mnras] {10.1093/mnrasl/slz086},
  \href {https://ui.adsabs.harvard.edu/abs/2019MNRAS.488L..35S} {488, L35}

\bibitem[\protect\citeauthoryear{{Scargle}}{{Scargle}}{1982}]{Scargle1982ApJ...263..835S}
{Scargle} J.~D.,  1982, \mn@doi [\apj] {10.1086/160554}, \href
  {https://ui.adsabs.harvard.edu/abs/1982ApJ...263..835S} {263, 835}

\bibitem[\protect\citeauthoryear{{Shakura}, {Postnov}, {Kochetkova}  \&
  {Hjalmarsdotter}}{{Shakura} et~al.}{2012}]{shakura12}
{Shakura} N.,  {Postnov} K.,  {Kochetkova} A.,   {Hjalmarsdotter} L.,  2012,
  \mn@doi [\mnras] {10.1111/j.1365-2966.2011.20026.x}, \href
  {https://ui.adsabs.harvard.edu/abs/2012MNRAS.420..216S} {420, 216}

\bibitem[\protect\citeauthoryear{{Shaposhnikov} \& {Titarchuk}}{{Shaposhnikov}
  \& {Titarchuk}}{2006}]{shaposhnikov06}
{Shaposhnikov} N.,  {Titarchuk} L.,  2006, \mn@doi [\apj] {10.1086/503272},
  \href {https://ui.adsabs.harvard.edu/abs/2006ApJ...643.1098S} {643, 1098}

\bibitem[\protect\citeauthoryear{{Shibazaki}, {Murakami}, {Shaham}  \&
  {Nomoto}}{{Shibazaki} et~al.}{1989}]{shibazaki89}
{Shibazaki} N.,  {Murakami} T.,  {Shaham} J.,   {Nomoto} K.,  1989, \mn@doi
  [\nat] {10.1038/342656a0}, \href
  {https://ui.adsabs.harvard.edu/abs/1989Natur.342..656S} {342, 656}

\bibitem[\protect\citeauthoryear{{Sidoli}, {Esposito}, {Motta}, {Israel}  \&
  {Rodr{\'\i}guez Castillo}}{{Sidoli} et~al.}{2016}]{sidoli16}
{Sidoli} L.,  {Esposito} P.,  {Motta} S.~E.,  {Israel} G.~L.,   {Rodr{\'\i}guez
  Castillo} G.~A.,  2016, \mn@doi [\mnras] {10.1093/mnras/stw1246}, \href
  {https://ui.adsabs.harvard.edu/abs/2016MNRAS.460.3637S} {460, 3637}

\bibitem[\protect\citeauthoryear{{Sobolewska}, {Papadakis}, {Done}  \&
  {Malzac}}{{Sobolewska} et~al.}{2011}]{sobolewska11}
{Sobolewska} M.~A.,  {Papadakis} I.~E.,  {Done} C.,   {Malzac} J.,  2011,
  \mn@doi [\mnras] {10.1111/j.1365-2966.2011.19209.x}, \href
  {https://ui.adsabs.harvard.edu/abs/2011MNRAS.417..280S} {417, 280}

\bibitem[\protect\citeauthoryear{{Stella} \& {Vietri}}{{Stella} \&
  {Vietri}}{1998}]{stella98}
{Stella} L.,  {Vietri} M.,  1998, \mn@doi [\apjl] {10.1086/311075}, \href
  {https://ui.adsabs.harvard.edu/abs/1998ApJ...492L..59S} {492, L59}

\bibitem[\protect\citeauthoryear{{Stella}, {White}  \& {Rosner}}{{Stella}
  et~al.}{1986}]{stella86}
{Stella} L.,  {White} N.~E.,   {Rosner} R.,  1986, \mn@doi [\apj]
  {10.1086/164538}, \href
  {https://ui.adsabs.harvard.edu/abs/1986ApJ...308..669S} {308, 669}

\bibitem[\protect\citeauthoryear{{Takahashi} \& {Ohsuga}}{{Takahashi} \&
  {Ohsuga}}{2017}]{takahashi17}
{Takahashi} H.~R.,  {Ohsuga} K.,  2017, \mn@doi [\apjl]
  {10.3847/2041-8213/aa8222}, \href
  {https://ui.adsabs.harvard.edu/abs/2017ApJ...845L...9T} {845, L9}

\bibitem[\protect\citeauthoryear{{Templeton}}{{Templeton}}{2004}]{Templeton2004JAVSO..32...41T}
{Templeton} M.,  2004, Journal of the American Association of Variable Star
  Observers (JAAVSO), \href
  {https://ui.adsabs.harvard.edu/abs/2004JAVSO..32...41T} {32, 41}

\bibitem[\protect\citeauthoryear{{Tetarenko}, {Dubus}, {Lasota}, {Heinke}  \&
  {Sivakoff}}{{Tetarenko} et~al.}{2018}]{tetarenko18}
{Tetarenko} B.~E.,  {Dubus} G.,  {Lasota} J.~P.,  {Heinke} C.~O.,   {Sivakoff}
  G.~R.,  2018, \mn@doi [\mnras] {10.1093/mnras/sty1798}, \href
  {https://ui.adsabs.harvard.edu/abs/2018MNRAS.480....2T} {480, 2}

\bibitem[\protect\citeauthoryear{{Traulsen} et~al.,}{{Traulsen}
  et~al.}{2020}]{traulsen20}
{Traulsen} I.,  et~al., 2020, \mn@doi [\aap] {10.1051/0004-6361/202037706},
  \href {https://ui.adsabs.harvard.edu/abs/2020A&A...641A.137T} {641, A137}

\bibitem[\protect\citeauthoryear{{Tse}, {Galloway}, {Chou}, {Heger}  \&
  {Hsieh}}{{Tse} et~al.}{2021}]{tse21}
{Tse} K.,  {Galloway} D.~K.,  {Chou} Y.,  {Heger} A.,   {Hsieh} H.-E.,  2021,
  \mn@doi [\mnras] {10.1093/mnras/staa3224}, \href
  {https://ui.adsabs.harvard.edu/abs/2021MNRAS.500...34T} {500, 34}

\bibitem[\protect\citeauthoryear{{Tsygankov}, {Mushtukov}, {Suleimanov}  \&
  {Poutanen}}{{Tsygankov} et~al.}{2016}]{tsygankov16}
{Tsygankov} S.~S.,  {Mushtukov} A.~A.,  {Suleimanov} V.~F.,   {Poutanen} J.,
  2016, \mn@doi [\mnras] {10.1093/mnras/stw046}, \href
  {https://ui.adsabs.harvard.edu/abs/2016MNRAS.457.1101T} {457, 1101}

\bibitem[\protect\citeauthoryear{{Tsygankov}, {Doroshenko}, {Lutovinov},
  {Mushtukov}  \& {Poutanen}}{{Tsygankov} et~al.}{2017}]{tsigankov17}
{Tsygankov} S.~S.,  {Doroshenko} V.,  {Lutovinov} A.~A.,  {Mushtukov} A.~A.,
  {Poutanen} J.,  2017, \mn@doi [\aap] {10.1051/0004-6361/201730553}, \href
  {https://ui.adsabs.harvard.edu/abs/2017A&A...605A..39T} {605, A39}

\bibitem[\protect\citeauthoryear{{Ueda}, {Yamaoka}  \& {Remillard}}{{Ueda}
  et~al.}{2009}]{ueda09}
{Ueda} Y.,  {Yamaoka} K.,   {Remillard} R.,  2009, \mn@doi [\apj]
  {10.1088/0004-637X/695/2/888}, \href
  {https://ui.adsabs.harvard.edu/abs/2009ApJ...695..888U} {695, 888}

\bibitem[\protect\citeauthoryear{{Urquhart} \& {Soria}}{{Urquhart} \&
  {Soria}}{2016}]{urquhart16}
{Urquhart} R.,  {Soria} R.,  2016, \mn@doi [\mnras] {10.1093/mnras/stv2293},
  \href {https://ui.adsabs.harvard.edu/abs/2016MNRAS.456.1859U} {456, 1859}

\bibitem[\protect\citeauthoryear{{Vasilopoulos} et~al.,}{{Vasilopoulos}
  et~al.}{2020}]{vasilopoulos20}
{Vasilopoulos} G.,  et~al., 2020, \mn@doi [\mnras] {10.1093/mnras/staa991},
  \href {https://ui.adsabs.harvard.edu/abs/2020MNRAS.494.5350V} {494, 5350}

\bibitem[\protect\citeauthoryear{{Walton} et~al.,}{{Walton}
  et~al.}{2018}]{walton18}
{Walton} D.~J.,  et~al., 2018, \mn@doi [\apj] {10.3847/1538-4357/aab610}, \href
  {https://ui.adsabs.harvard.edu/abs/2018ApJ...856..128W} {856, 128}

\bibitem[\protect\citeauthoryear{{Wang}, {Liu}, {Qiu}, {Bai}, {Yang}, {Guo}  \&
  {Zhang}}{{Wang} et~al.}{2016}]{wang16}
{Wang} S.,  {Liu} J.,  {Qiu} Y.,  {Bai} Y.,  {Yang} H.,  {Guo} J.,   {Zhang}
  P.,  2016, \mn@doi [\apjs] {10.3847/0067-0049/224/2/40}, \href
  {https://ui.adsabs.harvard.edu/abs/2016ApJS..224...40W} {224, 40}

\bibitem[\protect\citeauthoryear{{Webb} et~al.,}{{Webb} et~al.}{2020}]{webb20}
{Webb} N.~A.,  et~al., 2020, \mn@doi [\aap] {10.1051/0004-6361/201937353},
  \href {https://ui.adsabs.harvard.edu/abs/2020A&A...641A.136W} {641, A136}

\bibitem[\protect\citeauthoryear{{Wilkins}}{{Wilkins}}{1972}]{wilkins72}
{Wilkins} D.~C.,  1972, \mn@doi [\prd] {10.1103/PhysRevD.5.814}, \href
  {https://ui.adsabs.harvard.edu/abs/1972PhRvD...5..814W} {5, 814}

\bibitem[\protect\citeauthoryear{{Wilson-Hodge} et~al.,}{{Wilson-Hodge}
  et~al.}{2018}]{wilson-hodge18}
{Wilson-Hodge} C.~A.,  et~al., 2018, \mn@doi [\apj] {10.3847/1538-4357/aace60},
  \href {https://ui.adsabs.harvard.edu/abs/2018ApJ...863....9W} {863, 9}

\bibitem[\protect\citeauthoryear{{Yang}, {Xie}, {Yuan}, {Zdziarski},
  {Gierli{\'n}ski}, {Ho}  \& {Yu}}{{Yang} et~al.}{2015}]{yang15}
{Yang} Q.-X.,  {Xie} F.-G.,  {Yuan} F.,  {Zdziarski} A.~A.,  {Gierli{\'n}ski}
  M.,  {Ho} L.~C.,   {Yu} Z.,  2015, \mn@doi [\mnras] {10.1093/mnras/stu2571},
  \href {https://ui.adsabs.harvard.edu/abs/2015MNRAS.447.1692Y} {447, 1692}

\bibitem[\protect\citeauthoryear{{van den Eijnden} et~al.,}{{van den Eijnden}
  et~al.}{2021}]{vandeneijnden21}
{van den Eijnden} J.,  et~al., 2021, \mn@doi [\mnras] {10.1093/mnras/stab1995},
  \href {https://ui.adsabs.harvard.edu/abs/2021MNRAS.507.3899V} {507, 3899}

\makeatother
\end{thebibliography}



\appendix

\section{Spectral and timing properties of selected other {\it Chandra} observations}

\begin{figure*}
    \centering
    \begin{tabular}{cc}
        \begin{subfigure}{0.48\textwidth}
            \hspace{-0.3cm}
            \includegraphics[height=0.99\textwidth, angle=270]{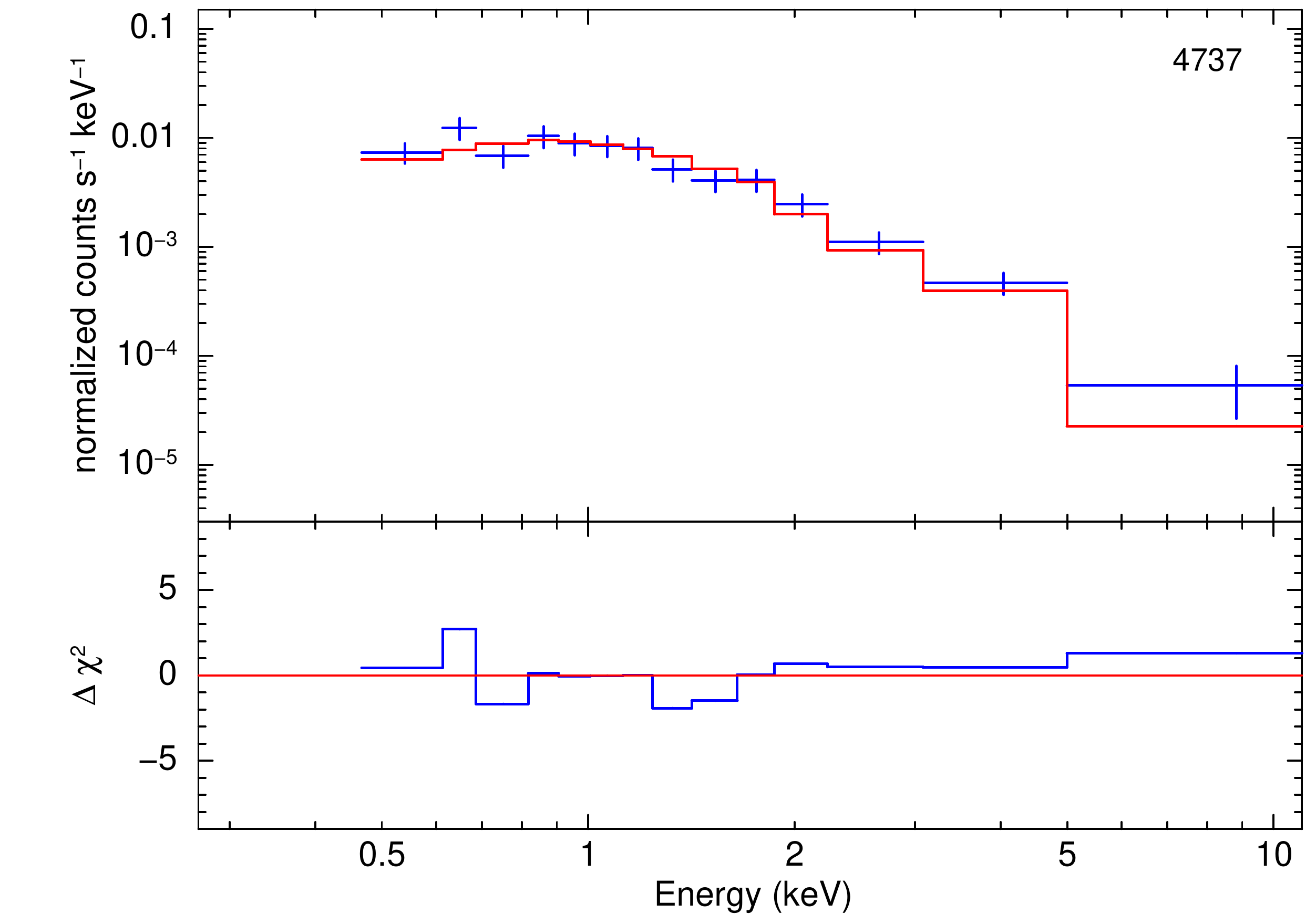}
        \end{subfigure} &
        \begin{subfigure}{0.48\textwidth}
            \hspace{-0.8cm}
            \includegraphics[height=0.99\textwidth, angle=270]{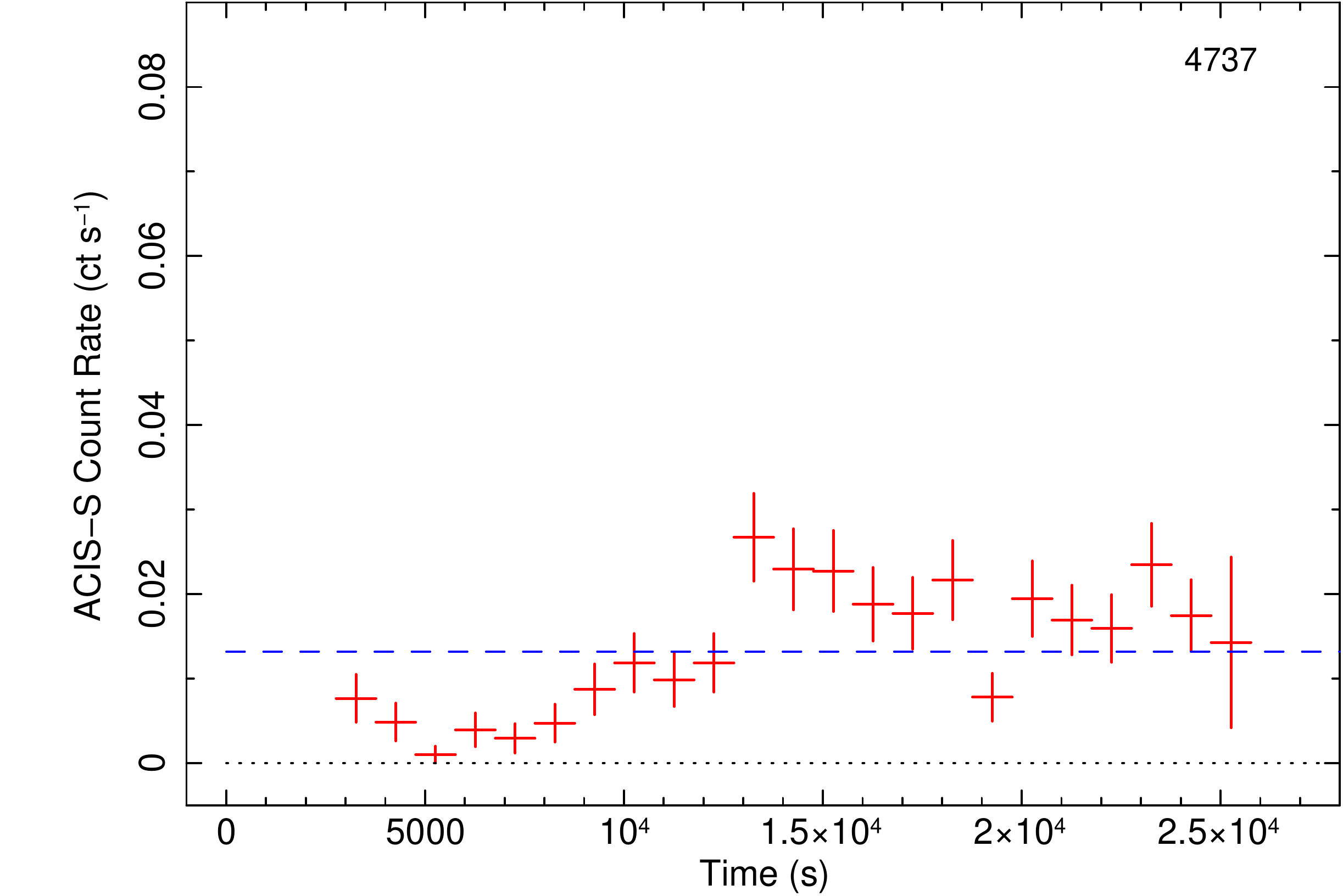}
        \end{subfigure}   \\
        \begin{subfigure}{0.48\textwidth}
            \vspace{1.3cm}
            \includegraphics[width=0.97\textwidth, angle=0]{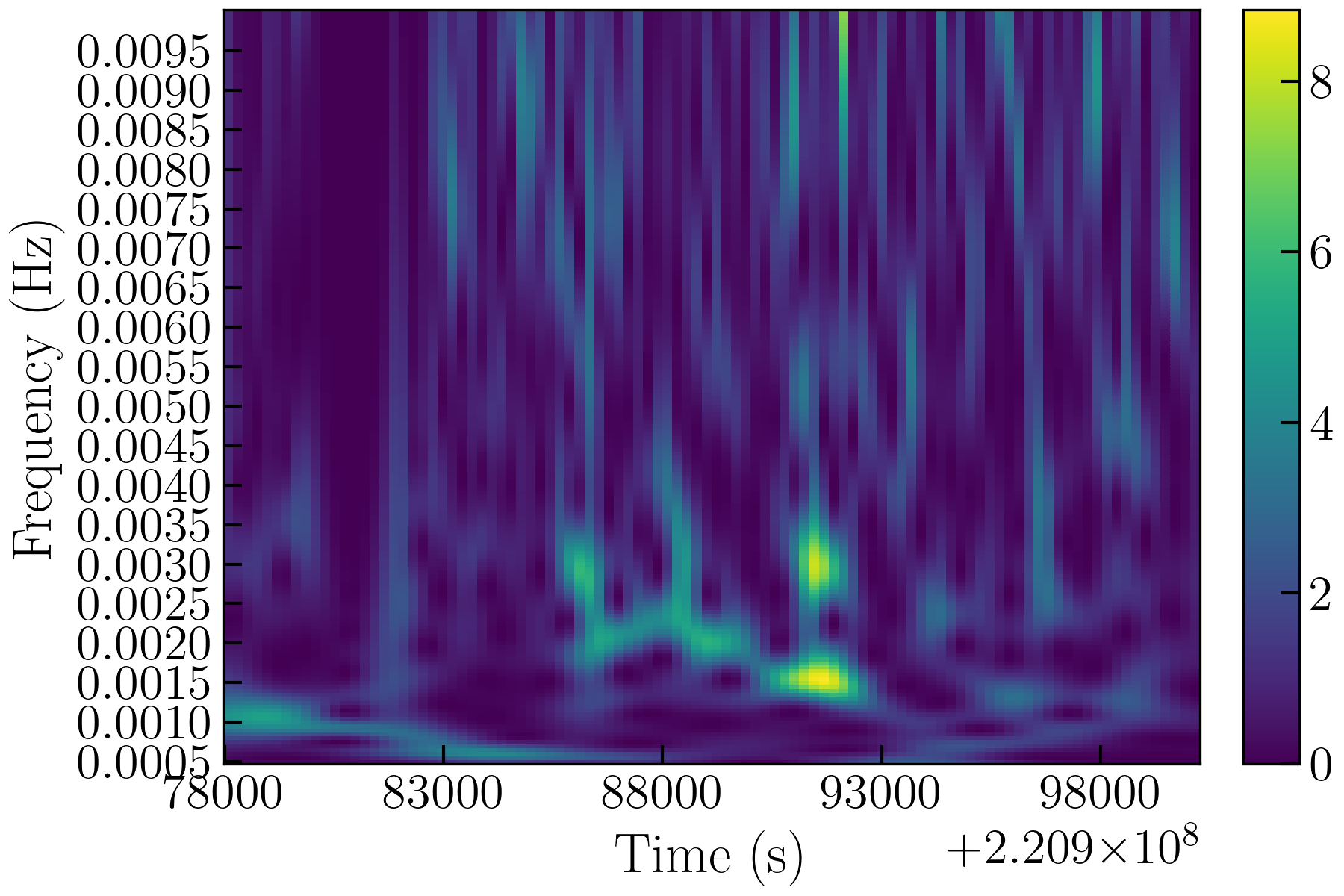}
        \end{subfigure} &
        \multirow{2}{*}{
            \begin{subfigure}{0.5\textwidth}
                \vspace{-1.5cm}
                \hspace{-0.5cm}
                \includegraphics[width=0.98\textwidth, angle=0]{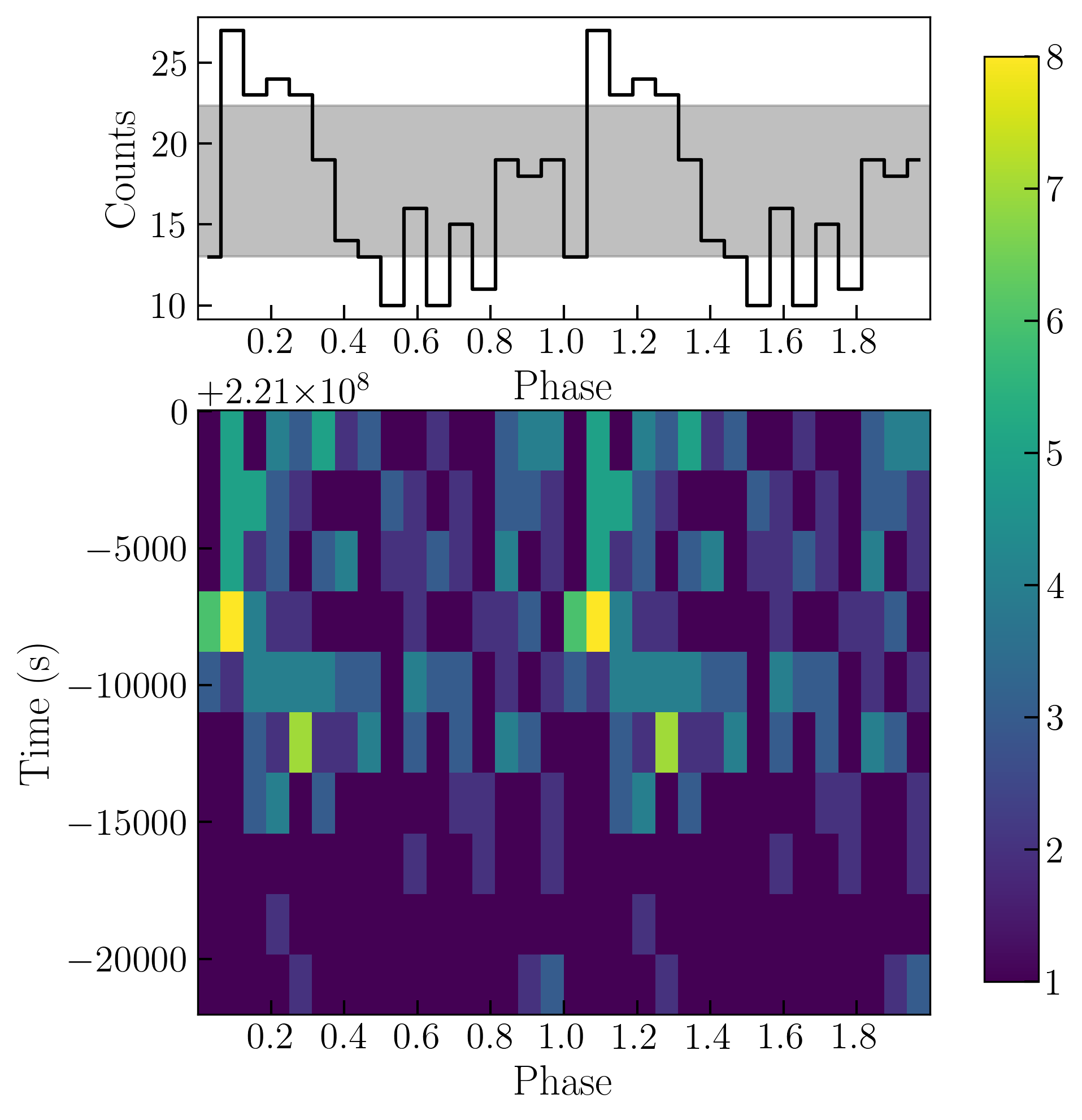}
            \end{subfigure}
        }                           \\
        \begin{subfigure}{0.48\textwidth}
            \includegraphics[width=0.95\textwidth, angle=0]{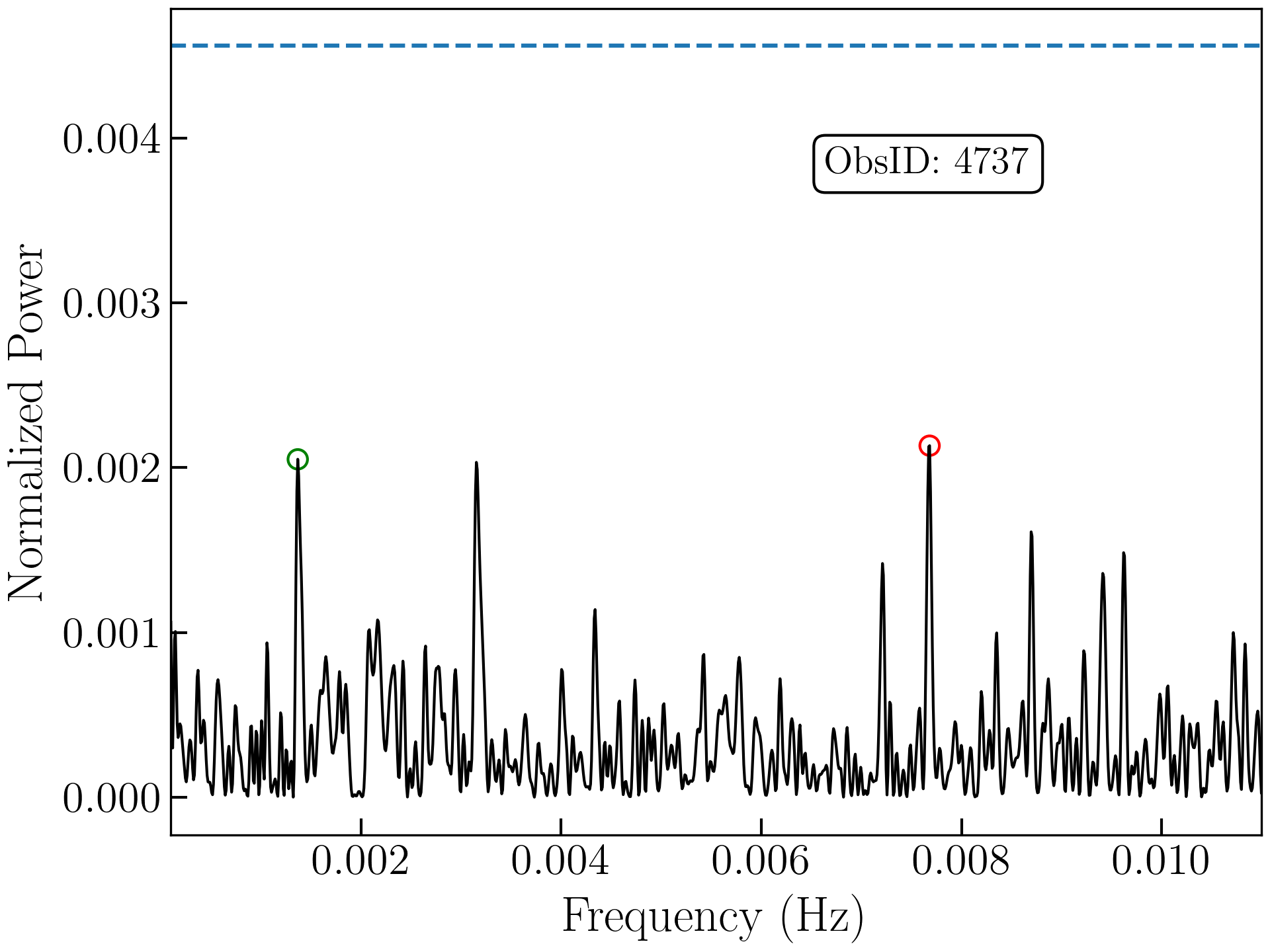}
        \end{subfigure}
                                  &
    \end{tabular}
    \caption{Summary of the main spectral and timing properties of J1403 during {\it Chandra} ObsID 4737. Left column, top panel: X-ray spectrum fitted with an absorbed double-thermal model ({\it bbodyrad} $+$ {\it diskpbb} in {\sc xspec}), and corresponding $\chi^2$ residuals (see Table 4 for the fit parameters, flux and luminosity). Right column, top panel: background-subtracted count rate in the 0.5--7 keV band (binned to 1000 s). The dashed blue line corresponds to an approximate unabsorbed 0.3--10 keV luminosity of $10^{39}$ erg s$^{-1}$, based on the best-fitting spectral model. Left column, middle panel: WWZ dynamical power spectrum, showing a strong QPO over a short interval of the observation. Left column, bottom panel: Lomb--Scargle analysis of the most significant frequencies. The dashed blue line represents a 99\% significance level over the whole observation. The open red circle is the most significant frequency and the green circle indicates a local peak around the 1.5\,mHz. Right column, bottom panel: phaseogram and folded lightcurve with using local frequency peak indicated by the open green circle.}
    \label{fig:4737_all}
\end{figure*}

\begin{figure*}
    \centering
    \begin{tabular}{cc}
        \begin{subfigure}{0.48\textwidth}
            \hspace{-0.3cm}
            \includegraphics[height=0.99\textwidth, angle=270]{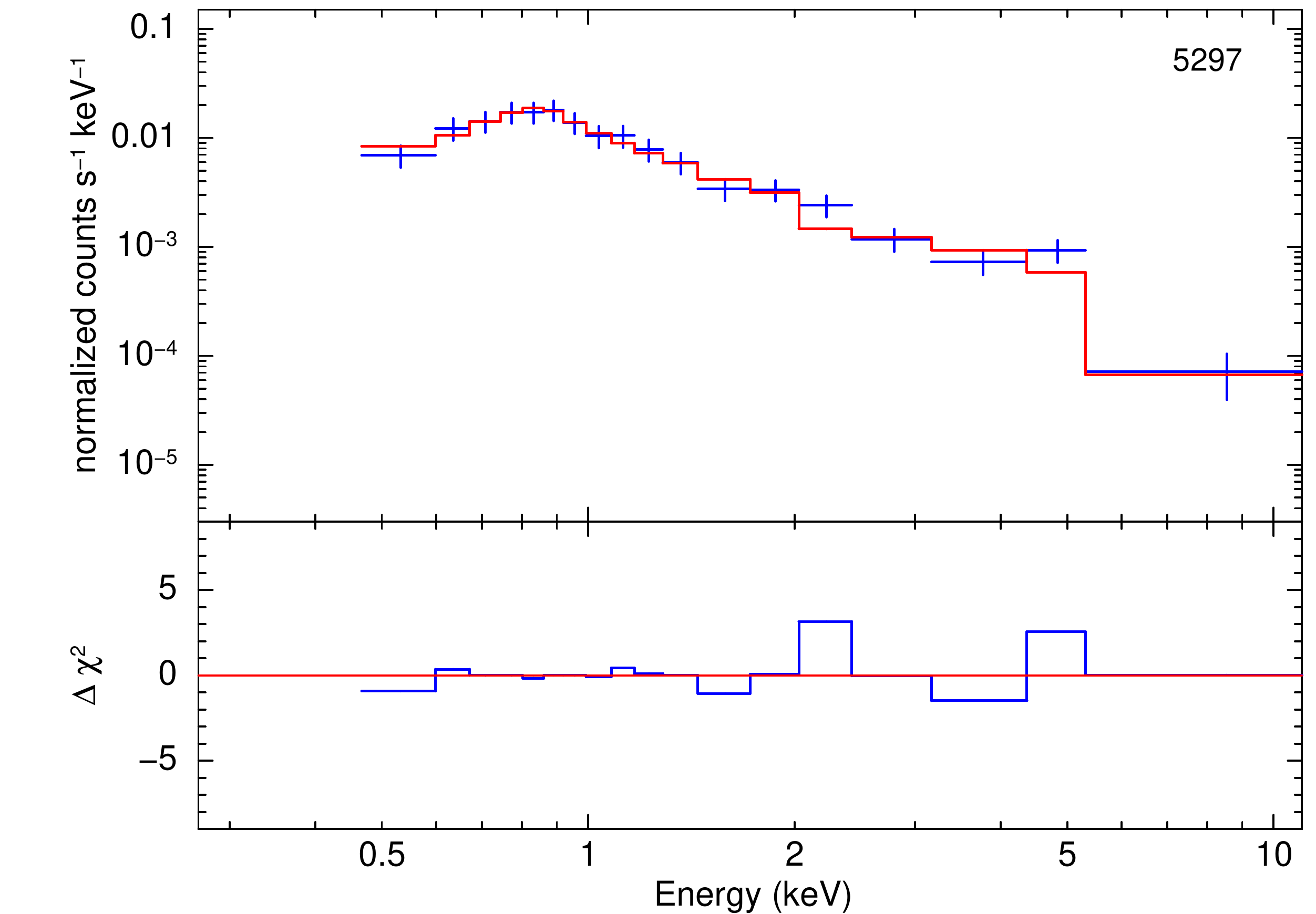}
        \end{subfigure} &
        \begin{subfigure}{0.48\textwidth}
            \hspace{-0.8cm}
            \includegraphics[height=0.99\textwidth, angle=270]{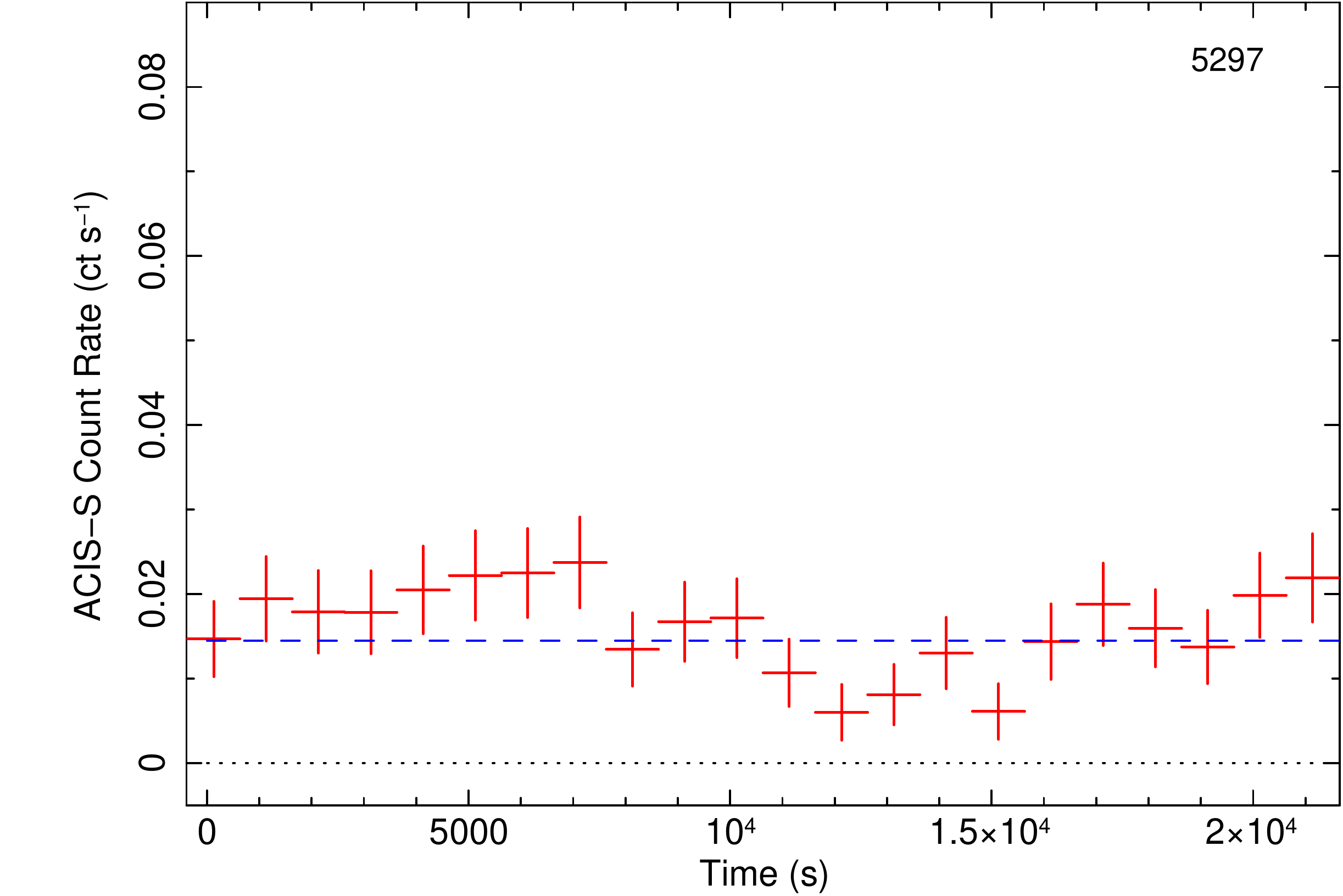}
        \end{subfigure}   \\
        \begin{subfigure}{0.48\textwidth}
            \vspace{1.3cm}
            \includegraphics[width=0.97\textwidth, angle=0]{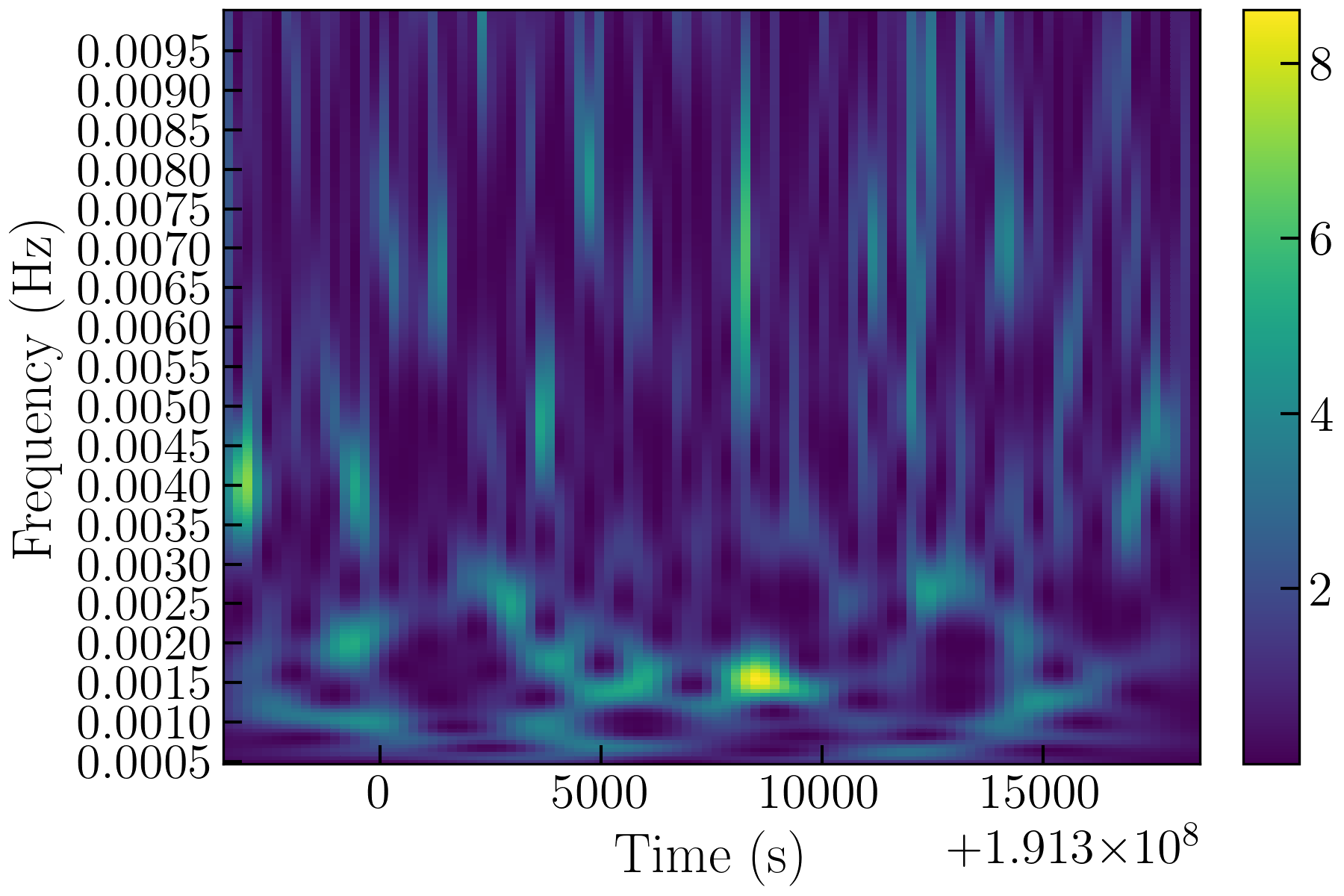}
        \end{subfigure} &
        \multirow{2}{*}{
            \begin{subfigure}{0.5\textwidth}
                \vspace{-1.5cm}
                \hspace{-0.5cm}
                \includegraphics[width=0.98\textwidth, angle=0]{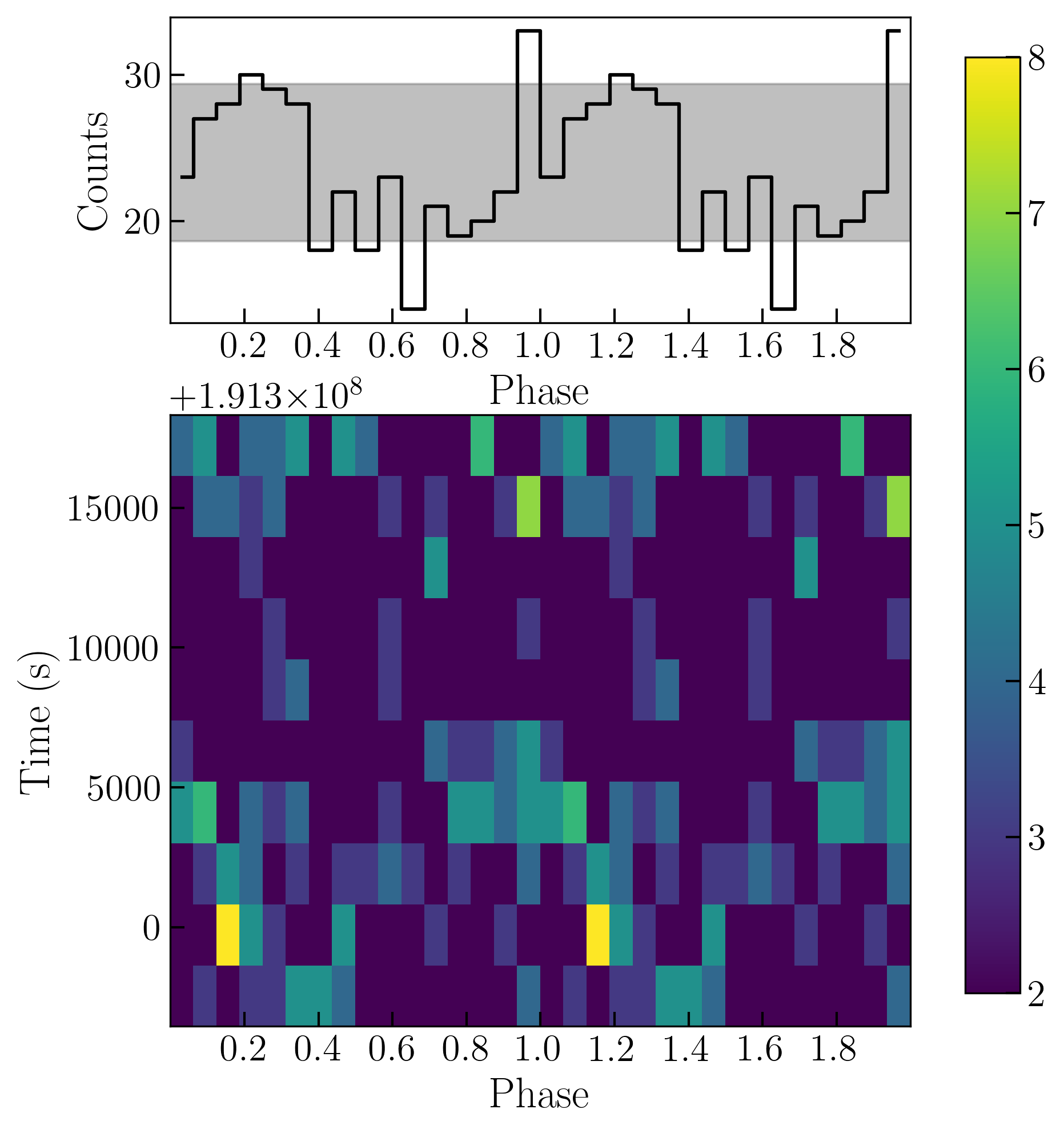}
            \end{subfigure}
        }                            \\
        \begin{subfigure}{0.48\textwidth}
            \includegraphics[width=0.95\textwidth, angle=0]{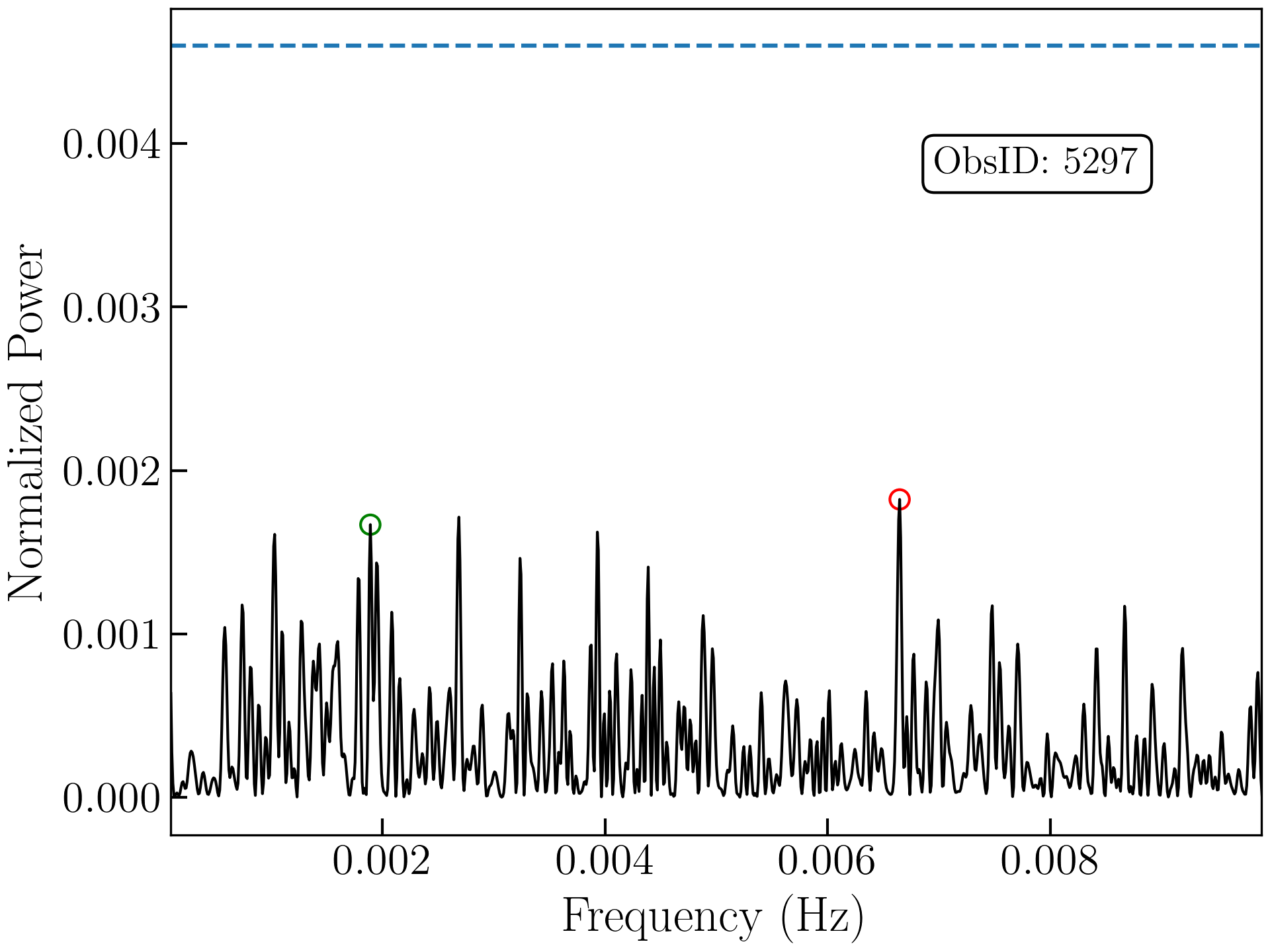}
        \end{subfigure}
                                   &
    \end{tabular}
    \caption{As in Figure A1, for {\it Chandra} ObsID 5297. The spectral model is an absorbed double-thermal model plus optically thin plasma ({\it bbodyrad} $+$ {\it diskpbb} $+$ {\it apec} in {\sc xspec}). See Table 5 for the fit parameters.}
    \label{fig:5297_all}
\end{figure*}

\begin{figure*}
    \centering
    \begin{tabular}{cc}
        \begin{subfigure}{0.48\textwidth}
            \hspace{-0.3cm}
            \includegraphics[height=0.99\textwidth, angle=270]{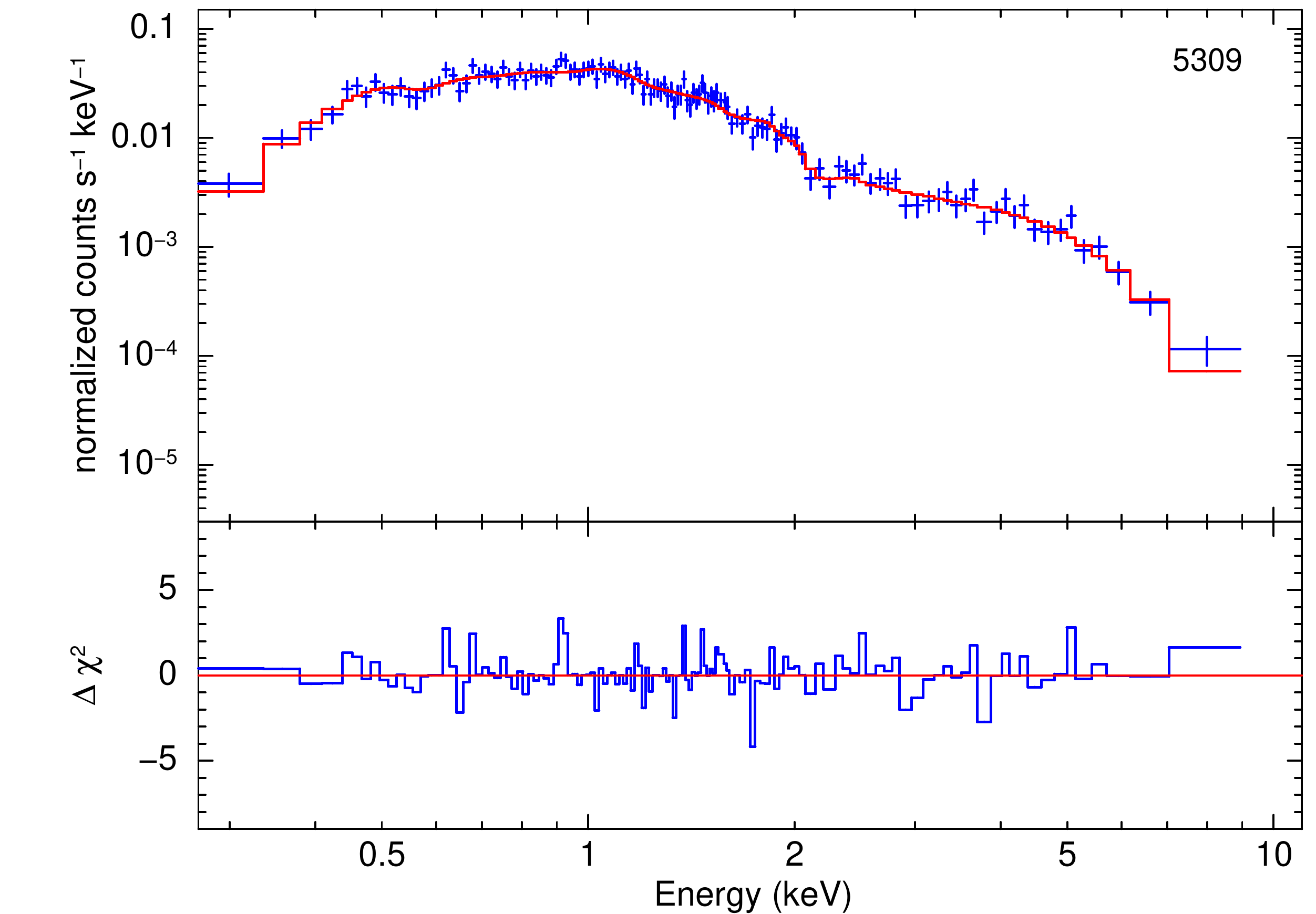}
        \end{subfigure} &
        \begin{subfigure}{0.48\textwidth}
            \hspace{-0.8cm}
            \includegraphics[height=0.99\textwidth, angle=270]{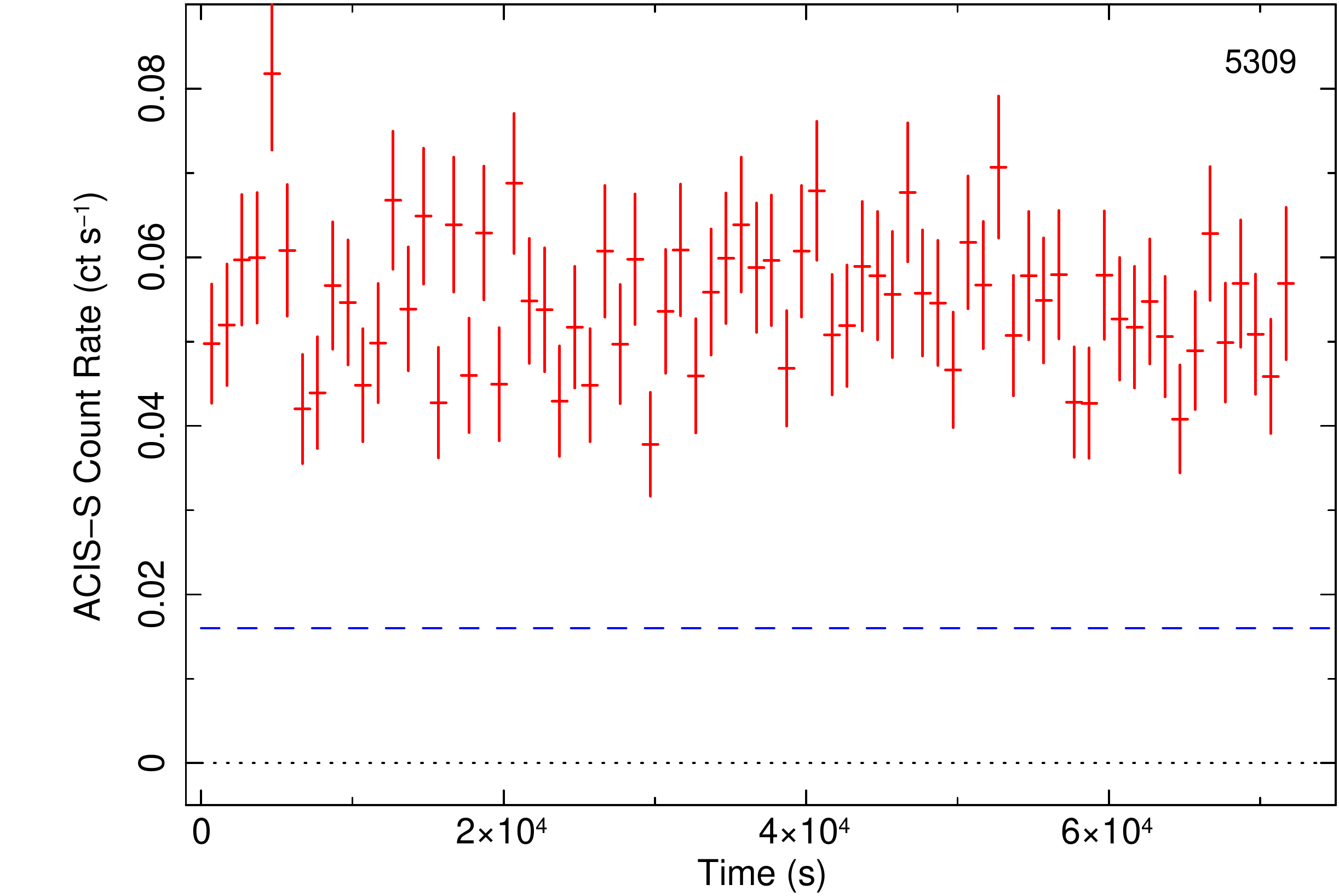}
        \end{subfigure}   \\
        \begin{subfigure}{0.48\textwidth}
            \vspace{1.3cm}
            \includegraphics[width=0.97\textwidth, angle=0]{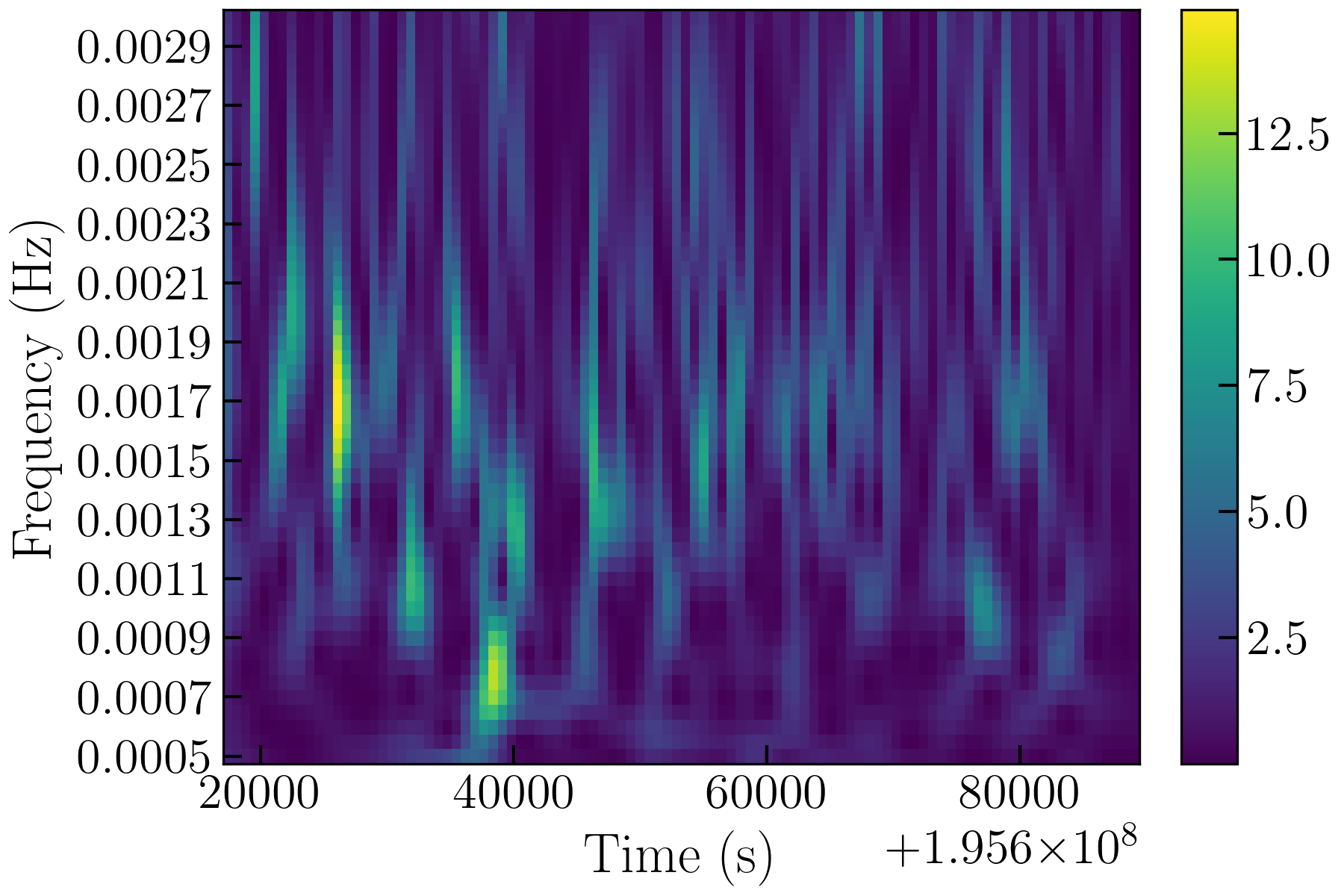}
        \end{subfigure} &
        \multirow{2}{*}{
            \begin{subfigure}{0.5\textwidth}
                \vspace{-1.5cm}
                \hspace{-0.5cm}
                \includegraphics[width=0.98\textwidth, angle=0]{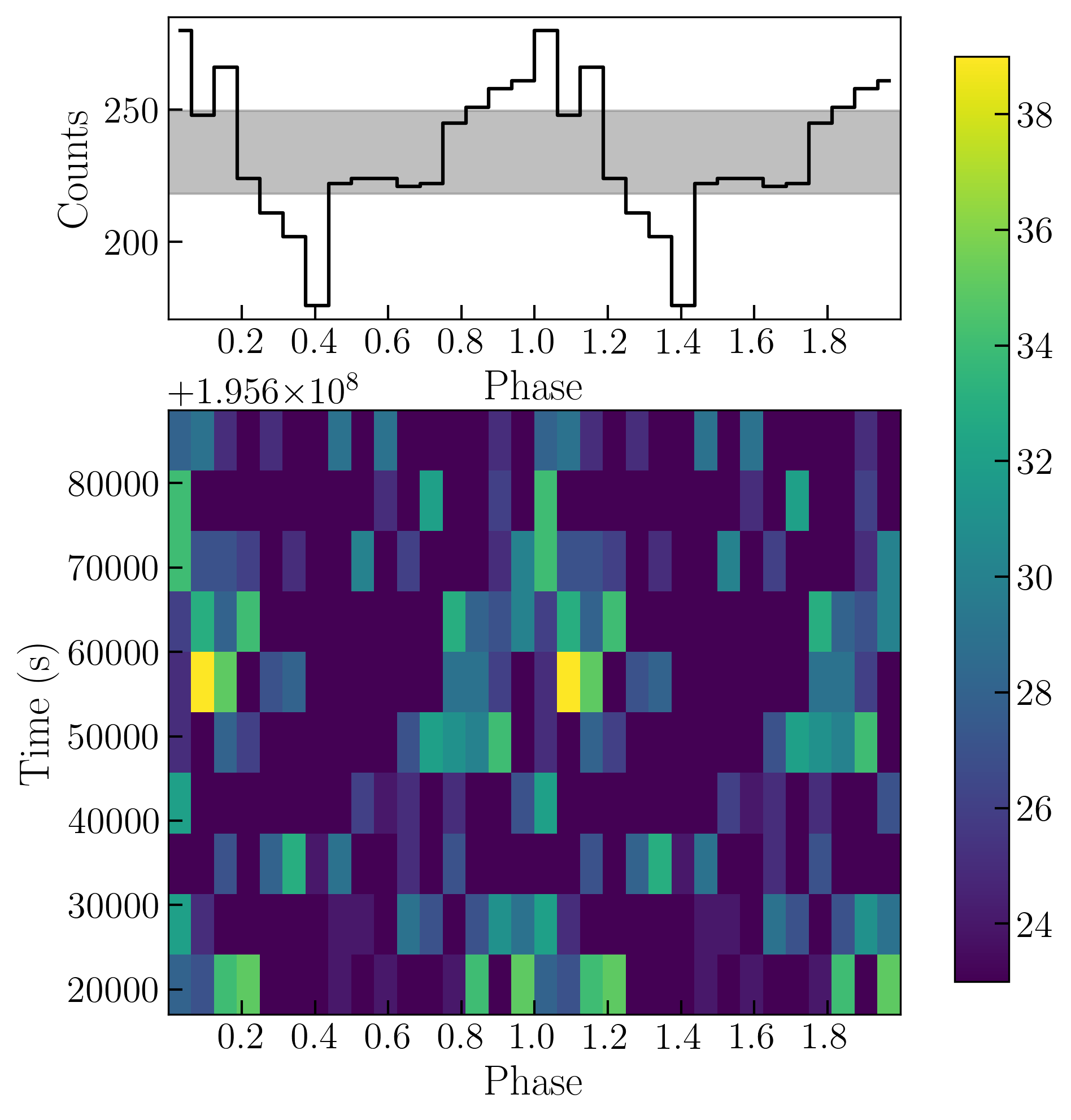}
            \end{subfigure}
        }                            \\
        \begin{subfigure}{0.48\textwidth}
            \includegraphics[width=0.95\textwidth, angle=0]{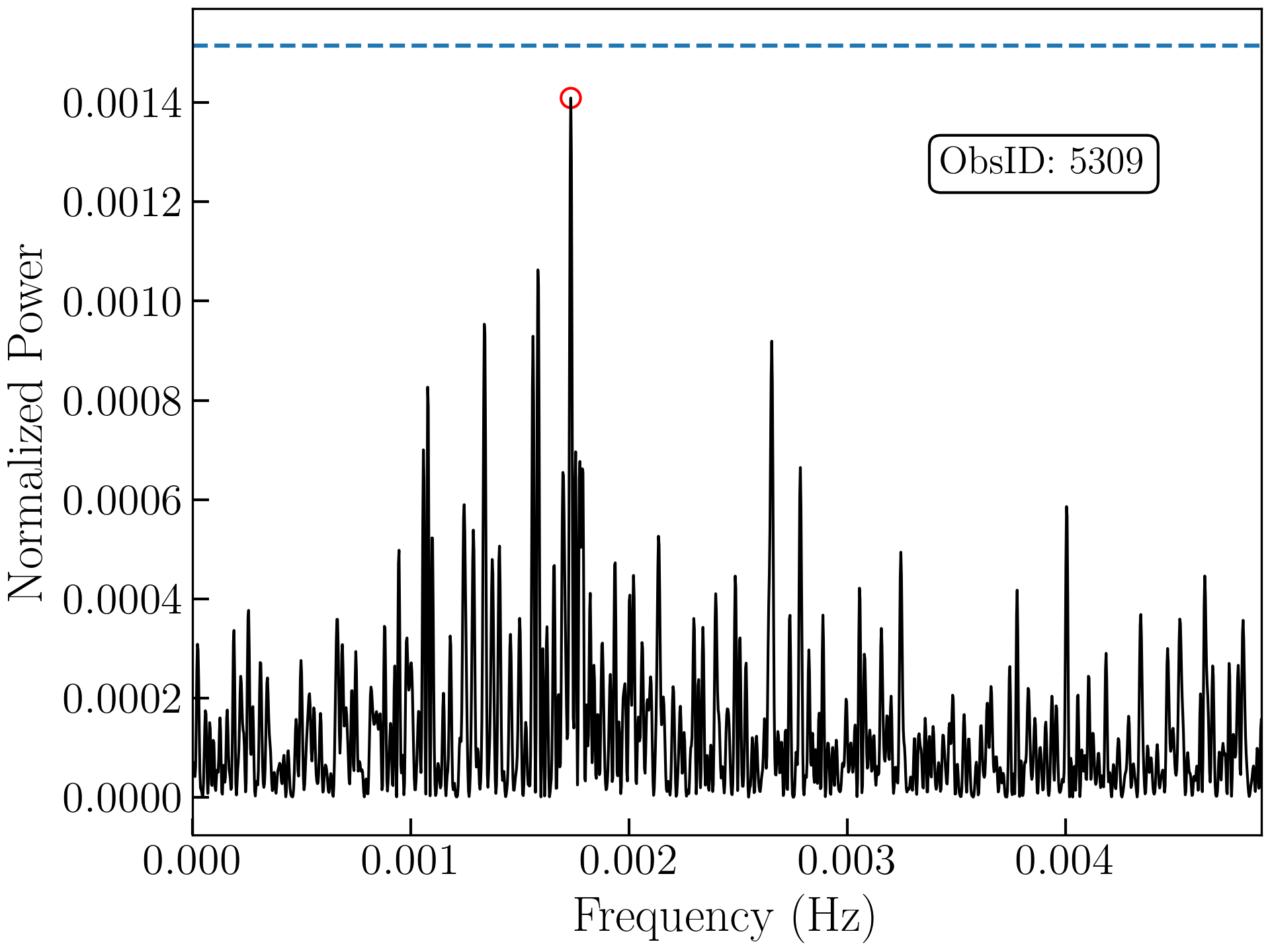}
        \end{subfigure}
                                   &
    \end{tabular}
    \caption{As in Figure A1, for {\it Chandra} ObsID 5309. The spectral model is an absorbed double-thermal model plus optically thin plasma ({\it bbodyrad} $+$ {\it diskpbb} $+$ {\it apec} in {\sc xspec}). See Table 5 for the fit parameters.}
    \label{fig:5039_all}
\end{figure*}
    
\begin{figure*}
    \centering
    \begin{tabular}{cc}
        \begin{subfigure}{0.48\textwidth}
            \hspace{-0.3cm}
            \includegraphics[height=0.99\textwidth, angle=270]{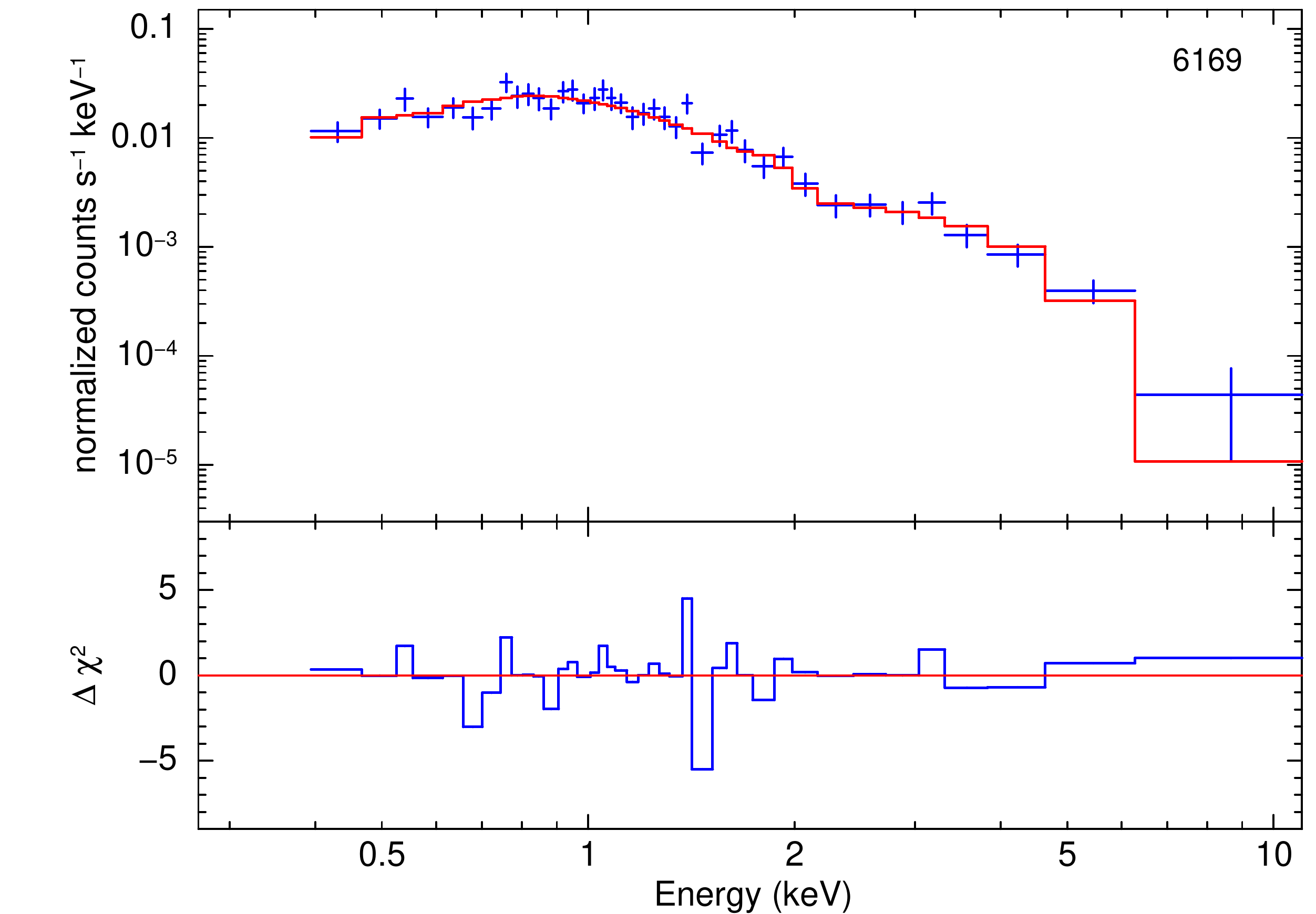}
        \end{subfigure} &
        \begin{subfigure}{0.48\textwidth}
            \hspace{-0.8cm}
            \includegraphics[height=0.99\textwidth, angle=270]{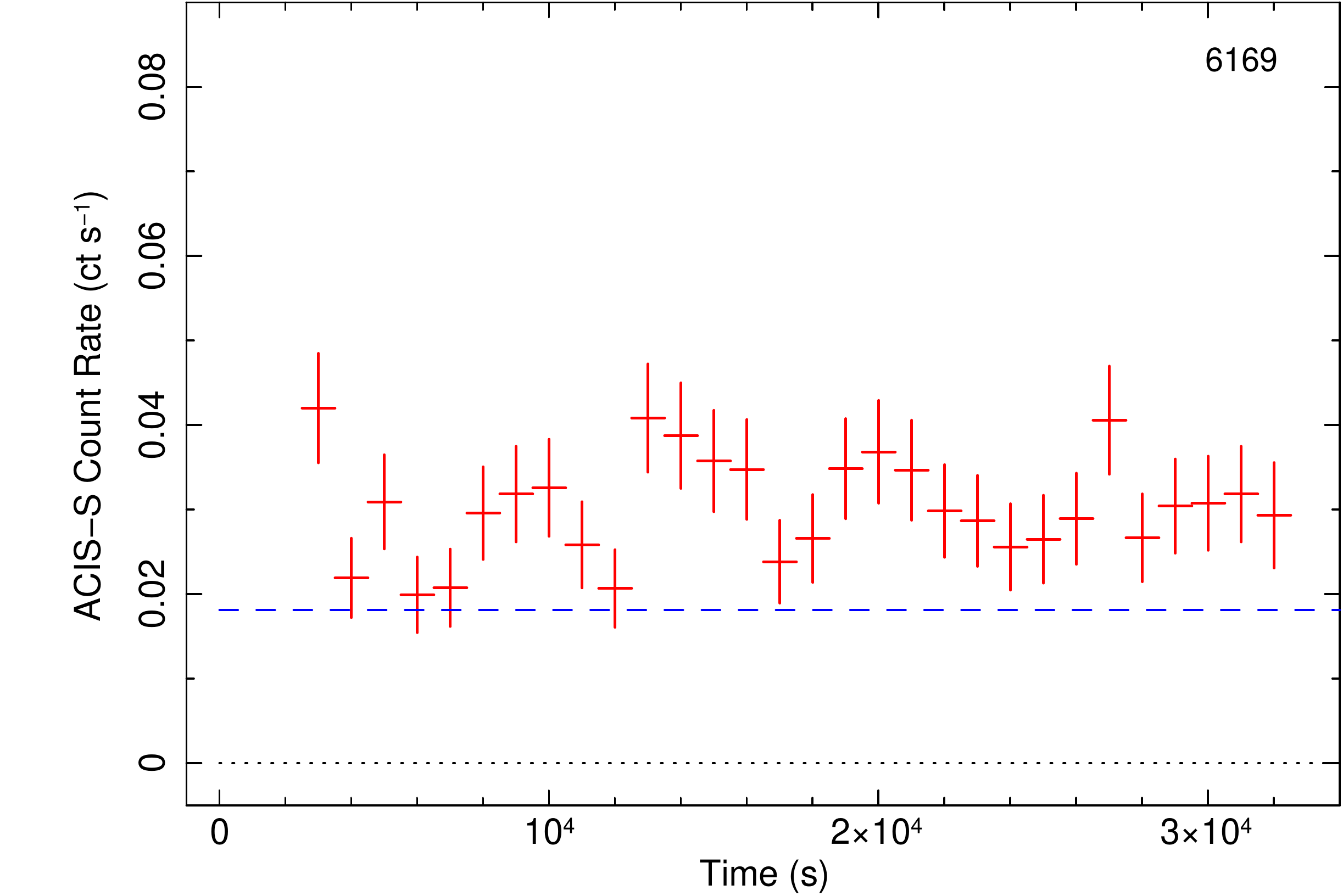}
        \end{subfigure}   \\
        \begin{subfigure}{0.48\textwidth}
            \vspace{1.3cm}
            \includegraphics[width=0.97\textwidth, angle=0]{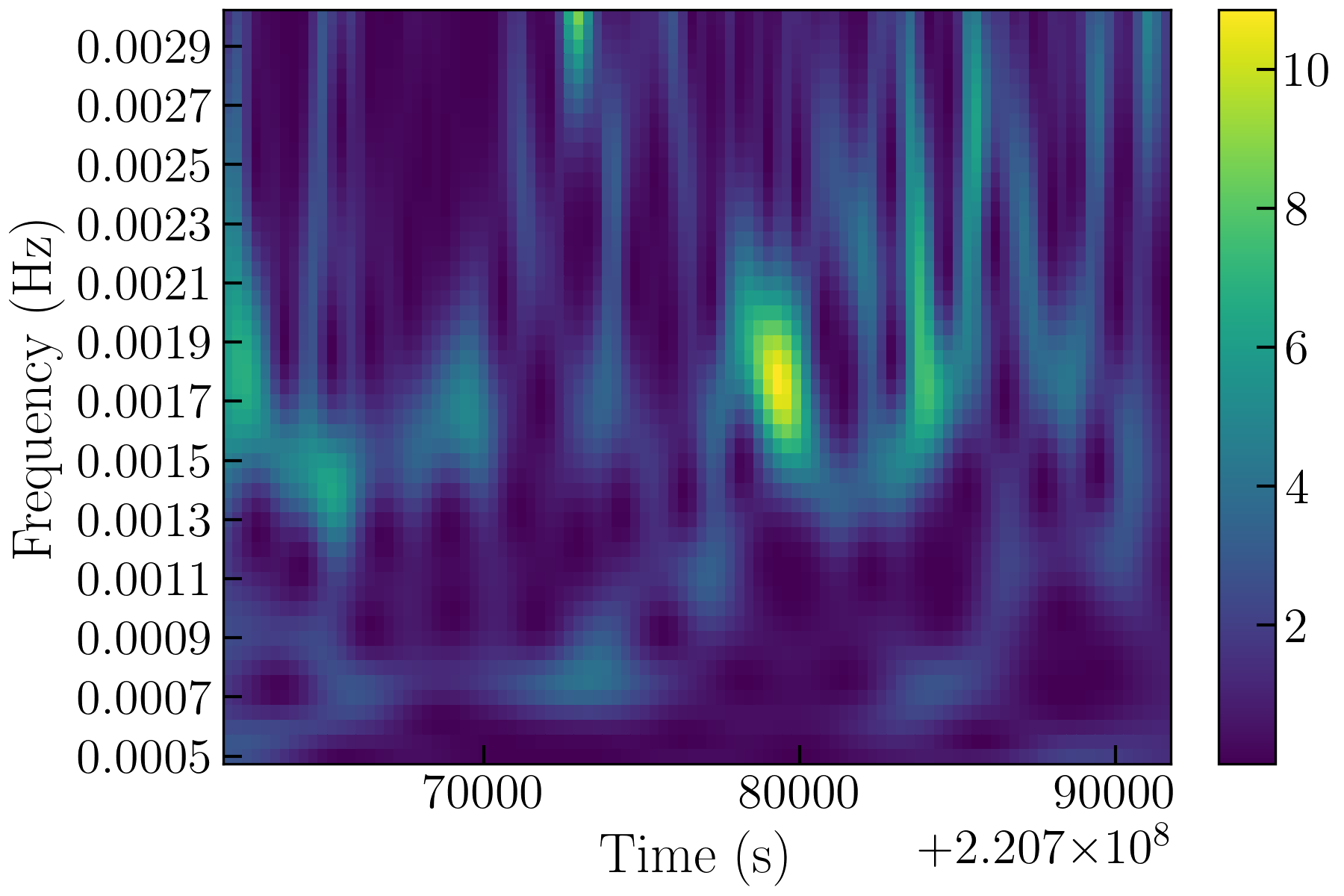}
        \end{subfigure} &
        \multirow{2}{*}{
            \begin{subfigure}{0.5\textwidth}
                \vspace{-1.5cm}
                \hspace{-0.5cm}
                \includegraphics[width=0.98\textwidth, angle=0]{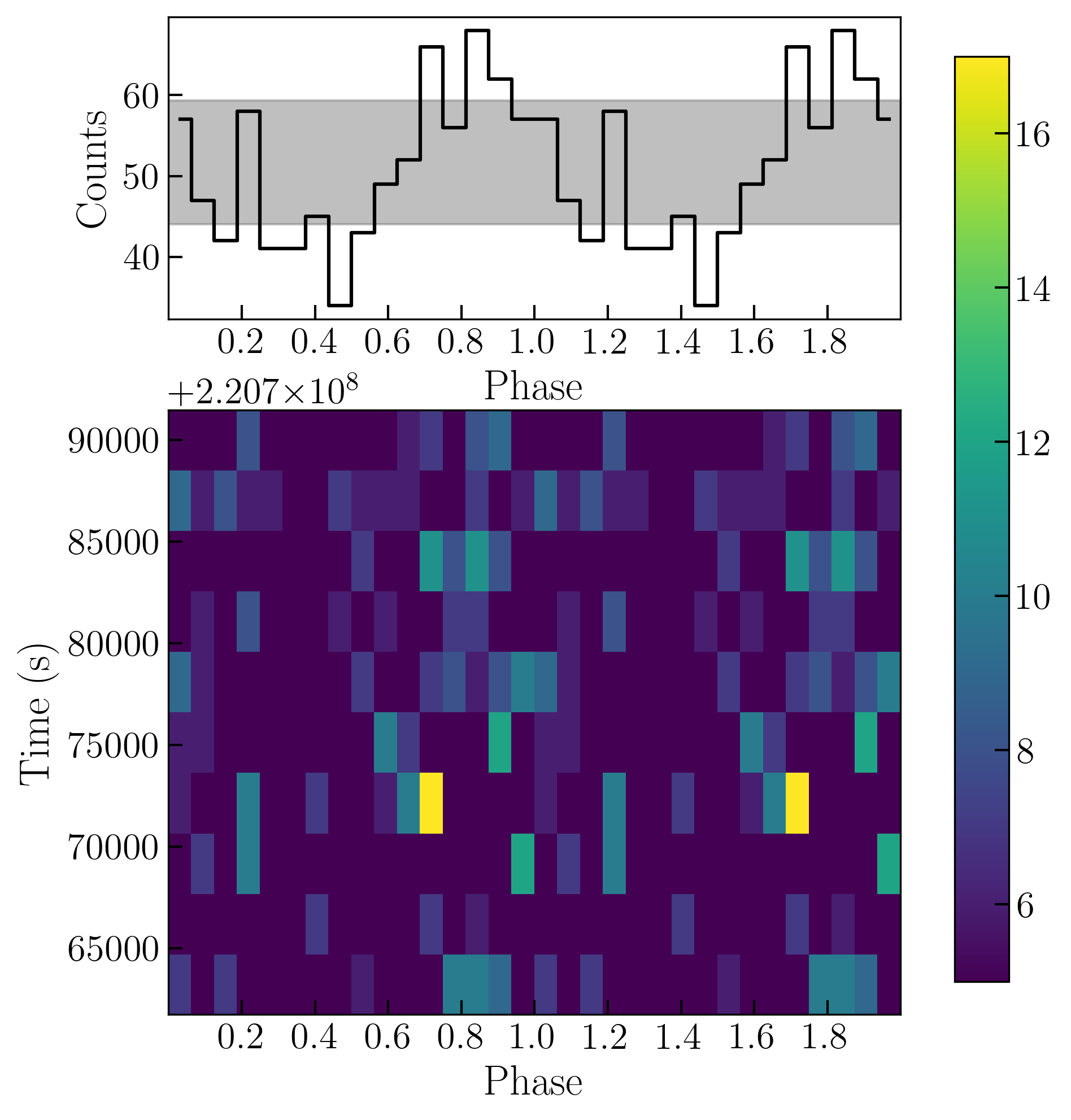}
            \end{subfigure}
        }                            \\
        \begin{subfigure}{0.48\textwidth}
            \includegraphics[width=0.95\textwidth, angle=0]{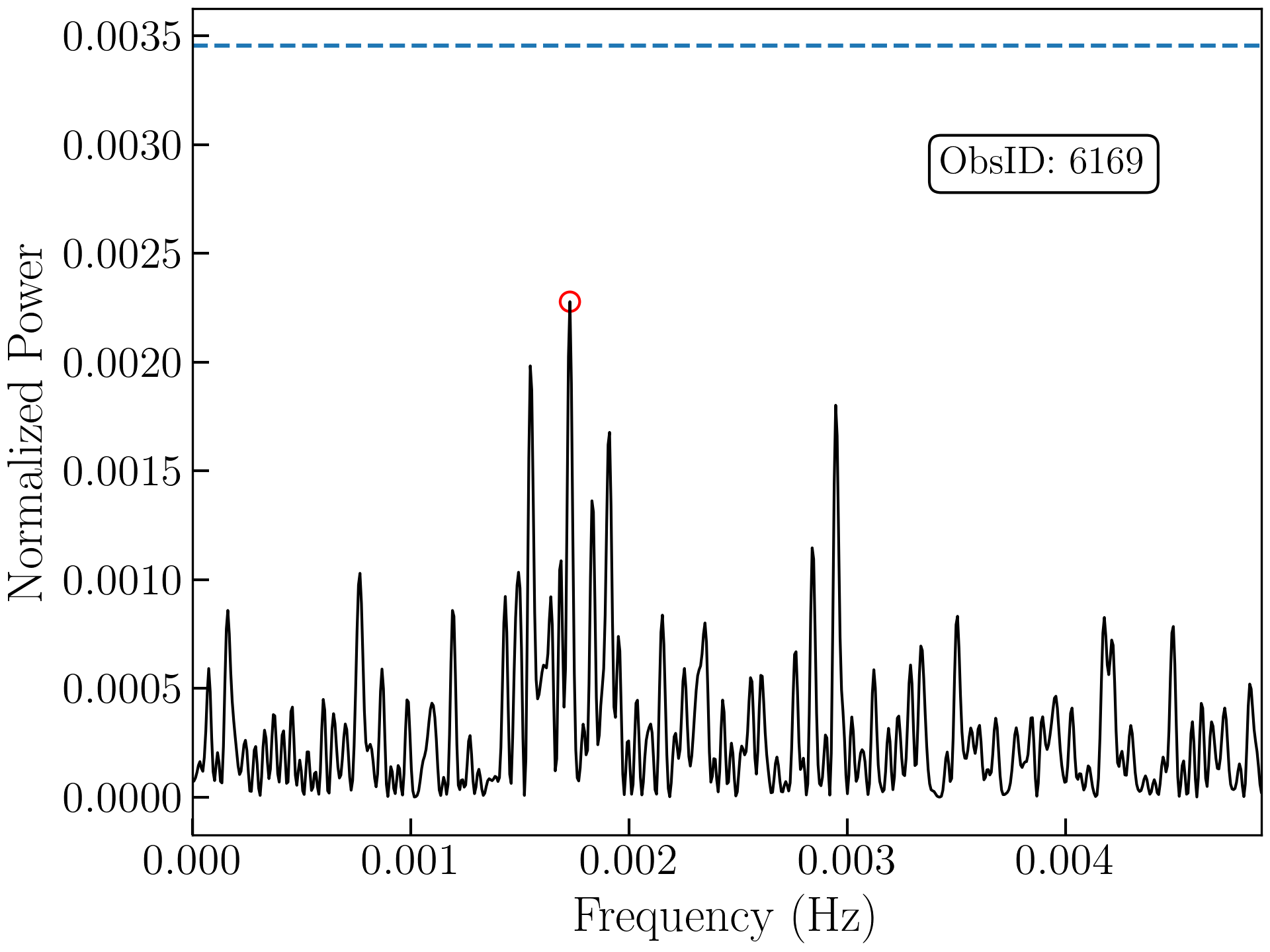}
        \end{subfigure}
                                   &
    \end{tabular}
    \caption{As in Figure A1, for {\it Chandra} ObsID 6169. The spectral model is an absorbed double-thermal model ({\it bbodyrad} $+$ {\it diskpbb} in {\sc xspec}). See Table 5 for the fit parameters.}
    \label{fig:6169_all}
\end{figure*}

\begin{figure*}
    \centering
    \begin{tabular}{cc}
        \begin{subfigure}{0.48\textwidth}
            \hspace{-0.3cm}
            \includegraphics[height=0.99\textwidth, angle=270]{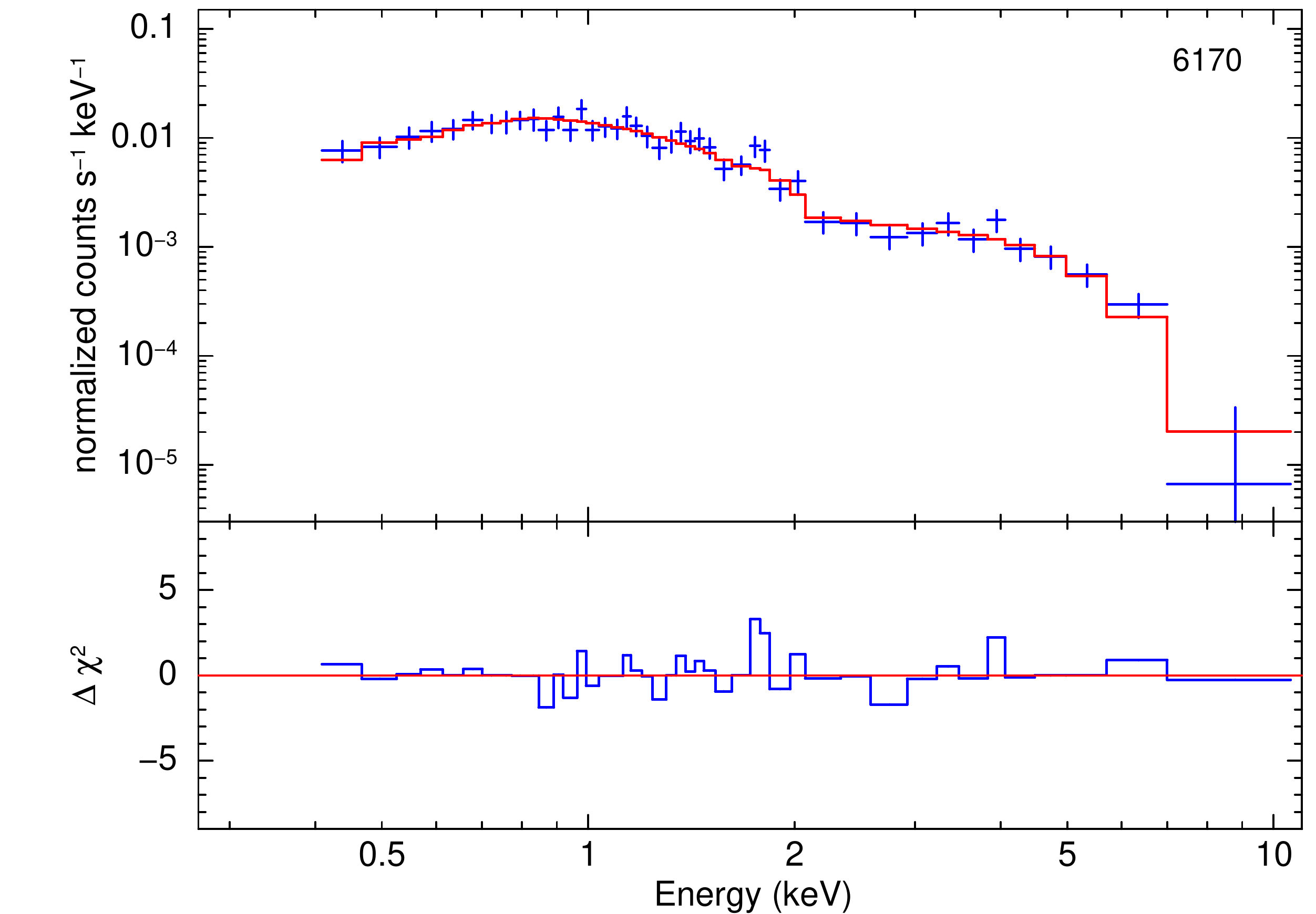}
        \end{subfigure} &
        \begin{subfigure}{0.48\textwidth}
            \hspace{-0.8cm}
            \includegraphics[height=0.99\textwidth, angle=270]{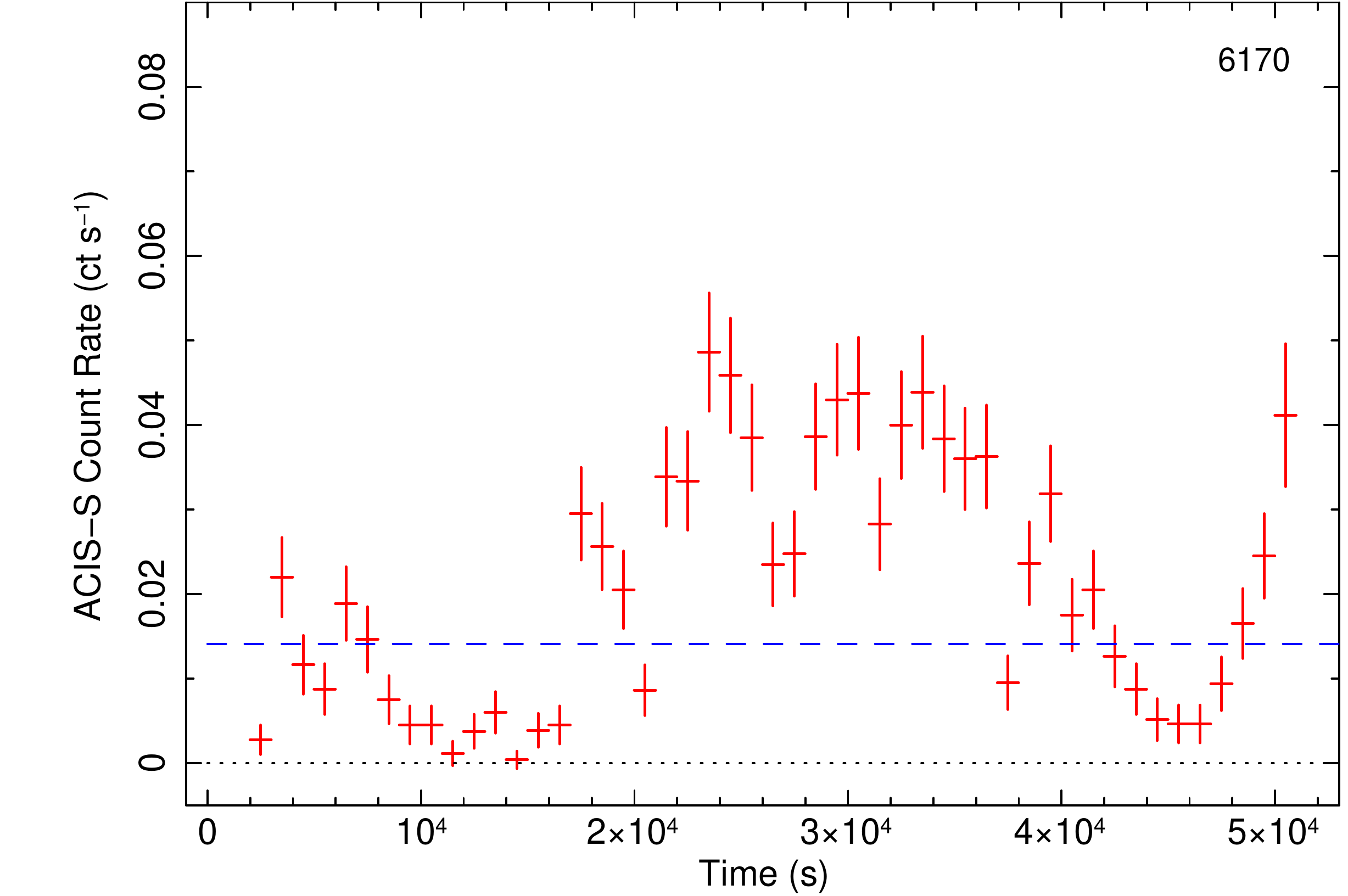}
        \end{subfigure}   \\
        \begin{subfigure}{0.48\textwidth}
            \vspace{1.3cm}
            \includegraphics[width=0.97\textwidth, angle=0]{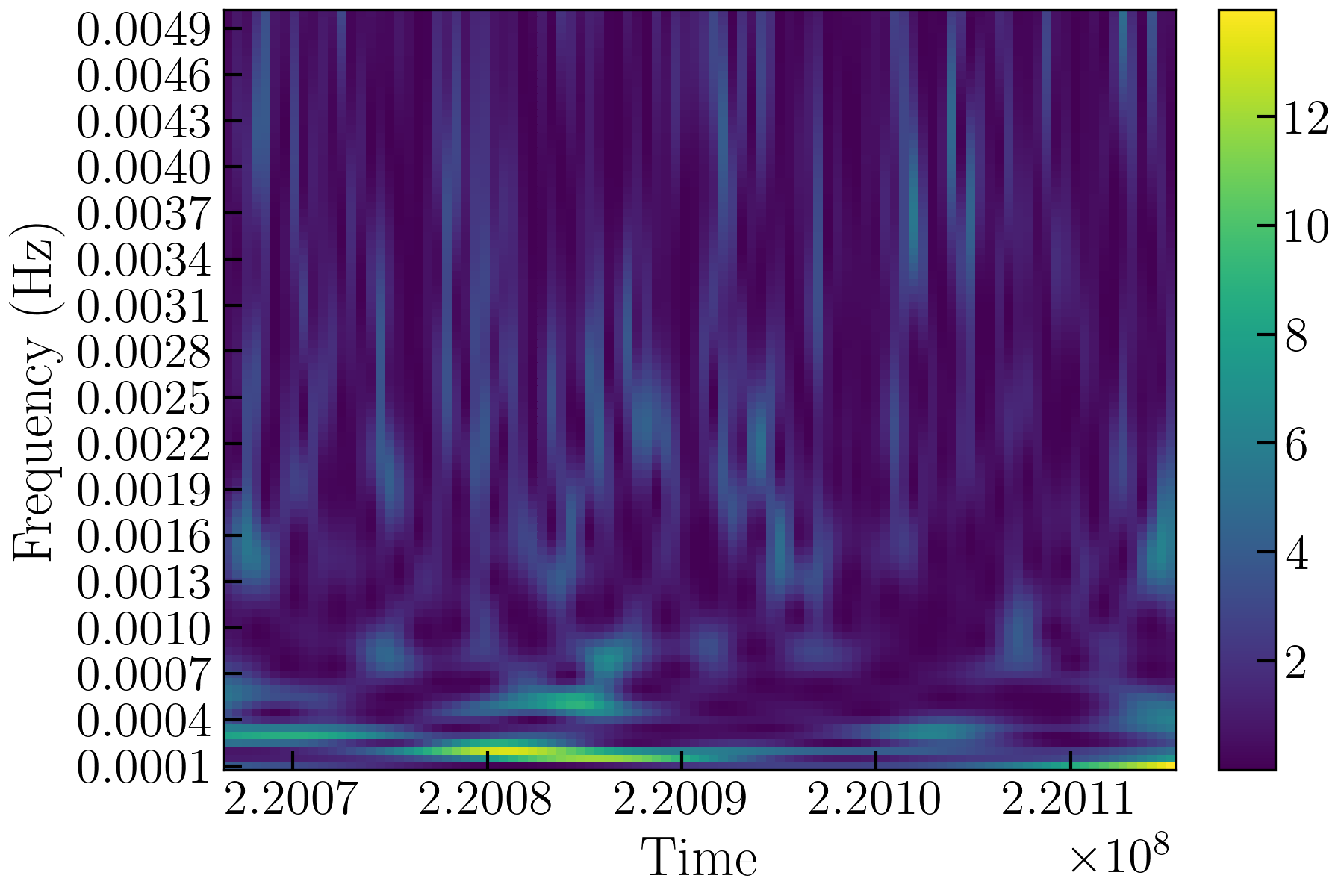}
        \end{subfigure} &
        \multirow{2}{*}{
            \begin{subfigure}{0.5\textwidth}
                \vspace{-1.5cm}
                \hspace{-0.5cm}
                \includegraphics[width=0.98\textwidth, angle=0]{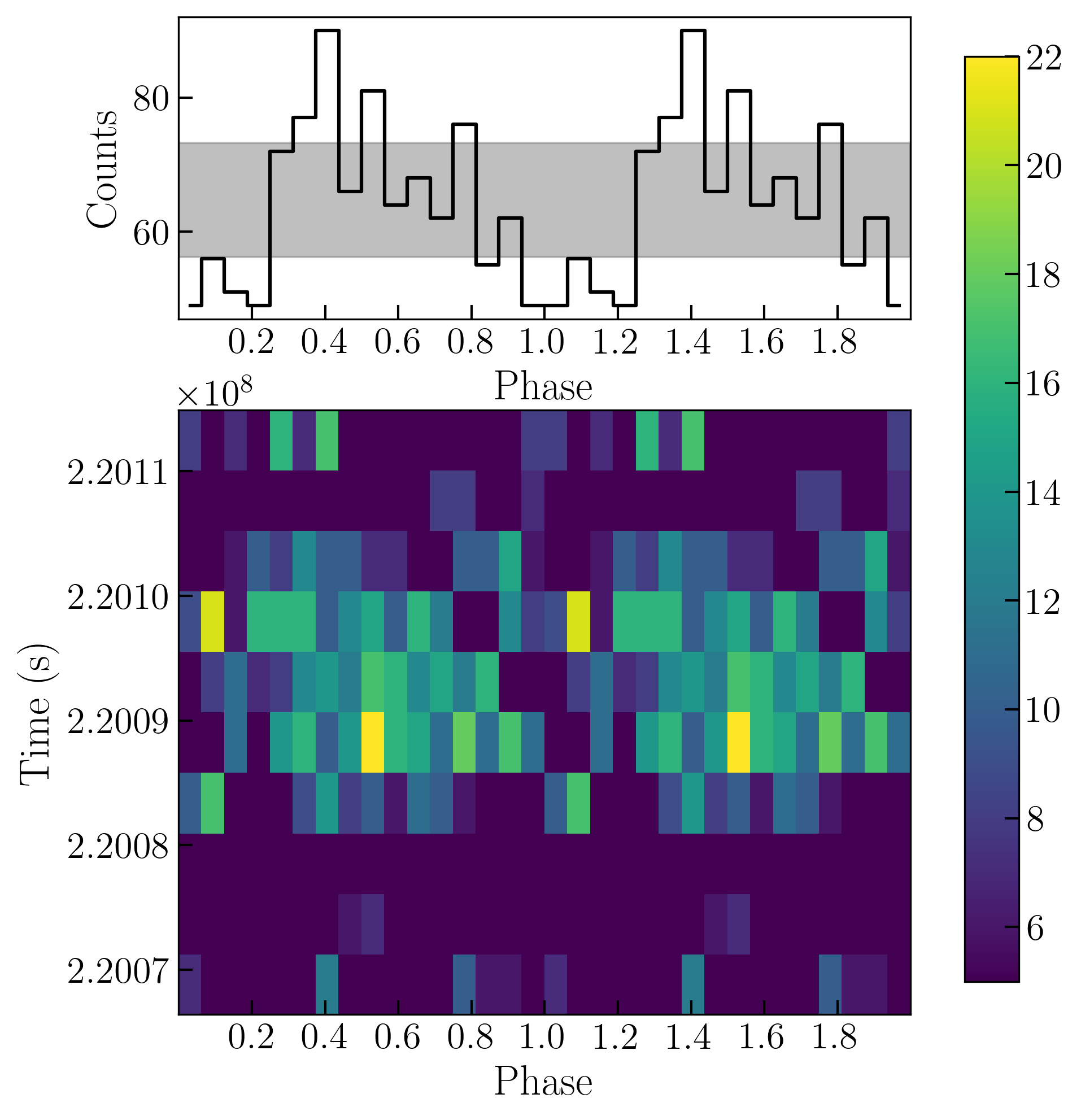}
            \end{subfigure}
        }                            \\
        \begin{subfigure}{0.48\textwidth}
            \includegraphics[width=0.95\textwidth, angle=0]{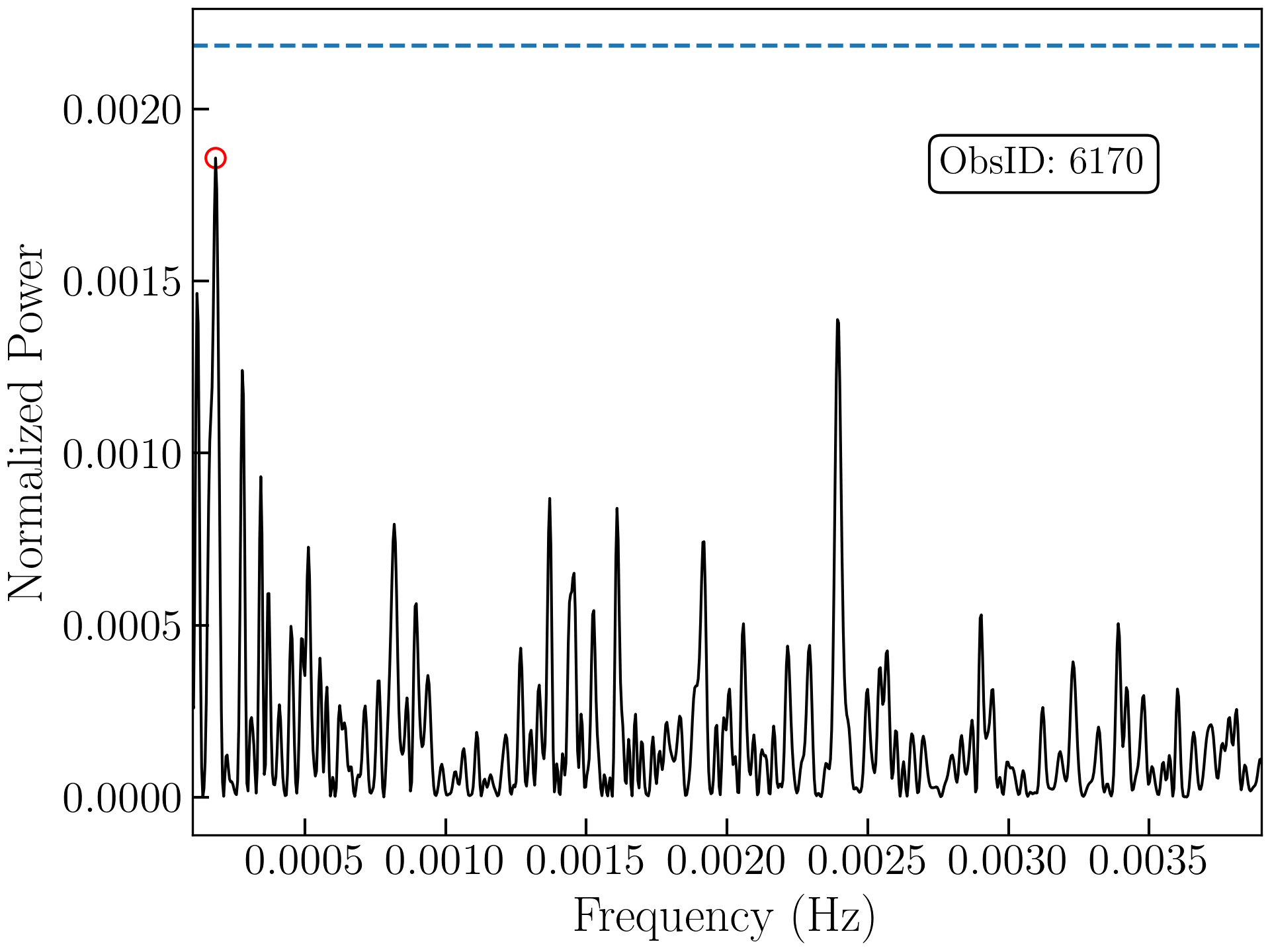}
        \end{subfigure}
                                   &
    \end{tabular}
    \caption{As in Figure A1, for {\it Chandra} ObsID 6170. The spectral model is an absorbed double-thermal model ({\it bbodyrad} $+$ {\it diskpbb} in {\sc xspec}). See Table 5 for the fit parameters.}
    \label{fig:6170_all}
\end{figure*}

\begin{figure*}
    \centering
    \begin{tabular}{cc}
        \begin{subfigure}{0.48\textwidth}
            \hspace{-0.3cm}
            \includegraphics[height=0.99\textwidth, angle=270]{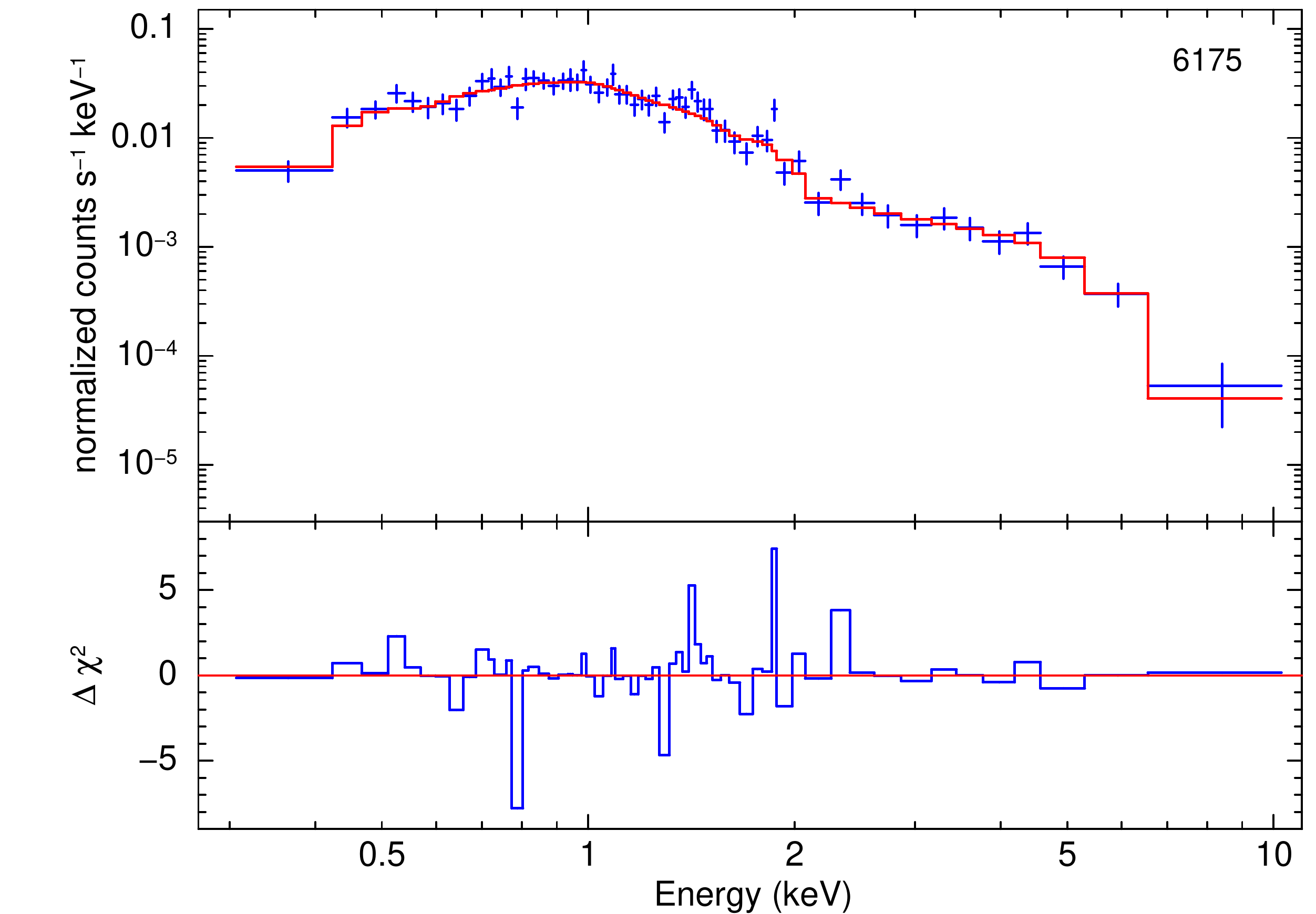}
        \end{subfigure} &
        \begin{subfigure}{0.48\textwidth}
            \hspace{-0.8cm}
            \includegraphics[height=0.99\textwidth, angle=270]{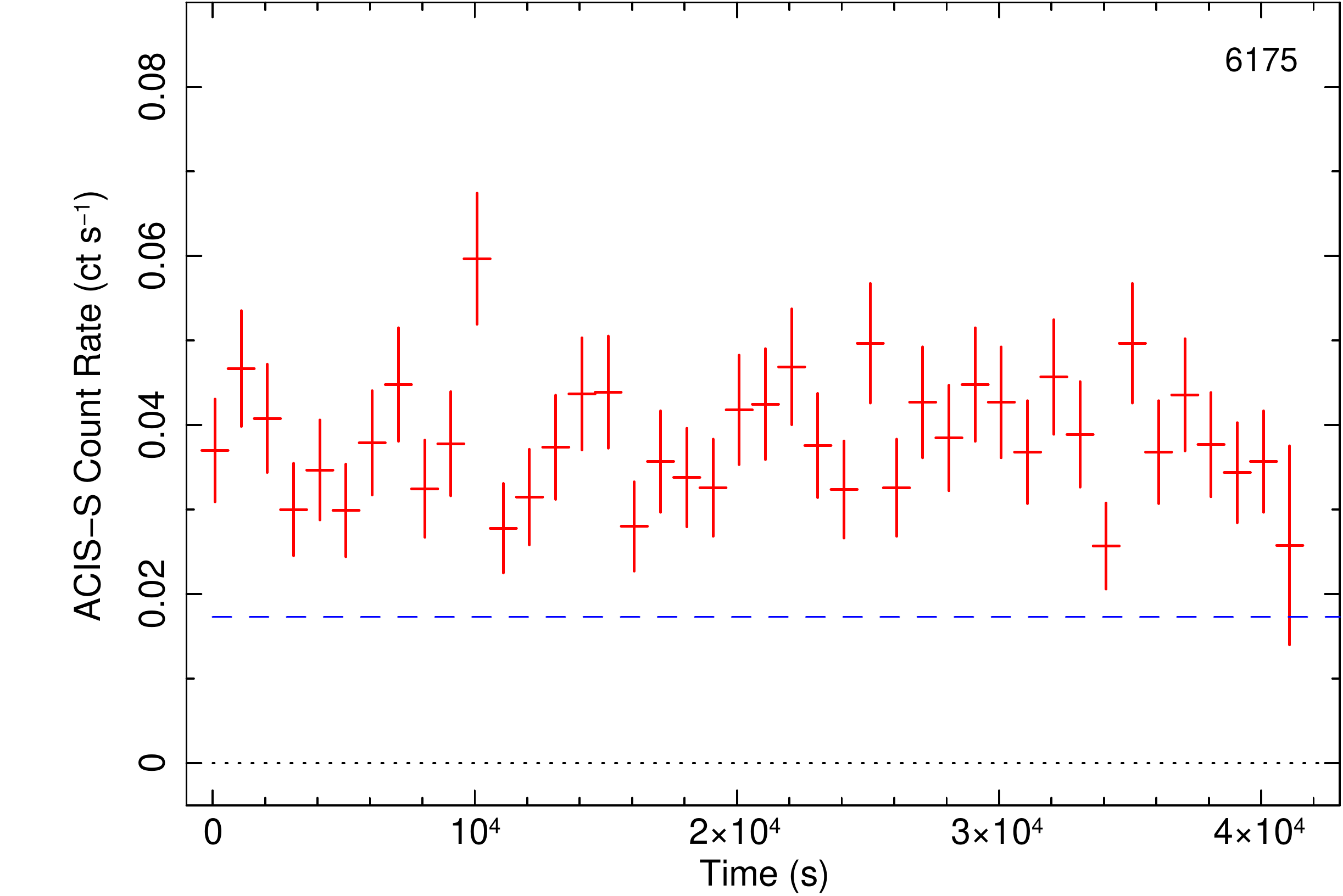}
        \end{subfigure}   \\
        \begin{subfigure}{0.48\textwidth}
            \vspace{1.3cm}
            \includegraphics[width=0.97\textwidth, angle=0]{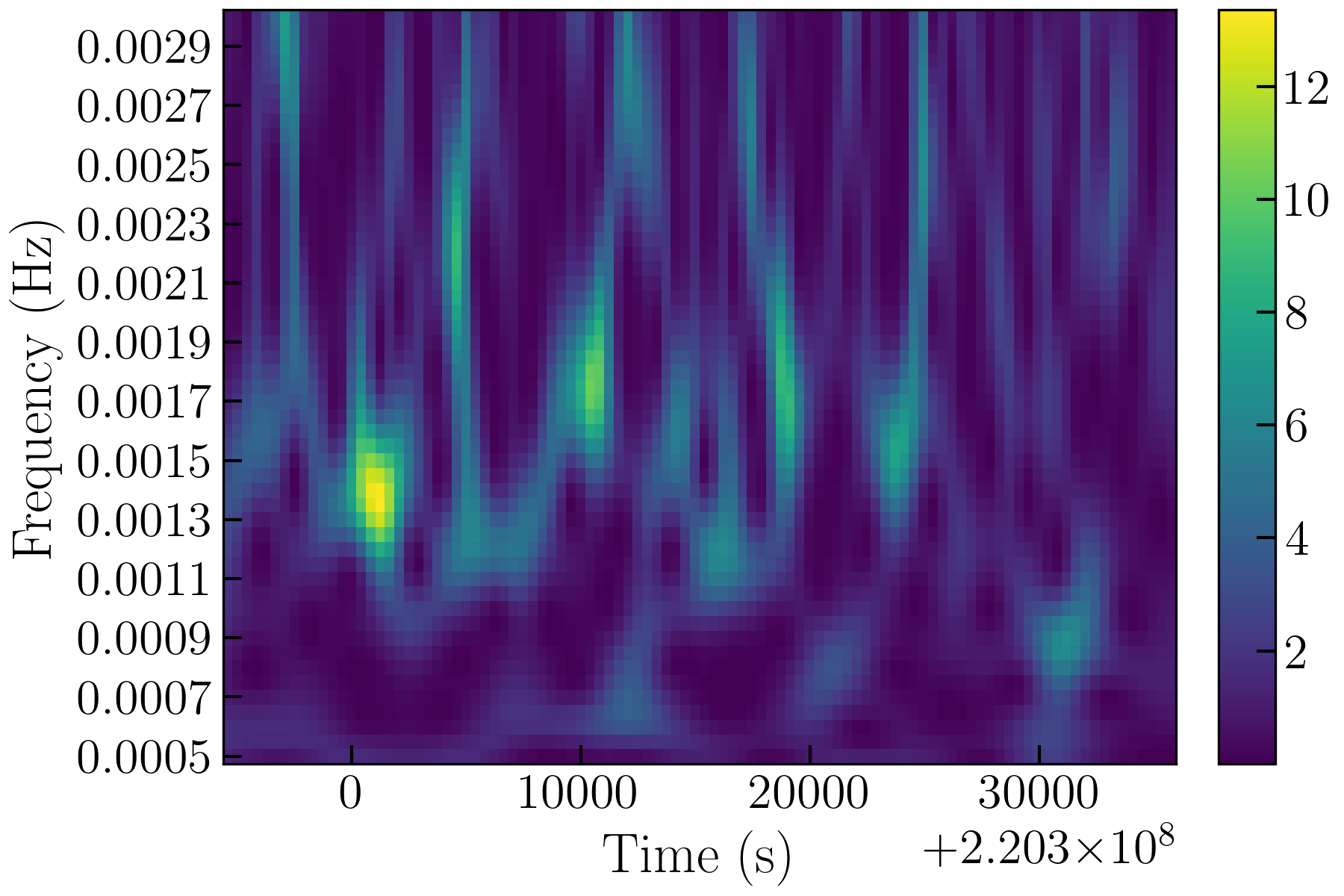}
        \end{subfigure} &
        \multirow{2}{*}{
            \begin{subfigure}{0.5\textwidth}
                \vspace{-1.5cm}
                \hspace{-0.5cm}
                \includegraphics[width=0.98\textwidth, angle=0]{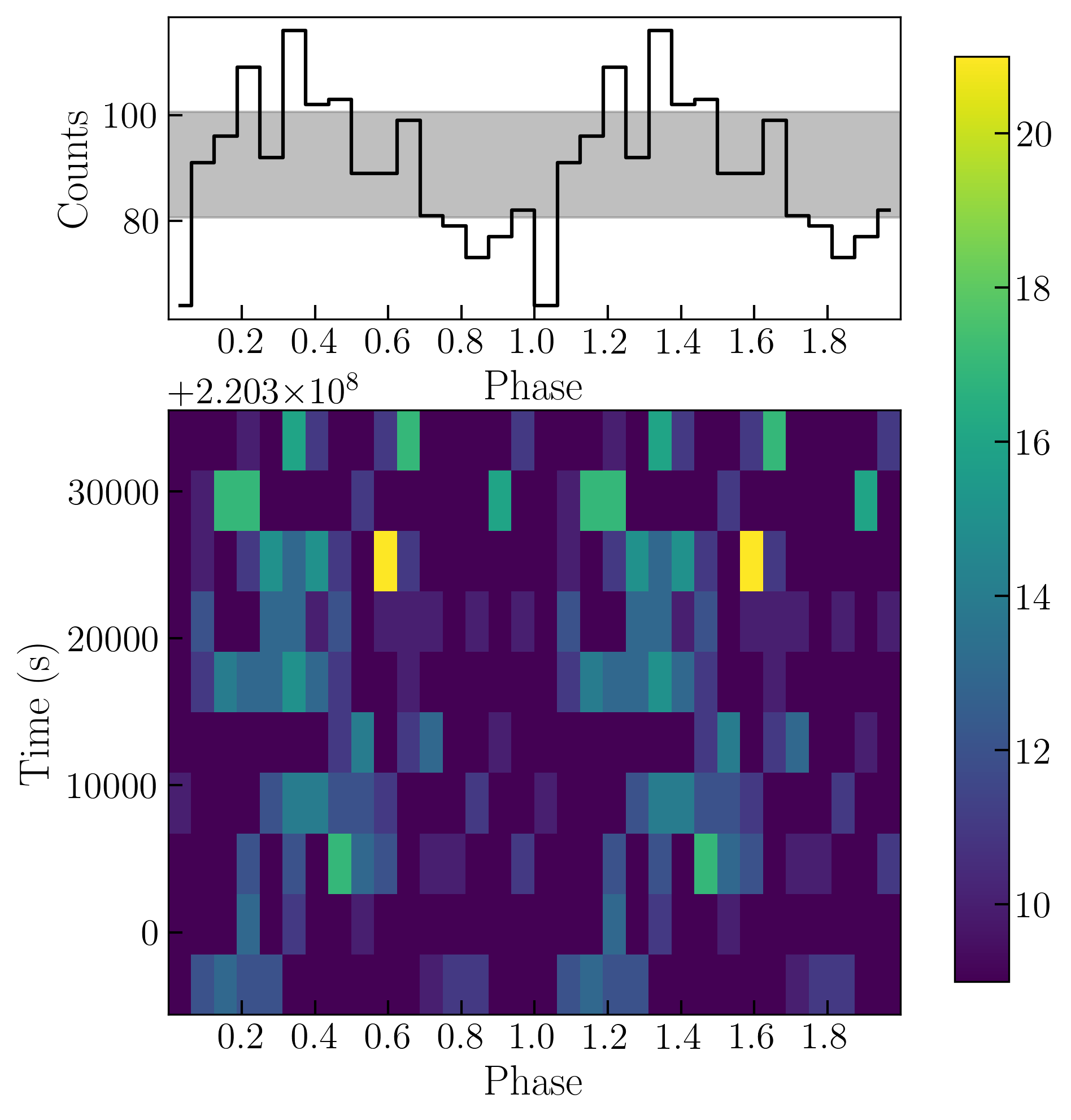}
            \end{subfigure}
        }                            \\
        \begin{subfigure}{0.48\textwidth}
            \includegraphics[width=0.95\textwidth, angle=0]{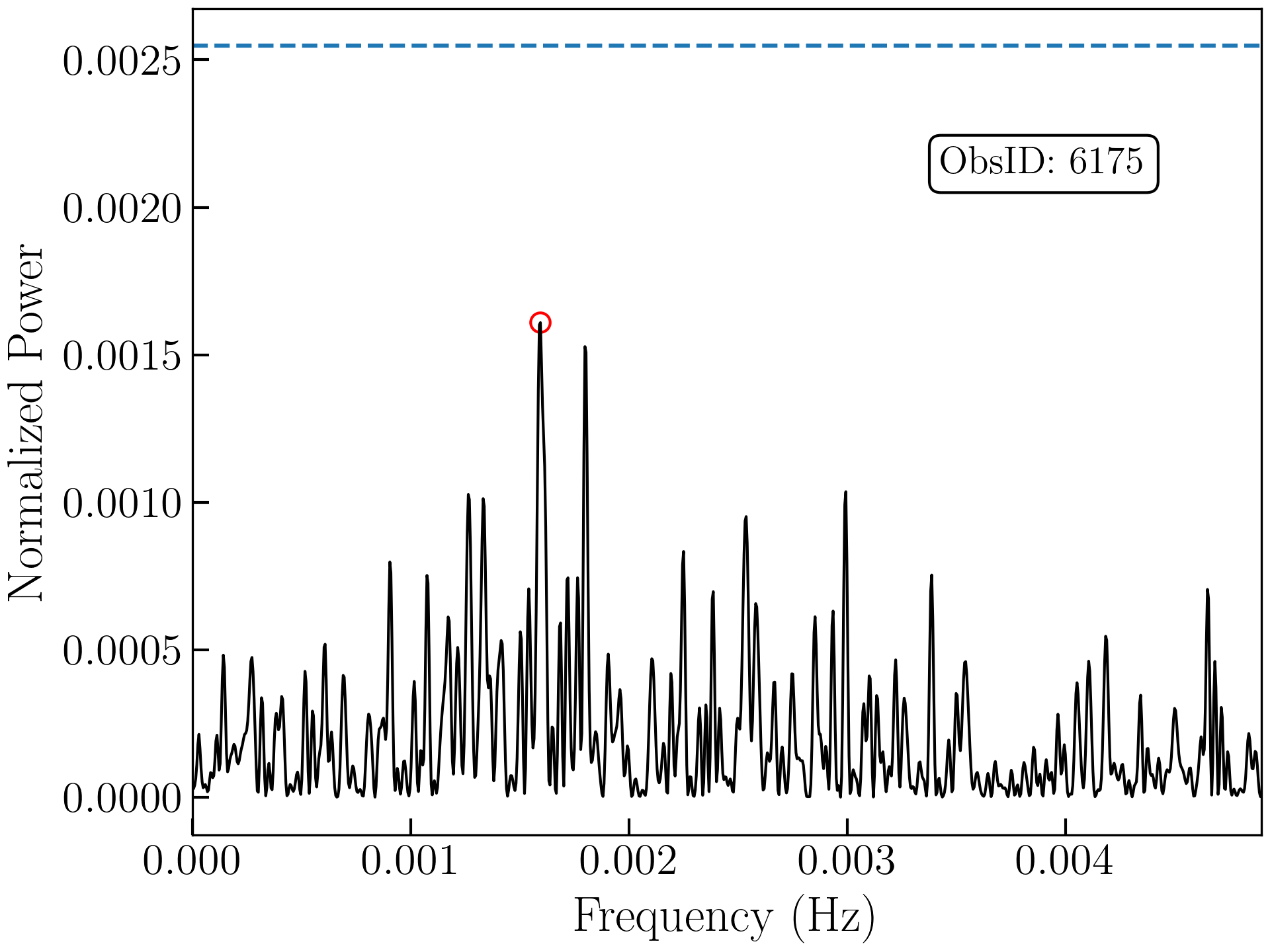}
        \end{subfigure}
                                   &
    \end{tabular}
    \caption{As in Figure A1, for {\it Chandra} ObsID 6175. The spectral model is an absorbed double-thermal model ({\it bbodyrad} $+$ {\it diskpbb} in {\sc xspec}). See Table 5 for the fit parameters.}
    \label{fig:6175_all}
\end{figure*}

\bsp	
\label{lastpage}
\end{document}